%% file: paper.tex
\pdfoutput=1
\documentclass[12pt]{article}

\usepackage{graphicx,psfrag,epsf,color}
\usepackage{amsmath,amssymb,amsfonts}
\usepackage{array}
\usepackage{cite}
\usepackage{multirow}
\usepackage{rotating}
\usepackage{slashed}
\usepackage{bm}

\usepackage[colorinlistoftodos]{todonotes} 
\presetkeys{todonotes}{fancyline, color=blue!15}{}

\renewcommand{\arraystretch}{1.25}

\newcounter{MBQ}

\newcounter{KUQ}

\newcounter{RSQ}

\newcommand{\grtsim}{\mbox{\raisebox{-3pt}{$\stackrel{>}{\sim}$}}}

\newcommand{\be}{\begin{equation}}
\newcommand{\ee}{\end{equation}}
\newcommand{\bea}{\begin{eqnarray}}
\newcommand{\eea}{\end{eqnarray}}
\newcommand{\bi}{\begin{itemize}}
\newcommand{\ei}{\end{itemize}}
\newcommand{\ben}{\begin{enumerate}}
\newcommand{\een}{\end{enumerate}}
\newcommand{\bt}{\begin{tabular}}
\newcommand{\et}{\end{tabular}}

\newcommand{\mchi}{m_\chi}

\newcommand{\mt}{m_t}
\newcommand{\mh}{m_H}
\newcommand{\mW}{m_W}
\newcommand{\mZ}{m_Z}
\newcommand{\mi}{m_i}
\newcommand{\mj}{m_j}

\newcommand{\xiW}{\ensuremath{\xi_W}}

\newcommand{\xiZ}{\ensuremath{\xi_Z}}

\newcommand{\xii}{\ensuremath{\xi_i}}
\newcommand{\xij}{\ensuremath{\xi_j}}

\newcommand{\xmW}{\ensuremath{\xi_W^{1/2}m_W}}
\newcommand{\xmZ}{\ensuremath{\xi_Z^{1/2} m_Z}}

\newcommand{\xmi}{\ensuremath{\xi_i^{1/2}m_i}}
\newcommand{\xmj}{\ensuremath{\xi_j^{1/2} m_j}}

\newcommand{\delW}{\ensuremath{\Delta_W}}
\newcommand{\delA}{\ensuremath{\Delta_A}}
\newcommand{\delZ}{\ensuremath{\Delta_Z}}

\newcommand{\deli}{\ensuremath{\Delta_i}}
\newcommand{\delj}{\ensuremath{\Delta_j}}

\numberwithin{equation}{section}

\begin{document}
\allowdisplaybreaks

\begin{titlepage}

\begin{flushright}
{\small
TUM-HEP-1281/20\\
CERN-TH-2020-144\\
September 01, 2020
}
\end{flushright}

\vskip1cm
\begin{center}
{\Large \bf Sommerfeld-corrected relic abundance of wino 
dark\\[0.2cm]  matter
with NLO electroweak potentials}
\end{center}

  \vspace{0.5cm}
\begin{center}
{\sc Martin~Beneke,$^{a}$ Robert~Szafron,$^{b}$} and {\sc Kai~Urban$^{a}$}
\\[6mm]
{\it ${}^a$Physik Department T31,\\
James-Franck-Stra\ss e~1, 
Technische Universit\"at M\"unchen,\\
D--85748 Garching, Germany}\\[6mm]
{\it ${}^b$Theoretical Physics Department, CERN, \\
    CH--1211 Geneva 23, Switzerland}
\\[0.3cm]
\end{center}

\vspace{0.6cm}
\begin{abstract}
\vskip0.2cm\noindent
Extending previous work, we calculate the electroweak 
potentials for all co-an\-ni\-hi\-la\-tion channels of wino 
dark matter at the one-loop order and obtain the wino relic 
abundance including the Sommerfeld effect at the  
next-to-leading order (NLO).  
\end{abstract}
\end{titlepage}



\section{Introduction}
\label{sec:introduction}

A weakly interacting massive particle (WIMP) is one of the best 
motivated dark matter (DM) candidates. Despite their simplicity, 
WIMP extensions of the standard model (SM) exhibit a rich and 
interesting phenomenology. Hisano et al. 
\cite{Hisano:2003ec,Hisano:2004ds,Hisano:2006nn} recognized that 
despite the fundamentally weak coupling, the annihilation 
cross section of DM particles with mass $m_\chi$ above a TeV is 
substantially enhanced by attractive forces that become effectively 
strong between slowly moving DM particles, the 
so-called Sommerfeld effect.
For minimal DM models and the minimal 
supersymmetric standard model, this non-perturbative effect is 
by now routinely included at leading order in the calculation of 
the forces in both, the prediction of signals of annihilating 
DM  in cosmic ray fluxes \cite{Hryczuk:2011vi,Fan:2013faa,Cohen:2013ama,Hryczuk:2014hpa,Beneke:2016jpw,Rinchiuso:2020skh}, and computations 
of relic abundance \cite{Cirelli:2007xd,Hryczuk:2010zi,Beneke:2014hja,Beneke:2016ync}.  The effect also appears in non-WIMP DM  
models, as long as there exists a suitable hierarchy between the 
DM mass and the mass of a light boson coupled to 
it \cite{ArkaniHamed:2008qn}. 

Other loop effects are often important as well.
In scenarios, where the DM particle originates from an electroweak multiplet, the Sommerfeld enhancement depends strongly on the mass differences \cite{Slatyer:2009vg} among the members of the multiplet. 
For the simplest multiplets, the mass splittings are known up to 
two loops \cite{Yamada:2009ve,Ibe:2012sx,McKay:2017xlc}. 
Further, in annihilation to final states with identified particles 
with electroweak charge, perturbation theory also breaks down, 
because the radiative corrections to the Born 
cross section are further enhanced by large logarithms of the ratio 
of the DM particle mass to the electroweak gauge boson 
mass \cite{Bauer:2014ula, Baumgart:2014vma, Ovanesyan:2014fwa}. Consequently, the fixed-order computations must be complemented by all-order resummation of the dominant logarithmic corrections. This has been 
achieved for high-energy cosmic photons with the help of  
soft-collinear effective field theory 
for the electroweak fermionic triplet (``wino'') \cite{Ovanesyan:2016vkk,Baumgart:2017nsr,Baumgart:2018yed,Beneke:2018ssm,Beneke:2019vhz} and doublet (``Higgsino'') DM model \cite{Beneke:2019gtg}. 
Overall, the most advanced computations of the high-energy photon 
yields from DM annihilation reach one to few percent accuracy, 
depending on the DM mass and model. This motivates a closer 
scrutiny of the calculation of the Sommerfeld effect, which 
is usually calculatedwith the leading order (LO) potential generated 
by electroweak gauge boson and photon exchange. 

In the previous letter \cite{Beneke:2019qaa}, we discussed, for the first time, the Sommerfeld effect for wino DM with one-loop, next-to-leading order (NLO) corrections to the electroweak Yukawa potential.
More precisely, we considered the pair annihilation of the DM 
particle $\chi^0$ into $\gamma +X$, and found the NLO potential to 
give a sizeable correction to the LO Sommerfeld effect, 
including a shift of the Sommerfeld resonance positions by 
about 6$\%$. For state-of-the-art theoretical predictions of 
high-energy photon signals for indirect detection experiments, 
such as the Cherenkov Telescope Array, the NLO computation of the 
Sommerfeld enhancement is therefore indispensable. 
The present paper serves two purposes. First, we extend the 
one-loop computation to the potentials in the co-annihilation 
channels $\chi^0 \chi^\pm$, $\chi^\pm\chi^\pm$,
which were not given in \cite{Beneke:2019qaa}, and perform the 
first computation of a DM relic abundance with the NLO Sommerfeld 
effect. Second, we provide analytic results for the potentials 
in momentum space, and the technical details of the NLO 
computation. We provide all the one-loop integrals relevant for the evaluation of the NLO correction and discuss the properties of the 
NLO potential functions.  

The outline of the paper is as follows. In Section~\ref{sec:EFTsetup}, we
discuss the construction of the EFT for the WIMPs, introduce the
power-counting, and review the calculation of the tree-level potentials.
Subsequently, in Section~\ref{sec:NLOpot} we discuss the computation of the
one-loop correction to the potential for the wino model in all channels
including details on renormalization schemes, asymptotic behaviours, and the
parameter dependence. In Section~\ref{sec:relic}, we calculate the relic abundance  and analyze the size of the correction. We conclude in Section~\ref{sec:conclusions}. In a
series of appendices, we collect additional technical details on the one-loop
calculations in Feynman and general covariant $R_\xi$-gauge, 
Fourier transforms between momentum and position space, 
and relevant expressions for the asymptotic behaviours.


\section{EFT of non-relativistic WIMPs}
\label{sec:EFTsetup}

The low-energy effective field theory (EFT) of non-relativistic 
WIMPs was constructed in 
\cite{Beneke:2012tg,Hellmann:2013jxa,Beneke:2014gja} in 
analogy with the respective non-relativistic EFTs of QED and QCD 
\cite{Pineda:1997bj,Beneke:1998jj,Beneke:1999qg,Brambilla:1999xf} 
for onium systems \cite{Bodwin:1994jh}. In this
section, we review the structure of the potential non-relativistic 
effective theory for wino DM, establish a
consistent power-counting, and identify the leading corrections to the
potential. The essence does not depend on the particular 
wino DM model and is applicable to general multiplets and cases that include
hypercharge. Our starting point is the SM supplemented with
the wino Lagrangian
\begin{align}
    \mathcal{L}_{\rm DM} &= \frac{1}{2} \, \overline{\chi}(x) \left(i
    \slashed{D} - m_\chi\right) \chi(x)\,,
\end{align}
where $\chi$ denotes an SU(2)-triplet of Majorana fermions and $D_\mu$ is the
covariant derivative, $D_\mu = \partial_\mu - i g_2 W_\mu^a T^a$. The modes
relevant to construct the non-relativistic WIMP EFT are 
(i) hard  ($k^0 \sim \mchi,\, \mathbf{k} \sim \mchi$),
(ii) soft  ($k^0 \sim \mathbf{k} \sim \mW $),
(iii) potential  ($k^0 \sim \mW^2/\mchi ,\, \mathbf{k} \sim \mW$),
and (iv) ultrasoft  ($k^0 \sim \mathbf{k} \sim \mW^2/\mchi $).
We introduce the power-counting parameter in terms of the $W$-boson
mass, and assume that the DM  mass is such that $\mW/\mchi \sim v \sim
\alpha_2$, where $v$ is the small velocity of the DM particles. A 
different relative size of $v$ compared to $\alpha_2$ or 
$\mW/\mchi$ does not affect the construction of the non-relativistic 
EFT, but leads to different regimes (e.g., Coulombic if  
$\alpha_2 \sim v \gg \mW / \mchi$).

In the first step, we match to the 
non-relativistic EFT, i.e., we integrate out the hard modes. This step is performed in the
unbroken phase where the electroweak symmetry is still manifest. The theory resembles NRQCD with an SU(2) gauge group, hence, the 
Feynman rules are known \cite{Beneke:2013jia} upon appropriately adjusting the group
factors.  The non-relativistic 
Lagrangian terms that we need in this paper are simply given by
\begin{align}
    \mathcal{L}_{\rm NRDM} &= \chi^\dagger(x) \left(i D^0 +
    \frac{\mathbf{D}^2}{2 \mchi}\right) \chi(x)\,,
    \label{eq:LNRDM}
\end{align}
where only the soft, potential, and ultrasoft modes are the dynamical degrees of freedom. 

Next, electroweak symmetry breaking is implemented. The
resulting Lagrangian is of the same form as above; however, the relevant
degrees of freedom change from weak eigenstates to mass eigenstates. The fields
$\chi_a$, where $a=1,2,3$ in the unbroken theory, are combined to mass
eigenstates with $\chi^\pm =(\chi_1 \mp i\chi_2)/\sqrt{2}$ and $\chi_3 =
\chi^0$. The electrically charged states $\chi^\pm$ acquire a radiatively induced mass
splitting $\delta \mchi = 164.1\,{\rm MeV}$ \cite{Ibe:2012sx}
with respect to the neutral $\chi^0$ state.\footnote{In the 
non-relativistic theory the mass correction is obtained from the 
soft $Z$- and $W$-boson correction to the DM field propagator, see Appendix~\ref{app:heavyDM}.}  

Finally, we integrate out the soft fields and potential gauge bosons. We obtain
non-local (in space) four-fermion operators whose matching coefficients coincide at tree-level with the classical static potential. Loops of soft fields induce quantum corrections to the
DM potential. We are left with a theory of potential fermions and ultrasoft
gauge bosons, which is described by the potential non-relativistic 
DM (PNRDM) Lagrangian \cite{Beneke:2012tg}
\begin{align}
\label{eq:LPNRDM}
\mathcal{L}_{\rm PNRDM} &= \sum_{i= \pm,0} \chi_{vi}^\dagger(x) \left(i
D^0_i(t,\mathbf{0}) - \delta m_i+ \frac{\bm{\partial}^2}{2\mchi} - e_i\, e\,
\mathbf{x} \cdot \mathbf{E}(t,\mathbf{0})\right)\chi_{vi}(x) \nonumber\\ 
                        &-\,\sum_{\{i,j\},\{k,l\}} \int d^3\mathbf{r}\,
V_{\{ij\}\{kl\}}(r)\, \chi_{vk}^\dagger(t,\mathbf{x})
\chi_{vl}^\dagger(t,\mathbf{x}+\mathbf{r}) \chi_{vi}(t,\mathbf{x})
\chi_{vj}(t,\mathbf{x}+\mathbf{r}) \, ,
\end{align}
where $e_i$ is the electric charge of fermion $i$ in units of 
the positron charge $e$. The electromagnetic covariant derivative is given by $i D^0_i = i
\partial^0 + e_i e A^0$.  In our convention, all DM fields are particle fields (cf.
\cite{Beneke:2012tg}). Structurally the above Lagrangian is the same as for  QED or QCD. The phenomenology, however, is different as the
gauge symmetry is broken. For example, the potentials can be off-diagonal for
the mass eigenstates, unlike in QCD, where
quarkonium states decompose into gauge eigenstates, which 
are simultaneously 
mass eigenstates (singlet/octet), and where the potentials are 
diagonal in the space of quarkonium mass eigenstates.  

The ultrasoft fields in the gauge-covariant derivative $D^0_i(t,\mathbf{0})$
and the electric field are multipole-expanded and only include the photon
field, as, in the broken theory, the $W$- and $Z$-bosons are too heavy to have
ultrasoft scaling.\footnote{For very large DM masses $\mchi \sim
\mathcal{O}(100\, {\rm TeV})$ there is a regime where $m_W/\mchi \sim
\alpha_2^2$. For this regime, the $W$- and $Z$-boson can contribute to the
ultrasoft radiation. However, in such a case, the potentials would be
effectively Coulombic unbroken-theory potentials, and the effective theory
would be very similar to PNRQCD for scattering states, 
as $\alpha_2 \sim v \gg
\mW/\mchi$.} The term $\mathbf{x} \cdot \mathbf{E}$ originates from the 
ultrasoft interaction of charged fermions with the photon field and the
additional coupling of the photon to the $W$-boson Yukawa potential after
application of equation-of-motion identities \cite{Beneke:1999zr}. It makes the
unbroken ultrasoft electromagnetic gauge symmetry manifest, and, relative to
the leading kinetic term, it is suppressed by $(\mW / \mchi)^{3/2}$.  Hence, it
will play no role in the determination of the NLO correction.\footnote{The leading ultrasoft correction comes from $A^0$ inside the
covariant derivative, which is only suppressed by $(\mW/\mchi)^{1/2}$ relative to the leading terms. For $S$-wave annihilation, the 
ultrasoft photons couple only to the total electric charge $Q$ of 
the wino two-particle state. Although $Q$ is non-vanishing for the 
$\chi^0\chi^\pm$ and $\chi^\pm\chi^\pm$ states, there is 
nevertheless no contribution to the total annihilation cross 
section, which is related to the forward-scattering amplitude, 
that is, the matrix element of a local four-fermion operator 
whose net charge vanishes. We checked this statement by performing 
an explicit one-loop computation for
the wino model.} 

The ultrasoft interactions, however, are relevant in determining 
the DM bound-state formation rates, which can modify indirect detection signals and the DM relic abundance. The effect is especially 
significant for large multiplets. However, 
for wino DM in the few TeV mass range it is important 
neither for indirect
detection \cite{Asadi:2016ybp} nor the relic abundance 
\cite{Mitridate:2017izz}.

The mass difference term $\delta m_i \sim \alpha_2 m_W \sim \mW^2/\mchi$ is of the same order as the kinetic terms by power-counting and
therefore contributes at leading order, even though  
$\delta m_i$ is a one-loop effect. Hence, to obtain NLO accuracy 
of the calculation, we include the two-loop result for the mass 
splitting \cite{Ibe:2012sx}.

The crucial new ingredient in obtaining NLO accuracy is the potential
term in \eqref{eq:LPNRDM}. NLO corrections to this term
could be twofold. First, from potentials that are more singular than
$1/r$, but the structure of the non-relativistic EFT vertices implies that such potentials can appear only from the next-to-next-to-leading order in the power-counting (similar to QED/QCD). Second, from the one-loop 
correction to the tree-level Coulomb and Yukawa potentials, which 
are the subject of this paper and \cite{Beneke:2019qaa}. 

For completeness, we recall that the LO potential is obtained from tree diagrams involving the exchange of an electroweak gauge bosons between two wino particles. In the neutral two-particle sector we
find (in momentum space)
\begin{align}
  \tilde{V}^{Q=0}(\mathbf{k}) &= i \, T^{\chi \chi \to \chi \chi}(\mathbf{k}) =
    \begin{pmatrix}
        0 & & -\frac{4 \pi \alpha_2}{\mathbf{k}^2 + m_W^2} & & -\frac{4 \pi
        \alpha_2}{\mathbf{k}^2 + m_W^2} \\
        -\frac{4 \pi \alpha_2}{\mathbf{k}^2 + m_W^2} & & -\frac{4 \pi
        \alpha}{\mathbf{k}^2}-\frac{4 \pi \alpha_2 c_W^2}{\mathbf{k}^2 + m_Z^2}
        & & 0 \\
        -\frac{4 \pi \alpha_2}{\mathbf{k}^2 + m_W^2} & & 0 & & -\frac{4 \pi
        \alpha}{\mathbf{k}^2}-\frac{4 \pi \alpha_2 c_W^2}{\mathbf{k}^2 + m_Z^2}
    \end{pmatrix} \, ,
    \label{eq:momwino}
\end{align}
where the entries refer to the $\chi^0 \chi^0,\, \chi^+ \chi^-, \, \chi^-
\chi^+$ states, respectively, and $T^{\chi \chi \to \chi \chi}$ denotes the
T-matrix in the specific scattering channel. In the single-charged and
double-charged sectors, one has
\begin{align}
  \tilde{V}^{Q=1}(\mathbf{k})=\begin{pmatrix}
    0 & \frac{4 \pi \alpha_2}{\mathbf{k}^2 + m_W^2} \\
    \frac{4 \pi \alpha_2}{\mathbf{k}^2 + m_W^2} & 0
  \end{pmatrix} \quad {\rm and} \quad \tilde{V}^{Q=2}(\mathbf{k}) = \frac{4 \pi
  \alpha}{\mathbf{k}^2} + \frac{4 \pi \alpha_2 c_W^2}{\mathbf{k}^2 + m_Z^2} \,
  ,
  \label{eq:momwino1}
\end{align}
where the entries refer to $\chi^0 \chi^+,\, \chi^+ \chi^0$ and $\chi^+
\chi^+$, respectively. The same expressions hold for the $Q=-1$ and $Q=-2$ charge sectors.
For solving the Schr\"{o}dinger equation, one transforms the
potentials to coordinate space, using
\begin{align}
    V(r =|\mathbf{x}|) &= \int\frac{d^3 \mathbf{k}}{(2 \pi)^3} \, e^{i
    \mathbf{k} \cdot \mathbf{x}}\, \tilde{V}(\mathbf{k}) \, .
    \label{eq:FTpotential}
\end{align}
At tree level we need the Fourier transform
\begin{align}
    \int \frac{d^3 \mathbf{k}}{(2 \pi)^3}\, \frac{e^{i \mathbf{k} \cdot
    \mathbf{x}}}{\mathbf{k}^2 + m^2} &= \frac{e^{-m\, r}}{4 \pi r} \, .
\end{align}
Therefore, at leading order we encounter Coulomb and Yukawa potentials only.

In the above basis, which we refer to as method-I, following
\cite{Beneke:2014gja}, the spin and angular momentum configuration of the
initial states do not play a role. It is more conventional \cite{Hisano:2006nn,Beneke:2014gja} to decompose the two-particle states 
into partial-wave configurations
${}^{2S+1}L_J$ of definite
total angular momentum $L$ and spin $S$. The resulting potential is referred to as method-II and in the
neutral sector reads (in coordinate space)
\begin{align}
\label{eq:singletLO}
  V^{Q=0}(r)({}^1S_0) &= \begin{pmatrix} 0 & & -\sqrt{2} \alpha_2 \frac{e^{-m_W
    r}}{r} \\
        - \sqrt{2} \alpha_2 \frac{e^{-m_W r}}{r} & & -\frac{\alpha}{r} -
        \alpha_2 c_W^2 \frac{e^{-m_Z r}}{r} 
    \end{pmatrix}\,, \\[0.2cm]
    V^{Q=0}(r)({}^3S_1) &= \begin{pmatrix} 0 & & 0 \\
        0 & & -\frac{\alpha}{r} - \alpha_2 c_W^2 \frac{e^{-m_Z r}}{r} 
    \end{pmatrix} \, .
\label{eq:tripletLO}
\end{align}
Similar decompositions hold for all other two-particle states and the $P$-wave
potentials. The decomposition into the method-II two-particle states removes
redundancies among the two-particle states. It automatically implements the
symmetry properties of the underlying states, e.g., that the identical Majorana
particles $\chi^0 \chi^0$ cannot exist in a ${}^3S_1$ spin configuration. A
detailed discussion of the correspondence between method-I and method-II can be
found in \cite{Beneke:2014gja}.

The calculation of the one-loop correction to the potential proceeds 
in an analogous fashion. First, the momentum-space potential is calculated 
in the form of~\eqref{eq:momwino}.
Then the Fourier transformation to position space is
performed. However, more complicated functions will lead to a
wider variety of potentials at the one-loop order. The 
transition from method-I to method-II
follows the same rules as the tree-level potential.


\section{The NLO potential}
\label{sec:NLOpot}

In this section, we turn to the calculation of the NLO potentials. 
We describe in detail the case of the $\chi^+ \chi^- \to \chi^+ 
\chi^-$ scattering channel, for which all possible diagram topologies contribute, and the effects of EWSB play an important role 
(particularly through $\gamma- Z$-mixing).  The results
for the other channels are discussed subsequently, and the differences are
highlighted. In doing so, we provide technical details and analytic expressions
not supplied in \cite{Beneke:2019qaa}. We further discuss the renormalization
scheme, the gauge invariance of the results,  
the large- and small-distance asymptotic behaviour, and the 
top-quark mass dependence of the NLO correction.

\subsection{The $\chi^+ \chi^- \to \chi^+ \chi^-$ channel} 

\begin{figure}[t]
    \centering
    \includegraphics[width=0.9\textwidth]{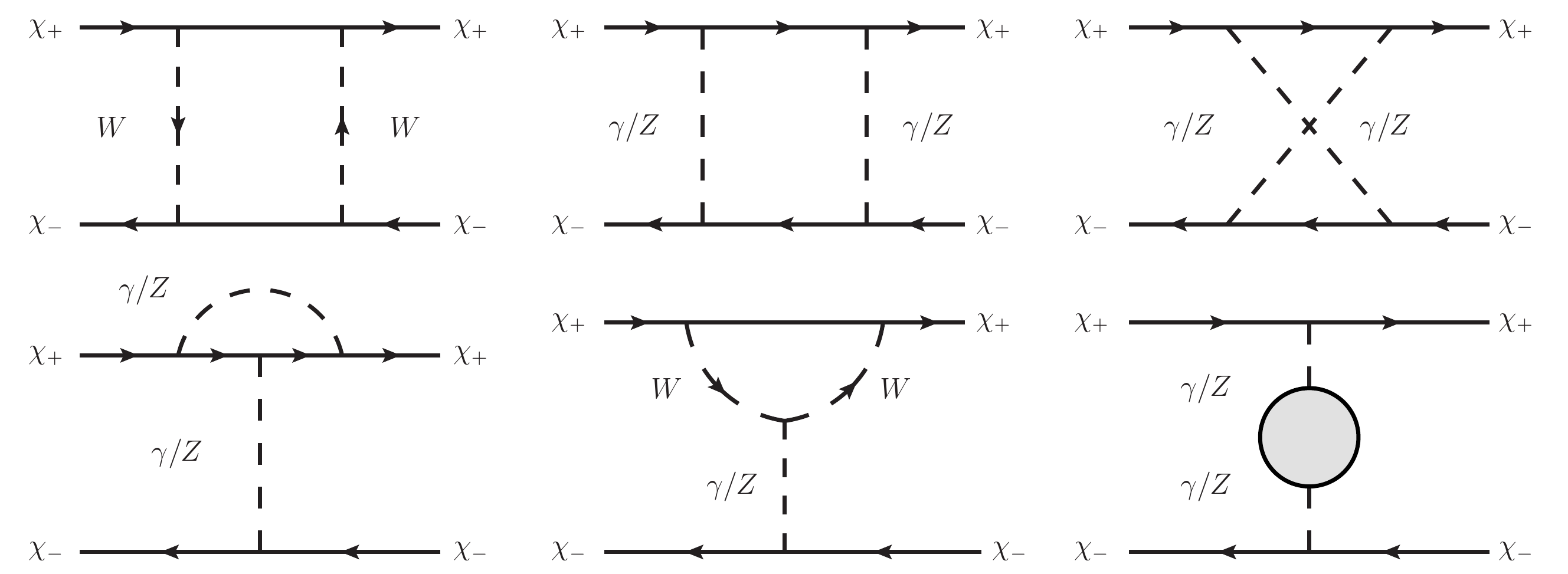}
    \caption{Diagram topologies for the $\chi^+ \chi^- \to \chi^+ \chi^-$
    channel excluding field renormalization, counterterm and tadpole
  topologies. Arrows on the lines indicate electric charge flow.}
  \label{fig:feynpmpm}
\end{figure}

The one-loop correction to the potential in
the $\chi^+ \chi^- \to \chi^+ \chi^-$ scattering channel originates 
from the non-relativistic EFT diagrams depicted in 
Fig.~\ref{fig:feynpmpm}. The relevant loop momentum for 
matching the potentials has the soft scaling in the framework 
of the threshold expansion \cite{Beneke:1997zp}. This amounts 
to replacing the DM propagators by static ones, as the 
soft momenta throw the non-relativistic propagators off-shell. In 
the calculation, it is important to ensure that the pinch poles 
at $k^0 = \pm i \varepsilon$ are not picked up, as they are 
reproduced in the EFT by iterations of the LO potentials and 
belong to the potential momentum region. The
results for the above diagrams are given in Appendix~\ref{app:loops}. 

The box and crossed box diagram involving photons and $Z$-bosons cancel each
other, similarly to the photon boxes in QED.  However, there is a box
contribution from $W$-boson exchange that has no crossed box counterpart. Furthermore,
the self-energy diagrams are not diagonal but mix the photon and the $Z$-boson. The
self-energies involve all SM particles and were evaluated in general covariant
gauge using \texttt{FeynArts} \cite{Hahn:2000kx}, \texttt{FORMCalc}
\cite{Hahn:2000kx} and \texttt{Package-X} \cite{Patel:2016fam} and checked
against the Feynman-gauge results (excluding tadpoles) from
\cite{Denner:1991kt}. Lastly, let us comment on the second diagram in
the second row of Fig.~\ref{fig:feynpmpm}.  This diagram vanishes in Feynman
gauge, as the vertices on the external fermion lines project out the
zero-component of the gauge-boson propagator. However, in general covariant
$R_\xi$-gauge this diagram is non-zero and required to obtain a gauge-parameter
independent result.  Before we assemble the full correction, we first discuss
the on-shell renormalization scheme that we used in our computation.

\subsubsection{The renormalization scheme}
\label{sec:onshell}
We choose to renormalize the ultraviolet (UV) divergences in the 
on-shell scheme following \cite{Denner:1991kt}.\footnote{In the electroweak literature,
several schemes are referred to as ``on-shell" scheme, which differ
in the choice of input parameters and have different applications. In
high-energy applications, e.g., at colliders or for the DM potential where the
energy scale relevant is $m_W$, it is customary to renormalize $\alpha(m_Z)$ to
avoid large logarithms of light fermion masses over the electroweak scale.}  As the
matching between the non-relativistic and the potential non-relativistic theory
is performed at the scale $m_Z$ it is natural
to choose the input parameters at this scale. As the input parameter set, we use
\begin{align}
    \alpha_{\rm os}(m_Z), \, m_W, \, m_Z, \, m_t, \, m_H
\end{align}
and set the CKM-matrix to the unit matrix.
The counterterms in this scheme are
\begin{equation}
\delta m_W^2 = {\rm Re}\, \Sigma_T^{WW}(m_W^2)\,,
\quad\qquad
\delta m_Z^2 = {\rm Re}\, \Sigma_T^{ZZ}(m_Z^2)
\label{eq:counterterms}
\end{equation}
for the gauge-boson masses and 
\begin{equation}
2 \delta Z_e \big|_{\alpha_{\text{os}}(m_Z)} = 
\frac{\partial \Sigma_T^{AA}(k^2)}{\partial k^2}\big|_{W,f=t}^{k^2=0} 
-\frac{2 s_W}{c_W} \frac{\Sigma_T^{AZ}(0)}{m_Z^2}
+ \text{Re}\frac{\left.\Sigma_T^{AA}(m_Z^2)\right|_{f\neq t}}{m_Z^2} 
\label{eq:DeltaZe}
\end{equation}
for the electromagnetic coupling at the $Z$-resonance. The 
self-energies on the right-hand side are evaluated in dimensional 
regularization with all fermions other than the top quark 
taken to be massless, and scaleless integrals are dropped.
For later convenience it is also useful to define the counterterms 
for the Weinberg angle 
\begin{eqnarray}
  s_W^{(0)} &=& s_W + \delta s_W\,, \qquad c_W^{(0)} = c_W + \delta c_W\,, \nonumber \\
  \frac{\delta c_W}{c_W} &=& \frac{1}{2} \left(\frac{\delta m_W^2}{m_W^2} -
  \frac{\delta m_Z^2}{m_Z^2}\right) = \frac{1}{2} \text{Re}
  \left(\frac{\Sigma_T^{WW}(m_W^2)}{m_W^2}-
  \frac{\Sigma_T^{ZZ}(m_Z^2)}{m_Z^2}\right)\,, \nonumber \\
  \frac{\delta s_W}{s_W} &=& - \frac{c_W^2}{s_W^2} \frac{\delta c_W}{c_W} \,.
  \label{eq:deltasW}
\end{eqnarray}
Equipped with these definitions, we can now assemble the potential
correction. Before doing so, let us comment on the treatment of
tadpole diagrams. Tadpoles in the electroweak theory are ubiquitous, 
and several treatments are possible.
In the end, regardless of the scheme adopted, their contribution 
cancels in physical observables \cite{Denner:1991kt}, and hence, in 
principle, we do not need to consider them.  However, keeping the 
tadpoles has its merits as the gauge boson self-energies, including 
tadpoles, are gauge-invariant on-shell \cite{Fleischer:1980ub}, 
and so are the mass and coupling counterterms, which 
helps to demonstrate the gauge-invariance of observables
\cite{Bardin:1999ak,Fleischer:1980ub}.  Therefore, we will keep the 
tadpole contributions to the self-energies as this will make the 
discussion of gauge-invariance more transparent.


\subsubsection{The correction to the momentum-space potential}
The previous considerations allow us to assemble the full one-loop correction
in the on-shell renormalization scheme. We obtain
\begin{align}
    \delta V_{\chi^+ \chi^- \to \chi^+ \chi^-} =& -\frac{4 \pi \alpha_2
    s_W^2}{\mathbf{k}^2} \left[2 \, I_{\text{vertex}}(\alpha_2 c_W^2,m_Z) +2\,
    I^{WW}_{\rm 3\, gauge} + \frac{\Sigma^{AA}_T(-\mathbf{k}^2)}{\mathbf{k}^2}
\right. \nonumber \\
  &\left. \phantom{-\frac{4 \pi \alpha_2 s_W^2}{\mathbf{k}^2}\quad} +2\, \delta Z_e
  +4\, \delta Z_{\chi^+} \right] \nonumber \\
  & -\frac{4 \pi \alpha_2 c_W^2}{\mathbf{k}^2 + m_Z^2} \left[2 \,
  I_{\text{vertex}}(\alpha_2 c_W^2,m_Z)+ 2\, I^{WW}_{\rm 3 \, gauge} +
 \frac{\Sigma^{ZZ}_T(-\mathbf{k}^2)}{\mathbf{k}^2+m_Z^2} +2\, \delta Z_e\right.
 \nonumber \\ 
&\hphantom{-\frac{4 \pi \alpha_2 c_W^2}{\mathbf{k}^2 + m_Z^2} \quad
 }\left.+4\,\delta Z_{\chi^+} -\frac{2}{c_W^2}\frac{\delta s_W}{s_W}  -
\frac{\delta m_Z^2}{\mathbf{k}^2 + m_Z^2}\right]\nonumber \\
         &-\frac{4 \pi \alpha_2}{\mathbf{k}^2 (\mathbf{k}^2+m_Z^2)} \left(-2
         s_W c_W\right) \Sigma^{AZ}_T(-\mathbf{k}^2)+
         I_\text{box}(\alpha_2,m_W;\alpha_2,m_W) 
\label{eq:potmompmpm}
\end{align}
in terms of box, vertex and self-energy functions, and the counterterms. The explicit results are lengthy and provided in Appendix~\ref{app:loops}.
The first large square bracket corresponds to the correction to the Coulomb potential,
namely the vertex corrections (\ref{eq:vertexFeyn}/\ref{eq:vertexRxi}) and
(\ref{eq:tripvertexFeyn}/\ref{eq:tripvertexWWRxi}), the photon
self-energy (\ref{eq:SigmaggLight}/\ref{eq:Sigmagg3rd}/\ref{eq:SigmaggYM}), 
the renormalization of the coupling \eqref{eq:DeltaZe} and the wave-function of the DM field \eqref{eq:DeltaZpm}. The equation numbers refer to the Feynman
and $R_\xi$-gauge results, respectively. The next large bracket corrects the
tree $Z$-exchange, which is analogous, apart from the additional term 
$\delta s_W/s_W$ from \eqref{eq:deltasW} due to the different 
coupling, and the 
$\delta m_Z^2$ mass counterterm \eqref{eq:counterterms}. The last line originates from $\gamma$-$Z$-mixing
(\ref{eq:SigmagZLight}/\ref{eq:SigmagZ3rd}/\ref{eq:SigmagZYM}) and the box term
due to the exchange of two $W$-bosons
(\ref{eq:IboxeqFeyn}/\ref{eq:boxuneqRxi}). These are the only terms that are
not directly associated with one of the tree terms (though the
$\gamma$-$Z$-mixing contribution could be partial-fractioned and grouped with the
tree terms). We checked that the UV and IR poles cancel in 
(\ref{eq:potmompmpm}) and that the expression is gauge-invariant, as discussed in 
detail below.


\subsubsection{Gauge-invariance of the potential}

The potential~\eqref{eq:potmompmpm} is explicitly gauge-invariant. 
As discussed above, the inclusion of tadpoles ensures that the 
on-shell self-energies, charge counterterm $\delta Z_e$, 
Weinberg angle counterterm $\delta s_W$, and the $Z$-mass 
counterterm $\delta m_Z^2$ are separately gauge-invariant.

The further cancellations between box, vertex, and self-energy 
diagrams are analogous to those for SM processes (see, e.g., 
\cite{Bardin:1999ak} for an extensive discussion of this issue), 
but with diagrams expanded in the 
non-relativistic/soft region. Contrary to e.g., HQET, the 
cancellation between wave-function renormalization and the vertex correction (lower left in topology in
Fig.~\ref{fig:feynpmpm}) is only partial, because we work in the mass eigenbasis and not in the weak eigenbasis.  The
remnant pieces are precisely the ones needed to complete the cancellation with the other diagrams.

Since the 
fermion self-energies are separately gauge-invariant, we can split the
potential into three separately meaningful corrections:  (1) A pure electroweak correction, which
includes all contributions of gauge and Higgs bosons, except the fermionic
self-energies, and also incorporates the parts of the counterterms that
originate from the respective topologies. (2) The light fermionic contribution
incorporates all massless fermion loops except (3) the third generation quarks,
which are separated for illustrative purposes (again also including the
respective parts of the counterterms). Note that although the bottom quark is
taken to be massless, its contribution is not separable from the top quark,
e.g., in the $W$-self-energy, as they form an SU(2) doublet, 
hence it belongs to (3). 


\subsection{The remaining channels}

The calculation of the potential correction in the other channels 
follows a similar logic. In the $\chi^0 \chi^0 \to \chi^0 \chi^0$ 
scattering channel, where the tree-level potential is vanishing, only the 
$W$-boson box and crossed box topologies are possible. They exactly 
cancel each other, such that
\begin{align}
    \delta V_{\chi^0 \chi^0 \to \chi^0 \chi^0} = 0\, .
\end{align}
The potential in the off-diagonal channel $\chi^0 \chi^0 \to \chi^+ \chi^-$ is also easily assembled. Except for the on-shell counter\-terms associated with the tree-level contributions, the other
topologies are similar  to the $\chi^+ \chi^- \to \chi^+ \chi^-$ channel (adapted to the exchanged $W$-boson). The only
topologies that again deserve a special comment are the boxes. Crossed boxes
are not possible, as the $\chi^0$ couples only to $W$-bosons. The boxes are comprised
of a $W$-boson and either a photon or a $Z$. The complete correction to the
off-diagonal potential reads
\begin{align}
    \delta V_{\chi^0 \chi^0 \to \chi^+ \chi^-} =& \quad \delta V_{\chi^+ \chi^-
      \to \chi^0 \chi^0}
\nonumber \\
      =& -\frac{4 \pi \alpha_2}{\mathbf{k}^2 + m_W^2}
      \left[\vphantom{\frac{\Sigma_T^{W}}{m_W^2}} 2 \,
      I_{\text{vertex}}(\alpha_2,m_W) + 2 \left( I_{\rm 3 gauge}^{W\gamma} +
  I_{\text{3 gauge}}^{W Z} \right) + 2\, \delta Z_{\chi^0} +2\, \delta
  Z_{\chi^+} \right.  \nonumber \\
&\hphantom{= -\frac{4 \pi \alpha_2}{\mathbf{k}^2 + m_W^2}\quad}
\left.+\frac{\Sigma^{WW}_T(-\mathbf{k}^2)}{ \mathbf{k}^2 + m_W^2} -
\frac{\delta m_W^2}{\mathbf{k}^2 + m_W^2} +2 \delta Z_e - 2 \frac{\delta
s_W}{s_W}\right] \nonumber \\
&+ I_\text{box}(\alpha_2,m_W;\alpha_2 c_W^2,m_Z) +
                                 I_\text{box}(\alpha_2,m_W;\alpha,0) \, .
                                 \label{eq:potmom00pm}
\end{align}
The individual terms are---as before equation numbers refer to the Feynman
and $R_\xi$-gauge result respectively---the vertex corrections
(\ref{eq:vertexFeyn}/\ref{eq:vertexRxi}), the triple gauge vertex diagrams
(\ref{eq:tripvertexFeyn}/\ref{eq:tripvertexWgRxi}/\ref{eq:tripvertexWZRxi}),
the DM wave function renormalization for $\chi^0 \chi^0$
(\ref{eq:DeltaZ00}/\ref{eq:DeltaZ00Rxi}) and $\chi^+ \chi^-$
(\ref{eq:DeltaZpm}/\ref{eq:DeltaZpmRxi}).  Furthermore, there is the $W$-boson
self-energy (\ref{eq:SigmaWWLight}/\ref{eq:SigmaWW3rd}/\ref{eq:SigmaWWYM}) and
the mass counterterm \eqref{eq:counterterms}, as well as the counterterms
associated with the tree-level coupling (\ref{eq:DeltaZe}/\ref{eq:deltasW}). In
the last line, the box topologies with unequal non-zero masses
(\ref{eq:IboxuneqFeyn}/\ref{eq:boxuneqRxi}) and one vanishing mass
(\ref{eq:IboxonemassFeyn}/\ref{eq:boxonemassRxi}) appear.

These channels are sufficient for the calculation of the indirect detection
cross section from $\chi^0 \chi^0$ annihilation as outlined in
\cite{Beneke:2019qaa}. For the DM relic abundance computation, the charged co-annihilation channels are also needed. For the singly charged channel
$\chi^0 \chi^\pm \to \chi^0 \chi^\pm$, the topologies are similar to the
$\chi^0 \chi^0 \to \chi^+ \chi^-$ channel as the tree-level potentials are the
same up to a minus sign. For the box diagrams, only a crossed box is possible
due to the charge flow, which in turn leads to the conclusion that the
correction in this channel is exactly the negative one of the $\chi^0 \chi^0
\to \chi^+ \chi^-$ channel: 
\begin{align}
    \delta V_{\chi^0 \chi^\pm \to \chi^0 \chi^\pm} = - \delta V_{\chi^0 \chi^0
    \to \chi^+ \chi^-} \, .
\end{align}
Finally, we consider the doubly charged channels. Similar to before the
correction is identical up to a minus sign to the $\chi^+ \chi^- \to \chi^+
\chi^-$ channel, as the same tree-level structures are involved, i.e.
\begin{align}
    \delta V_{\chi^\pm \chi^\pm \to \chi^\pm \chi^\pm} = - \delta V_{\chi^+
    \chi^- \to \chi^+ \chi^-} \, .
\end{align}
The only difference stems from the fact that only the crossed $W$-boson box is
possible, compared to the normal $W$-box topology in the charge-0 channel. This
leads to the overall minus sign. Gauge invariance for all
these channels can be checked as for the $\chi^+
\chi^- \to \chi^+ \chi^-$ channel above. Finally, we also checked for all
channels that in the limit $m_W \to m_Z$ (i.e., $s_W \to 0, c_W \to 1$) we
reproduce previously known results for the Higgsed SU(2) theory
\cite{Laine:1999rv,Schroder:1999sg}. More precisely, the unrenormalized
potential (Eq.~16 of \cite{Laine:1999rv}) was compared analytically. The
renormalized result was not compared, as the renormalization scheme was not
fully specified in \cite{Laine:1999rv}.

\subsection{Analysis of the channels}

\begin{figure}[t]
  \centering
  \includegraphics[width=0.7\textwidth]{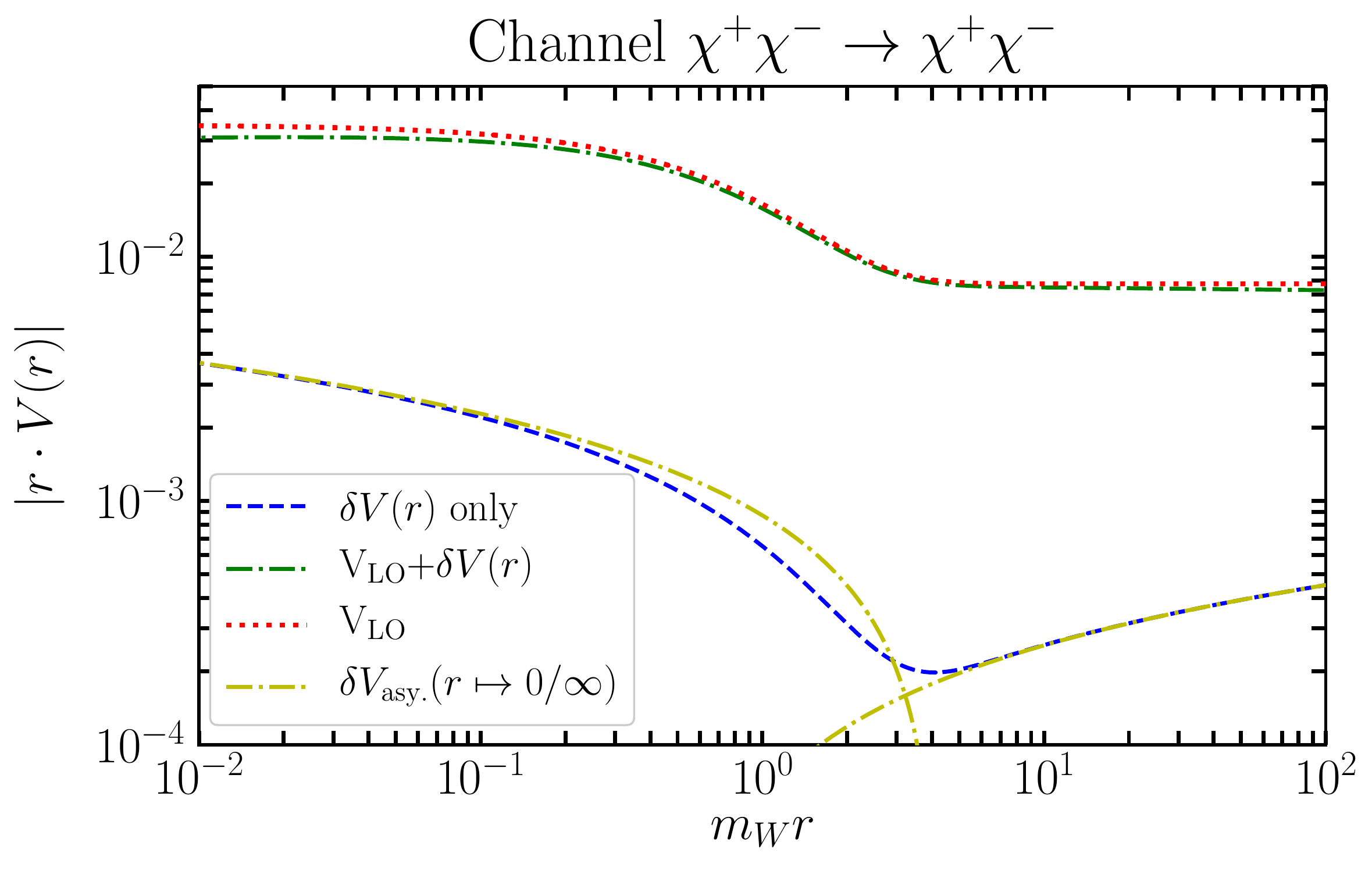}
\hspace*{-0.5cm}\includegraphics[width=0.7\textwidth]{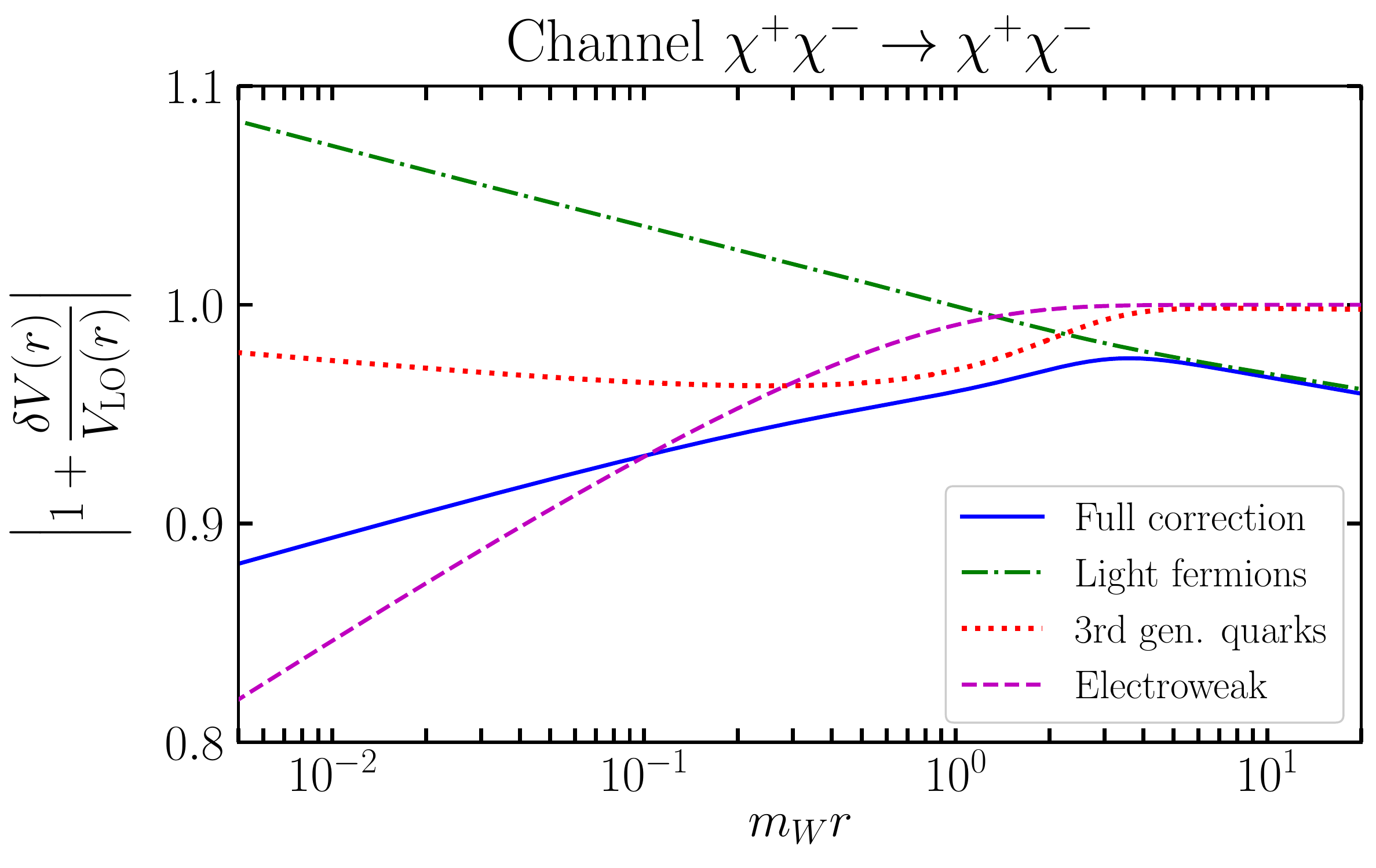}
  \caption{The NLO correction to the potential in the channel $\chi^+ \chi^- \to \chi^+
  \chi^-$. The upper panel shows the modulus of the potential $|r \cdot V(r)|$ for the LO and NLO potential, the NLO contribution only, 
and the small and large-distance asymptotic behaviour. In the lower panel, we show the ratio of the full NLO potential
to the LO potential (blue solid), and separately for the three 
gauge-invariant pieces identified in the text 
(other curves). }
  \label{fig:Potpmpm}
\end{figure}

\begin{figure}[t]
  \centering
  \includegraphics[width=0.7\textwidth]{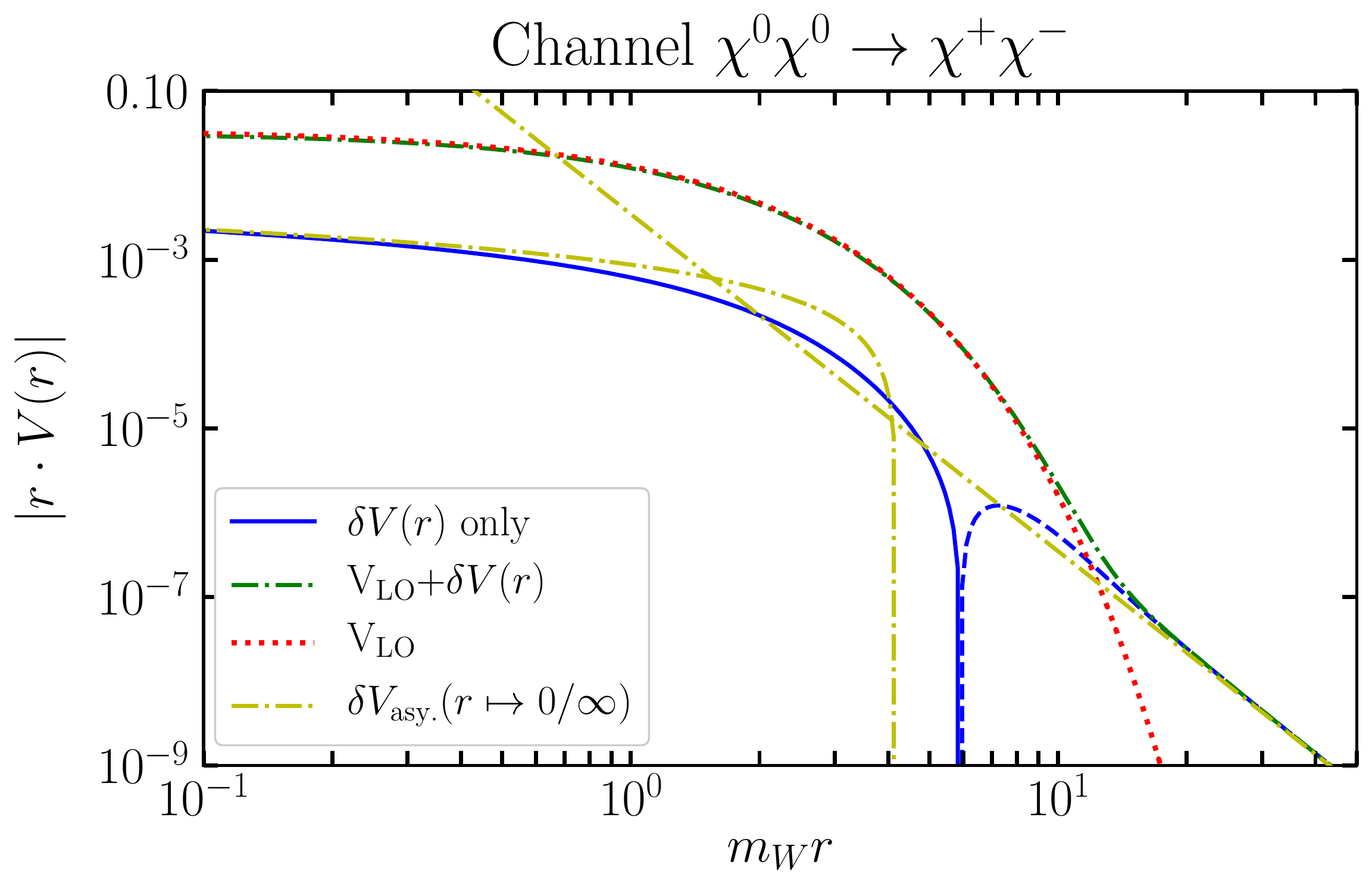}
\hspace*{-0.5cm} \includegraphics[width=0.725\textwidth]{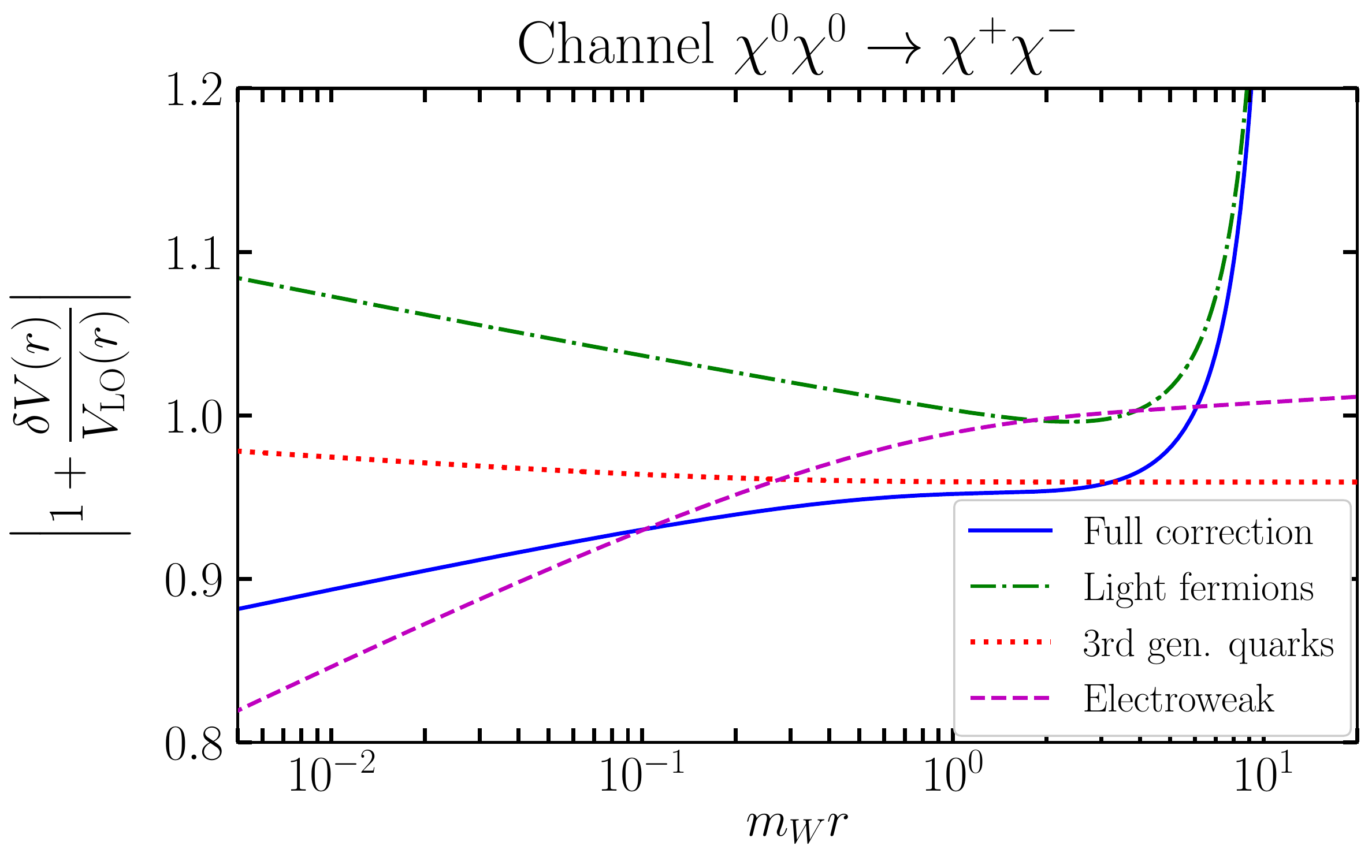}
\caption{The NLO correction to the potential in the channel 
$\chi^0 \chi^0 \to \chi^+ \chi^-$. The upper panel shows the modulus 
of the potential $|r \cdot V(r)|$ for the LO and NLO potential, the 
NLO contribution only and the asymptotic behaviours. The change from 
solid to dashed for the blue $\delta V(r)$ curve marks its change of 
sign. In the lower panel, we show the ratio of the NLO to the LO
potential for the full correction, and for the three 
gauge-invariant pieces, which illustrates their different behaviour.}
  \label{fig:Pot00pm}
\end{figure}

For an investigation of the Sommerfeld effect and other applications, one solves the position-space Schr\"odinger equation. The analytic and
numerical Fourier transforms required to obtain the position-space NLO potential are given in Appendix~\ref{app:Fourier}. Here we
discuss the charge-neutral channels, as the remaining  channels have
equal corrections up to a minus sign.  The results are shown in
Figs.~\ref{fig:Potpmpm} and \ref{fig:Pot00pm}. In the following, we 
discuss the asymptotic behaviours and relative importance of 
the NLO correction.

As input parameters for the numerics and plots, we use the following: 
the on-shell electromagnetic coupling $\alpha = \alpha_{\rm
os}(m_Z) = 1/128.943$ at the $Z$-mass, and the gauge boson masses $m_W =
80.385\,{\rm GeV}$ and $m_Z = 91.1876\,{\rm GeV}$. The SU(2) coupling and the
Weinberg angle are determined by the on-shell relations $\alpha_2
=\alpha_{\rm os}(m_Z) / s_W^2$ and $c_W =m_W/m_Z$. Furthermore, we need the
Higgs-boson and top-quark mass, for which we take the on-shell masses $m_H =
125\,{\rm GeV}$ and $m_t = 173.1\,{\rm GeV}$. The uncertainty of these
parameters is small enough to be ignored, except for the top-quark mass. 


\subsubsection{The asymptotic behaviour of the NLO potentials}
\label{sec:asymptotics}

We start with the discussion of the small- and large-distance 
asymptotics of the various channels. For
technical reasons (not all Fourier transforms are analytically available), we
discuss the results in $|\mathbf{k}|$ space, separately for the various gauge-invariant pieces to outline the
origin of the corrections and the dominant contributions.

\paragraph{The $r \to 0$ / $\mathbf{k}^2 \to \infty$ limit}\mbox{}\\
The short distance $|\mathbf{k}| \to \infty$
asymptotic behaviour is the same for the $\chi^0 \chi^0 \to \chi^+ \chi^-$ and
$\chi^+ \chi^-\to \chi^+ \chi^-$ channels, because for high energies, the SU(2) symmetry is
restored. We begin with the light-fermion contribution, and find that
\begin{align}
  \delta V_{\text{light ferm.}}(\mathbf{k}^2 \to \infty) = -\frac{3
  \alpha_2^2}{\mathbf{k}^2} \left(\ln \frac{\mathbf{k}^2}{m_Z^2} -
\frac{c_W^2}{s_W^2} \ln \frac{m_W^2}{m_Z^2}\right) \, .
\end{align}
This behaviour is similar to QED/QCD, namely, the prefactor of the
logarithmic term is proportional to the light-fermion contribution to the SU(2)
beta function.  For the contribution of the third generation quarks 
we find  
\begin{align}
  \delta V_{\rm{3rd\,gen.\,quarks}}(\mathbf{k}^2 \to \infty) =
  -\frac{\alpha_2^2}{\mathbf{k}^2} \left(\ln \frac{\mathbf{k}^2}{m_Z^2} +
  A(m_W,m_Z,m_t)\right)  \, .
\end{align}
Again the prefactor is proportional to the SU(2) beta function contribution of the
third generation quarks. The function $A(m_W,m_Z,m_t)$ is a complicated
function of the top, $W$ and $Z$ mass given in Appendix~\ref{app:asy}. To per mille
accuracy in the interval of $\pm 10\, \rm{GeV}$ around the on-shell top mass
$m_{t,{\rm os}} = 173.1\,{\rm GeV}$ it can be approximated by $A(m_W,m_Z,m_t) =
-17.1808- 4.99861 \cdot 10^{-4} \, {\rm GeV}^{-2} \times (m_t^2 - m_{t,{\rm
os}}^2)$.

The most complicated contribution originates from the electroweak 
corrections, i.e., the gauge and Higgs bosons. Contrary to the two 
former contributions, the non-self-energy diagrams also
contribute here. We find
\begin{align}
  \delta V_{\rm{electroweak}}(\mathbf{k}^2 \to \infty)
  =\frac{\alpha_2^2}{\mathbf{k}^2} \left(\frac{43}{6} \ln
  \frac{\mathbf{k}^2}{m_Z^2} + B(m_W,m_Z,m_H)\right)\, ,
\end{align}
where the logarithmic term is proportional to the non-fermionic part 
of the SU(2) beta function. $B(\mW,\mZ,\mh)$ is a 
function of the Higgs, $W$- and $Z$-mass given in 
Appendix~\ref{app:asy} and evaluates for on-shell parameters to 
$-1.03577$. The analytic result displays an interesting 
manifestation of the screening theorem \cite{Veltman:1976rt}. Even though individual terms are Higgs-mass dependent up to $m_H^6$, 
$B$ itself is only logarithmically dependent on the Higgs mass 
for large $m_H$. 

Adding all three separately gauge-invariant pieces, and defining
\begin{align}
  \Delta &= \frac{3 c_W^2}{s_W^2} \ln \frac{m_W^2}{m_Z^2}
  -A(m_W,m_Z,m_t) +B(m_W,m_Z,m_H) \, ,
\end{align}
the $r \to 0$ short-distance asymptotics of the 
position-space potential is ($\beta_{0,{\rm SU(2)}}
=43/6-1-3=19/6$)
\begin{align}
  \delta V^{r \to 0}_{\chi^+ \chi^- \to \chi^+ \chi^-}(r) &= \delta V^{r \to
  0}_{\chi^0 \chi^0 \to \chi^+ \chi^-}(r) \nonumber \\
&= \frac{\alpha_2^2}{2 \pi r}\left(-\beta_{0,{\rm SU(2)}} (\ln (m_Z r) +
  \gamma_E) +\frac{1}{2}\Delta \right) \nonumber \\
&\approx \frac{\alpha_2^2}{2 \pi r} \left(-\beta_{0,{\rm SU(2)}} \ln (m_Z r)
+ 4.92585 \right) \, , 
\label{eq:rto0asy}
\end{align}
where $\gamma_E$ is the Euler-Mascheroni constant. The last line 
provides the numerical value for the on-shell parameters as was already given in
\cite{Beneke:2019qaa}. Of the numerical coefficient, the light-fermion term
makes up $-1.3188$, the third generation quarks $8.59038$, the electroweak
terms $-0.51788$, and Euler-Mascheroni constant associated with the logarithm
$-1.82785$. The identical short-distance behaviour of the diagonal and off-diagonal channels is also visible by comparing the lower panels of 
Figs.~\ref{fig:Potpmpm} and \ref{fig:Pot00pm}.
 
The logarithmic behaviour implies a breakdown of
perturbation theory, as $\ln (m_Z r)$ grows arbitrarily large for 
small $r$. This is a consequence of renormalizing the parameters on-shell. The logarithmic behaviour can be absorbed by using $\overline{\rm MS}$
running couplings as will be discussed in Sec.~\ref{sec:MSbar}.  However, let us note
that using the on-shell renormalized potentials  is sufficient in the calculation of the
Sommerfeld effect, where the dominant contribution comes from the
region $m_W r \sim 1$, where the difference between various renormalization
schemes is of higher-order (which we also checked numerically).

\paragraph{The $r \to \infty$ / $\mathbf{k}^2 \to 0$ limit}\mbox{}\\
In the opposite limit $r \to \infty$, we have to distinguish between
the $\chi^0 \chi^0 \to \chi^+ \chi^-$ and $\chi^+ \chi^- \to \chi^+
\chi^-$ scattering potentials. We begin with the latter and the light-fermion contribution
\begin{align}
  \delta V^{(+-)(+-)}_{\rm light\, ferm.}(\mathbf{k}^2 \to 0) = - \frac{76}{9}
  \frac{\alpha^2}{\mathbf{k}^2} \ln \frac{\mathbf{k}^2}{m_Z^2} +
  \mathcal{O}(\mathbf{k}^0) \, ,
\end{align}
that scales according to the U(1)$_{\rm em}$ beta function 
contribution of the massless fermions. At large distances, 
respectively, small momenta, the potential is dominated by photon 
exchange, which explains the transition to the
electromagnetic beta function. This also holds for the 
third generation quarks, where the coefficient is determined by the
massless bottom-quark contribution to the U(1)$_{\rm em}$ beta 
function:
\begin{align}
  \delta V^{(+-)(+-)}_{\rm 3rd\,gen.\,quarks}(\mathbf{k}^2 \to 0) = -
  \frac{4}{9} \frac{\alpha^2}{\mathbf{k}^2} \ln \frac{\mathbf{k}^2}{m_Z^2}
  + \mathcal{O}(\mathbf{k}^0) 
\end{align}
The top-quark contribution is cut off due to the finite mass and therefore does
not contribute to the asymptotic behaviour.

The electroweak contribution does not play a role in the 
large-distance behaviour of
the $\chi^+ \chi^- \to \chi^+ \chi^-$ potential, as it is cut off 
by the boson masses. It starts with a constant term,
\begin{align}
  \delta V^{(+-)(+-)}_{\rm electroweak}(\mathbf{k}^2 \to 0)
  &=\frac{\alpha_2^2}{m_W^2}\, C(m_W,m_Z,m_H) \, ,
\end{align}
where the function $C(m_W,m_Z,m_H)$ is given in Appendix~\ref{app:asy} and for
on-shell parameters evaluates to $3.67219$. Even
though it does not contribute to the asymptotic behaviour, the result is a good
check of the calculation through its $m_H$ dependence. As required by the screening theorem
\cite{Veltman:1976rt}, it is logarithmic in $\mh$, even though the individual
terms depend on the Higgs mass with up to $m_H^6$.

Overall we find that the dominant $r \to \infty$ behaviour in the
$\chi^+ \chi^- \to \chi^+ \chi^-$ channel is given by the purely abelian
correction to the Coulomb potential due to massless fermions,
\begin{align}
  \delta V^{r \to \infty}_{\chi^+ \chi^- \to \chi^+ \chi^-}(r) &=
  \frac{\alpha^2}{2 \pi r} (-\beta_{0,{\rm em}})(\ln (m_Z r)+\gamma_E) \, ,
\end{align}
where $\beta_{0,{\rm em}} = -80/9$ is the electromagnetic beta function
coefficient for all SM fermions except the top quark. This 
asymptotic behaviour dominates the correction to the potential for 
$m_W r \geq 5$ as can be seen in Fig.~\ref{fig:Potpmpm}.

In the channel $\chi^0 \chi^0 \to \chi^+ \chi^-$, the large-distance 
asymptotics also originates from the light-fermion terms. The 
relevant terms are 
\begin{align}
  \delta V^{(00)(+-)}_{\rm light\, ferm.}(\mathbf{k}^2 \to 0) &= \frac{3}{5}\,
  n_{\rm ld}\, \alpha_2^2 \ln \frac{\mathbf{k}^2}{m_W^2}
  \left(\frac{m_W^2}{(\mathbf{k}^2 + m_W^2)^2} - \frac{1}{\mathbf{k}^2 +
  m_W^2}\right),
  \label{eq:momasy00pmlightferm}
\end{align}
which scales as $\mathbf{k}^2 \ln (\mathbf{k}^2/\mZ^2)$ for 
$\mathbf{k}^2 \to 0 $. $n_{\rm ld}$~denotes the number of massless 
fermion doublets, in our case $n_{\rm ld} =5$. The Fourier 
transforms for the individual terms are discussed in detail in 
Appendix~\ref{app:Fourier}. After expanding for large $r$, 
we find 
\begin{align}
  \delta V^{(00)(+-), r \to \infty}_{\rm light \, ferm.}(r) = -\frac{9 \,
  n_{\rm ld} \, \alpha_2^2 }{5 \pi m_W^4 r^5} \,.
  \label{eq:rtoinfasy00pm}
\end{align}
This power-like long-range behaviour is a consequence of taking the 
SM fermions to be massless.\footnote{A similar result is known for the
long-range force due to massless neutrinos in atomic physics
\cite{Feinberg:1968zz}. While the long-range force is universal, the
dependence on fermion mass is different for Dirac and Majorana fermions
\cite{Grifols:1996fk}.}  It is also a manifestation of a 
breakdown of perturbation theory, as for $r \gg 1/m_W$, the 
correction exceeds the exponentially decreasing tree-level 
potential. For the later physics applications this does not pose a
problem, since the Sommerfeld effect is governed by distances 
$m_W r \sim 1$. We checked this numerically and confirmed that the 
region where the power-like long-range
potential dominates does not affect the 
calculation of the Sommerfeld factors in any significant way.  

In reality, the SM fermions are, of course, not massless. A formal treatment of
the $r \to \infty$ limit would require a further matching procedure, where the
$W$-mass scale is integrated out. The resulting theory predicts the same
$r^{-5}$ asymptotics as above. In the next step, one would successively match on
theories where the individual fermions acquire mass $m_f$, which will then
cut off the contributions at distances $r \sim 1/m_f$, similar to the top-quark
contribution discussed below.

The third-generation quarks and the electroweak piece start with a constant in
the Taylor expansion around $\mathbf{k}^2 =0$ and are therefore
exponentially suppressed for large $r$. Explicitly, 
\begin{align}
  \delta V^{(00)(+-)}_{\rm 3rd\,gen.\,quarks}(\mathbf{k}^2 \to 0)
  &=\frac{\alpha_2^2}{m_W^2}\, D(m_W,m_Z,m_t) \, , \\
  \delta V^{(00)(+-)}_{\rm electroweak}(\mathbf{k}^2 \to 0)
  &=\frac{\alpha_2^2}{m_W^2}\, E(m_W,m_Z,m_H)\, ,
\end{align}
where the functions $D,E$ are given in Appendix~\ref{app:asy} and evaluate for
on-shell values to $14.6515$ and $2.76239$, respectively. Again the
screening theorem is fulfilled by these expressions. A breakdown of perturbation theory  at large $r$ manifests itself also in these
channels, as
the tree-level potential is exponentially suppressed. Terms such
as from the gauge-boson mass renormalization behave as
\begin{align}
    \frac{\delta m_W^2}{(\mathbf{k}^2 + m_W^2)^2} \quad \to \quad \frac{\delta
    m_W^2}{8 \pi \mW} \exp(-m_W r)\, ,
\end{align}
compared to the tree-level $\exp(-m_W r)/r$. However, these terms are
subdominant compared to the light-fermion tail and therefore contribute even
less to the Sommerfeld factor.  Dyson resummation would cure this behaviour
and result in potentials of the form $\exp(-(m_W^2+\delta m_W^2)^{1/2} r)/r$.

Overall the contribution in the $\chi^0 \chi^0 \to \chi^+ \chi^-$ channel for
large $r$ is given by the light-fermion contribution as discussed above and
reads
\begin{align}
  \delta V^{r \to \infty}_{\chi^0 \chi^0 \to \chi^+ \chi^-}(r) &=- \frac{9 \,
  n_{\rm ld} \, \alpha_2^2}{5 \pi m_W^4 r^5} \, ,
\end{align}
which explains the steep increase in Fig.~\ref{fig:Pot00pm} around $m_W r
\approx 10$. As also seen in this figure, the third generation quark and
electroweak contributions are exponentially suppressed for large $r$.


\subsubsection{The complete NLO corrections}
\label{sec:fitfn}

The exact NLO potential interpolates between the $r \to 0$ and $r \to
\infty$ asymptotics. The most significant deviations from the asymptotics are observed around $m_W r\sim 1$, which is the crucial region to
determine the Sommerfeld effect accurately. Therefore it is not sufficient to
simply glue the asymptotics together. For an accurate determination, either the
full numerically calculated potential or the fitting functions provided in
\cite{Beneke:2019qaa} have to be used.  They read for the off-diagonal potentials in \eqref{eq:singletLO} and \eqref{eq:tripletLO} 
\begin{align}
  \delta &V^{\text{fit}}_{\chi^0 \chi^0 \to \chi^+ \chi^-} = - \delta V^{\rm
  fit}_{\chi^0 \chi^\pm \to \chi^\pm \chi^0} \nonumber \\
         &= \frac{2595 \alpha^2_2}{\pi r} \times \left\{\begin{array}{r}
         \text{exp} \left[ -\frac{79 \left(L-\frac{787}{12}\right)
     \left(L-\frac{736}{373}\right)\left(L-\frac{116}{65}\right)
     \left(L^2-\frac{286L}{59}+\frac{533}{77}\right)}{34
     \left(L-\frac{512}{19}\right)\left(L-\frac{339}{176}\right)
     \left(L-\frac{501}{281}\right)\left(L^2-\frac{268
     L}{61}+\frac{38}{7}\right)}\right] ,\quad  x<x_0 \\[0.5cm]
 -\text{exp}\left[-\frac{13267 \left(L-\frac{76}{43}\right)
 \left(L-\frac{28}{17}\right)\left(L+\frac{37}{30}\right)
 \left(L^2-\frac{389L}{88}+\frac{676}{129}\right)}{5
 \left(L-\frac{191}{108}\right)\left(L-\frac{256}{153}\right)
 \left(L+\frac{8412}{13}\right) \left(L^2-\frac{457
 L}{103}+\frac{773}{146}\right)}\right] ,\quad  x>x_0 \end{array}\right.
      \end{align}
and for the diagonal ones
      \begin{align}
        \delta &V^{\text{fit}}_{\chi^+ \chi^- \to \chi^+ \chi^-} = - \delta
        V^{\rm fit}_{\chi^\pm \chi^\pm \to \chi^\pm \chi^\pm} \nonumber \\  
               &= \frac{\delta V^{r \to \infty}_{\chi^+ \chi^- \to \chi^+
    \chi^-} }{1+ \frac{32}{11 }x^{-\frac{22}{9}}} + \frac{\delta V^{r \to
0}_{\chi^+ \chi^- \to \chi^+ \chi^-}}{1+\frac{7}{59} x^{\frac{61}{29}}} +
\frac{\alpha}{r} \left[\frac{-\frac{1}{30} + \frac{4}{135} \ln x}{1 +
\frac{58}{79} x^{-\frac{17}{15}}+\frac{1}{30} x^{\frac{119}{120}} +
\frac{8}{177} x^{\frac{17}{8}}}\right]\,,
\end{align}
where $x = m_W r$, $x_0 = 555/94$ and $L = \ln x$. The fitting functions
provide per mille accuracy for the Sommerfeld factors \cite{Beneke:2019qaa}. In
general, the correction to the Coulomb and $Z$-Yukawa potential is closer to
the full numerical result. The reason is the sign change for the $W$-Yukawa
potential at $x_0 = 555/94$. The position of this sign change is set by the
distance where the light-fermion contribution starts to dominate the
correction.

Although the correction for very small (large) $r$ is significant 
due to the logarithmic (power-like) behaviour, 
these regions contribute little to the Sommerfeld factors. 
In the relevant region $m_W r \sim 1$, the NLO correction to the 
potentials is in the few percent range. The complete NLO 
result is determined
by the interplay of the various corrections. For example, for $r \to
0$ in both, the $\chi^+ \chi^- \to \chi^+ \chi^-$ and $\chi^0 \chi^0 \to \chi^+
\chi^-$ channels, the correction due to light fermions is of opposite sign to
the electroweak contribution.


\subsubsection{Scheme conversion to $\overline{\rm MS}$-couplings}
\label{sec:MSbar} 

The on-shell scheme employed for the calculations so far exhibits large short-distance logarithms related to the beta function. This 
behaviour originates from on-shell renormalization at the scale 
$m_Z$, which is suitable for the calculation of the Sommerfeld 
effect, but leads to logarithms of the form $\ln (m_Z r)$.

It is more appropriate to use running couplings at the scale
$\mu =e^{-\gamma_E}/r$ or $\mu^2 = \mathbf{k}^2$ 
in position or momentum space, respectively, if one is interested 
in the potential at short distances. To
this end, we convert the on-shell coupling to the $\overline{\rm MS}$-scheme
using \begin{align} \alpha_{\rm \overline{MS}}(m_Z) &= \alpha_{\rm os}(m_Z)
    \left[1+2 \left.  \delta Z_{e} \right|_{\alpha_{\rm os} (m_Z)} -2
            \left.\delta Z_{e} \right|_{\alpha_{\overline{\rm MS}(m_Z)}}\right]
            \nonumber \\ &= \alpha_{\rm os}(m_Z) \left[ 1+ \frac{\alpha_{\rm
        os}(m_Z)}{4 \pi} \left(\frac{382}{27} + 7 \ln \frac{m_W^2}{m_Z^2}-
        \frac{16}{9} \ln \frac{m_t^2}{m_Z^2} \right)\right]\nonumber \\ &=
    0.00780372 \,, \end{align} 
with on-shell renormalization
    factors given in Sec.~\ref{sec:onshell}.  Furthermore, we need the Weinberg
    angle in the $\overline{\rm MS}$ scheme. In the literature, one finds
    different definitions of the $\overline{\rm MS}$ Weinberg angle. We choose
    \cite{Degrassi:1990tu,Jegerlehner:1991dq} \begin{align} s^2_{W,\,\rm
            \overline{MS}}(m_Z) &= s^2_{W,\,{\rm os}}(m_Z) \left[1+ 2 \left.
                \frac{\delta s_W}{s_W} \right|_{\rm os} - 2 \left.\frac{\delta
                s_W}{s_W} \right|_{\rm \overline{MS}}\right] = 0.232486 \, ,
            \end{align} 
where $\delta s_W$ was defined in \eqref{eq:deltasW}
and on-shell parameters were used for all terms involved. For
            numerics in the $\overline{\rm MS}$ scheme, we use the
            $\overline{\rm MS}$ top mass $\overline{m}_t(\overline{m}_t) =
            163.35 \, {\rm GeV}$. To keep notation short, from here on
            couplings in the $\overline{ \rm MS}$-scheme are denoted by a hat.

With these ingredients, the issue of large logarithms in the 
$r \to 0$ asymptotics can be further investigated. To see the 
cancellation of the large logarithms, the $\overline{\rm MS}$-coupling 
at $\mZ$ is converted to the coupling at an 
arbitrary scale $\mu$ by expanding the running couplings to fixed 
order, which, e.g., for the tree-level
$W$-Yukawa potential in momentum space leads to 
\begin{align} -\left.\frac{4\pi
            \hat{\alpha}_2(\mu)}{\mathbf{k}^2 +
        m_W^2}\right|_{\mu^2=\mathbf{k}^2} = -\frac{4 \pi
    \hat{\alpha}_2(m_Z)}{\mathbf{k}^2 + m_W^2} \left(1 +
\frac{\hat{\alpha}_2(m_Z)}{4 \pi} \, \beta_{0,{\rm SU(2)}} \ln
\frac{\mathbf{k}^2}{m_Z^2} \right)\, .  \label{eq:YukawaWMSbar} \end{align} 
For $|\mathbf{k}| \to \infty$, this exactly cancels the logarithmic
contribution in the asymptotic behaviour \eqref{eq:rto0asy}.  Splitting the
beta function contribution into $\beta_{0,\,{\rm
SU(2)}} = 19/6 = 43/6 - 1 -3$, where terms correspond to the electroweak, third-generation quark, and light-fermion contributions, respectively, this can also
be done for each of the separately gauge-invariant pieces. 
In position space, a similar expansion can be performed using $\mu =
e^{-\gamma_E}/r$,
\begin{align}
  \left. -\frac{\hat{\alpha}_2(\mu)}{r} e^{-m_W r} \right|_{\mu =
e^{-\gamma_E}/r} &= - \frac{\hat{\alpha}_2(m_Z)}{r} e^{-m_W r} \left(1-
        \frac{\hat{\alpha}_2(m_Z)}{2 \pi} \beta_{0,{\rm SU(2)}} \ln\left(m_Z
      r e^{\gamma_E}\right)\right) \,,
      \label{eq:posconversion}
\end{align}
which cancels the logarithms for $r\to 0$. The $\overline{\rm MS}$
scheme presented here applies to momenta and distances of 
$\mathbf{k}^2 > \mW^2$ and $1/r >\mW$, 
respectively.\footnote{\label{fn:asyMSbar}
Although the expansions in position and momentum space are
    equivalent in the high-energy limit $r \to0 \,/\, \mathbf{k}^2 \to \infty$
    (up to higher-order constant terms), they differ fundamentally in the
    low-energy limit.  The reason is hidden in the fact, that the Fourier
    transform of $\hat{\alpha}_2(\mathbf{k}^2)$ is proportional to $r^{-3}$.
    For example, the Fourier transform of \eqref{eq:YukawaWMSbar} together with
    the NLO terms \eqref{eq:potmom00pm} produces the large-$r$ asymptotics
\begin{align}
    V^{r \to \infty , \overline{\rm MS}}_{\chi^0 \chi^0 \to \chi^+
    \chi^-} (r) = -\beta_{0,{\rm SU(2)}} \frac{ \hat{\alpha}_2^2}{2 \pi m_W^2
r^3} \, ,
    \label{eq:MSbarasy0pm}
\end{align}
while in position space the asymptotics is the same as for the on-shell
potential \eqref{eq:rtoinfasy00pm} (exchanging the couplings). A more detailed
discussion of the Fourier transform that leads to this behaviour is
found in Appendix~\ref{app:Fourier}. The $r^{-3}$ dependence is not a
conceptual problem for two reasons. First, similar to the on-shell case
\eqref{eq:rtoinfasy00pm} that shows an $r^{-5}$ behaviour, another EFT would
need to be constructed that integrates out the massive bosons and keeps only
light fermions dynamical. Secondly, the $\overline{\rm MS}$-scheme is designed
to absorb the logarithms that grow large for $r \to 0\, /\, \mathbf{k}^2 \to
\infty$ and is therefore not expected to work in the opposite limit anyway.}

For the tree-level Coulomb and $Z$-Yukawa potential, the logarithms 
for $r \to 0$ can be eliminated in a similar fashion using
\begin{align}
  -\left.\frac{4 \pi \hat{\alpha}(\mu)}{\mathbf{k}^2}\right|_{\mu^2 =
    \mathbf{k}^2} &= - \frac{4 \pi \hat{\alpha} (m_Z)}{\mathbf{k}^2} \left(1 +
  \frac{\hat{\alpha}(m_Z)}{4 \pi} \left(\beta_{0,{\rm SU(2)}} +
\beta_{0,Y}\right)\ln \frac{\mathbf{k}^2}{m_Z^2}\right), \\
  -\left. \frac{4 \pi \hat{\alpha}_2(\mu) \hat{c}_W^2(\mu)}{\mathbf{k}^2 +
    m_Z^2} \right|_{\mu^2 = \mathbf{k}^2} &= - \frac{4 \pi \hat{\alpha}_2(m_Z)
  \hat{c}_W^2(m_Z)}{\mathbf{k}^2 + m_Z^2} \left[1+ \frac{\hat{\alpha}_2(m_Z)}{4
\pi} \right. \nonumber \\
 &\hspace{0.5cm}\left.  \times  \left(\beta_{0,{\rm SU(2)}}
(1+\hat{s}_W^2(m_Z)) - \beta_{0,Y}
\frac{\hat{s}^4_W(m_Z)}{\hat{c}_W^2(m_Z)}\right) \ln
\frac{\mathbf{k}^2}{m_Z^2}\right]\,,
\end{align}
where we have used the beta function for the hypercharge $\beta_{0,Y} = -41/6=
-1/6 -11/9 -49/9$ (which can be split into electroweak, third-generation quarks
and light fermions, respectively).  As expected from the tree-level 
potential, since $\hat{\alpha} + \hat{\alpha}_2 \hat{c}_W^2 = 
\hat{\alpha}_2$, the hypercharge contribution drops out in the limit
$\mathbf{k}^2 \to \infty$ when the two above potentials are summed, 
and the logarithmic contribution cancels
in the asymptotic behaviour \eqref{eq:rto0asy}. A similar argument works in
position space.  

\begin{figure}[t]
  \centering
  \includegraphics[width=0.95\textwidth]{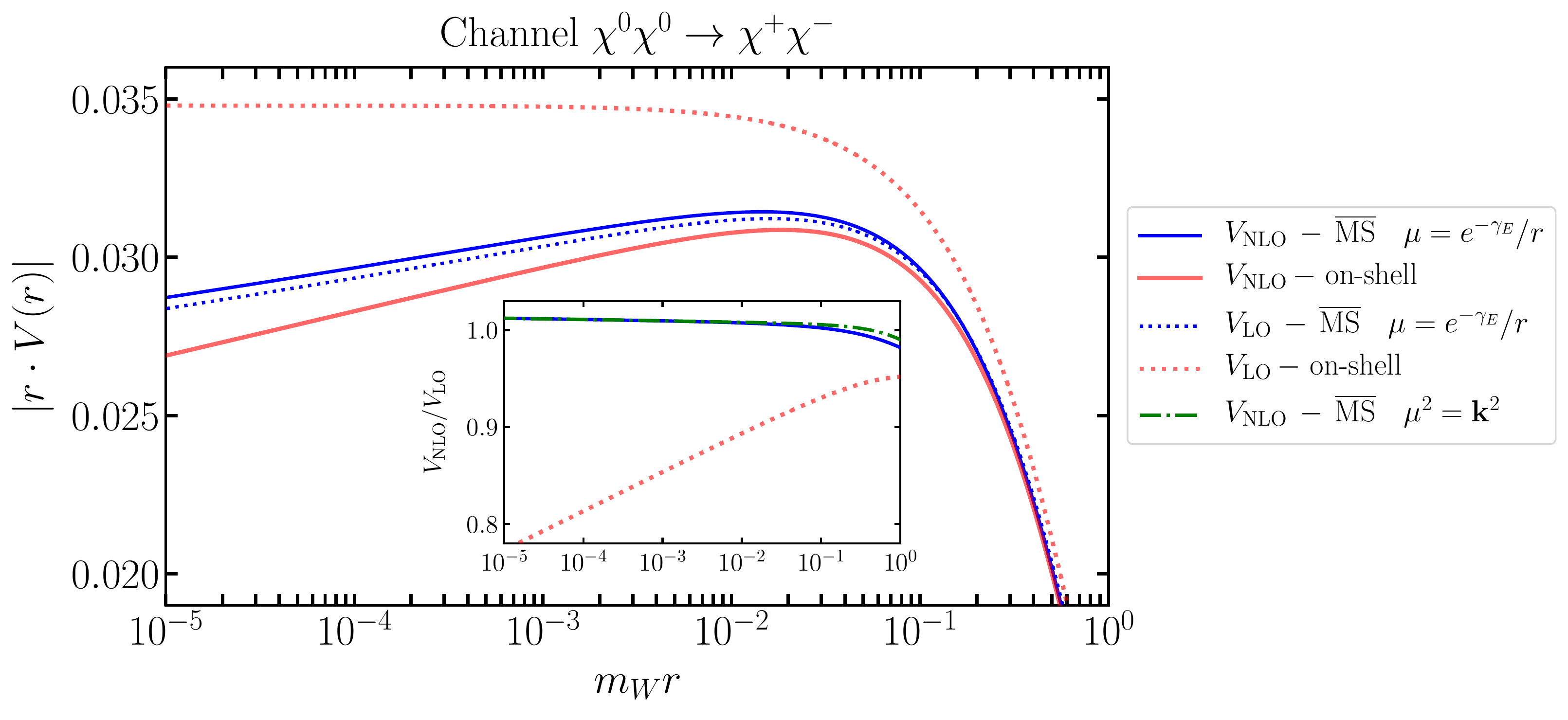}
\caption{The NLO (solid) and LO potential (dotted) 
$| r \cdot V(r)|$ in the $\overline{\rm MS}$-scheme
  using $\mu = e^{-\gamma_E}/r$ (blue) and in the on-shell scheme
  (red) for the channel $\chi^0 \chi^0 \to \chi^+ \chi^-$. The inset shows
  the ratio of the NLO potential to the LO potential for with same colour
  coding. In addition the dashed-dotted (green) line shows the 
$\overline{\rm MS}$-scheme potential with running coupling 
at $\mu^2= \mathbf{k}^2$ implemented before Fourier 
transformation to position space.}
 \label{fig:AbsValueMSbar}
\end{figure}

In Fig.~\ref{fig:AbsValueMSbar}, we show the absolute value of the 
potential in the $\overline{\rm MS}$-scheme using the position-space 
conversion \eqref{eq:posconversion} and one-loop running couplings 
at the scale $\mu=e^{-\gamma_E}/r$. 
While the NLO and LO potentials diverge for small $r$ in the 
on-shell scheme due to the breakdown of perturbation theory, 
the NLO correction remains always small in the   
$\overline{\rm MS}$-scheme and the correct short-distance 
behaviour is already attained at tree-level, due to the 
use of the running scale. This behaviour is also exemplified by 
the inset of Fig.~\ref{fig:AbsValueMSbar}, which shows the ratio of
 the NLO to the LO potential. It also shows that it does not 
matter whether the running coupling is implemented in position 
or in momentum space, as it should be. 

The $\overline{\rm MS}$-scheme is clearly the better scheme 
for large momenta. Solving the Schr\"o\-dinger equation to obtain 
the Sommerfeld effect technically probes all momentum
regions, but we find that the changes for the Sommerfeld factor 
for various $\overline{\rm MS}$ approximations and the
on-shell result are compatible with differences of the size 
of well-behaved electroweak corrections beyond the 
one-loop order considered here. 
We therefore stick with the on-shell scheme for the computation
of the relic abundance. However, let us note that the conceptual 
control over the $r \to 0$ logarithms
is an essential check of the calculation and demonstrates 
perturbative control over the potential correction.


\subsubsection{Top mass dependence}

The input parameter uncertainties have a negligible impact on the 
accuracy of the potential except for the top-quark mass. The 
top-quark mass first enters the potential at NLO, and the dependence 
on it is not only logarithmic, but also quadratic.  At this point it 
would be possible to use the pole mass $m_t = 173.1 \, {\rm GeV}$ or 
the corresponding $\overline{\rm MS}$-mass (at four
loops) $\overline{m}_t(\overline{m}_t) = 163.35 \,{\rm GeV}$. Since 
the scheme ambiguity is not fixed at NLO accuracy for the potential, 
both choices are legitimate input values. The difference of 
$10\,{\rm GeV}$ causes by far the largest uncertainty of all input 
parameters. For the other parameters, the dependence is negligible, 
as they either already enter at leading order ($W/Z$-mass and 
couplings) thus the scheme dependence is reduced
by the NLO correction, or the dependence is only logarithmic due to 
screening \cite{Veltman:1976rt} (Higgs mass), or they are known 
precisely anyway. 

To estimate the top-mass dependence, we investigate the function
$A(m_W,m_Z,m_t)$ given in Appendix~\ref{app:asy} that controls 
the size of the Coulomb term in the $r \to 0$ asymptotics. We find
\begin{align}
  \frac{A(m_W,m_Z,\overline{m}_t(\overline{m}_t))}{A(m_W,m_Z,m_t)} = 1 + 2.92
  \cdot 10^{-5} \frac{1}{\rm GeV^2} \left[ \overline{m}_t(\overline{m}_t)^2 -
  (173.1 \, {\rm GeV})^2\right] =0.904 \, ,
\end{align}
where the last number is given for $\overline{m}_t(\overline{m}_t) =
163.35\,{\rm GeV}$. Therefore we expect changes of the order of 
$10\, \%$ in the third quark-generation part of the potential. For the
non-logarithmic term in \eqref{eq:rto0asy} this means
\begin{align}
  m_t = 173.1 \,{\rm GeV} : \quad  4.92585 \quad \to \quad
  \overline{m}_t(\overline{m}_t) = 163.35\, {\rm GeV} : \quad 4.09563
  \, ,
\end{align}
which is a $17\,\%$ decrease of the coefficient of the Coulombic 
behaviour for $r \to 0$ for the full correction to the potential. However, in this 
region also the logarithmic term contributes, which is of similar 
size, decreasing the effect of the correction to roughly $10\,\%$ of 
the NLO correction.

\begin{figure}[t]
    \centering
    \includegraphics[width=0.85\textwidth]{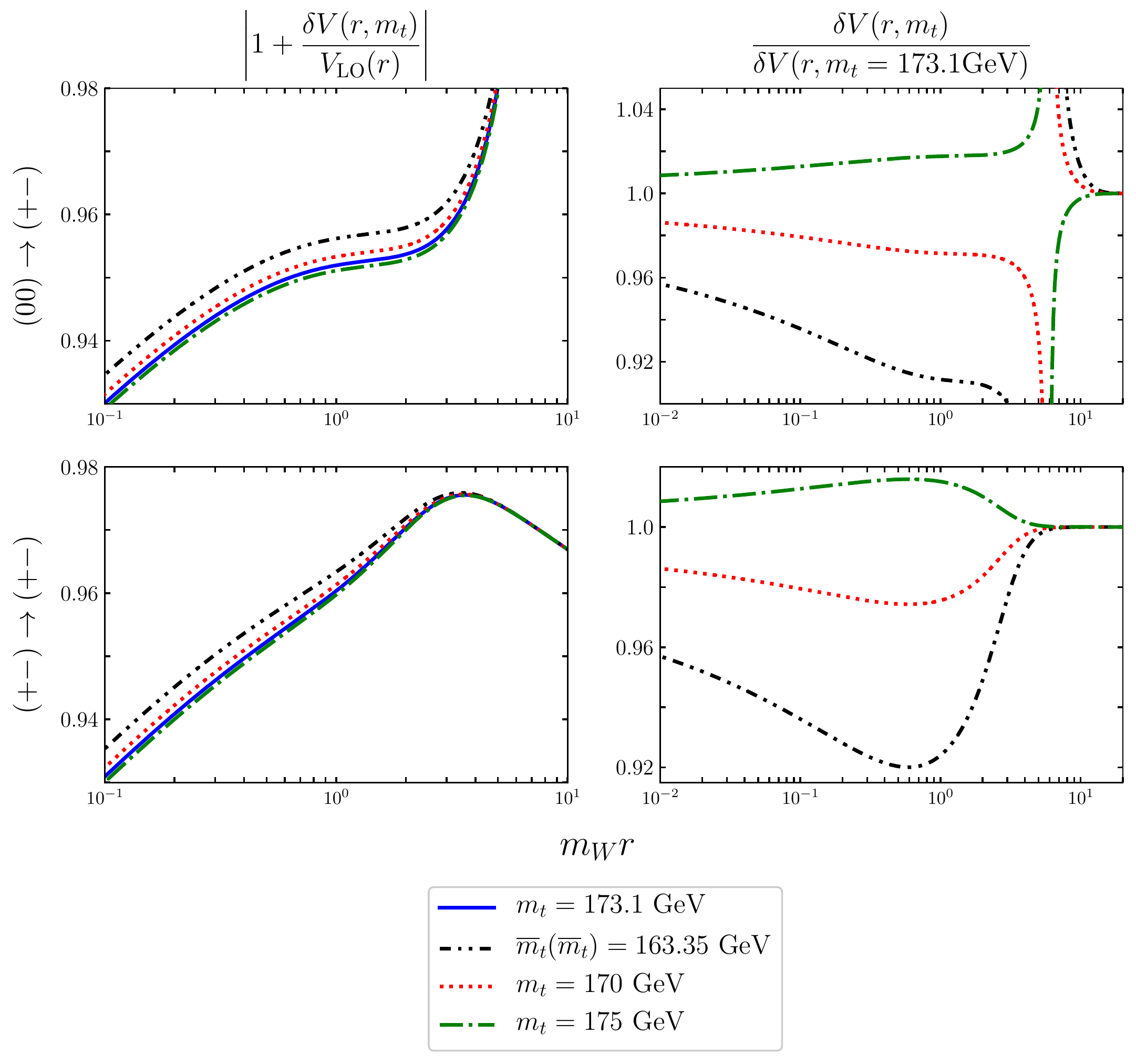}
    \caption{Ratio of the NLO potential to the LO potential (left panel) and
    the ratio of the correction vs. the correction for the reference value 
$m_t = 173.1\, {\rm GeV}$ (right panel) for various values of top mass. The
upper panel gives the channel $(00) \to (+-)$ and the lower panel $(+-) \to
(+-)$. The lines refer to $m_t=173.1$ GeV (blue/solid), $163.35$ GeV
(black/dot-dot-dashed), $170$ GeV (red/dotted), and $175$ GeV (green/dot-dashed) all in the on-shell renormalization scheme for the couplings.}
    \label{fig:TopMassRatio}
\end{figure}

In Fig.~\ref{fig:TopMassRatio} we show the ratio of the full 
NLO potential to the LO potential for a range of $r$ centred 
around $1/m_W$ for different values for the 
top mass (left panel). The figure shows that the top mass dependence 
is more relevant in some regions than in others, 
but it does not change the gross features of the NLO 
correction. 
In the right panel of the figure, the ratio to the
default value $m_t =173.1 \, {\rm GeV}$ is depicted. In both
diagonal $(+-) \to (+-)$ and off-diagonal $(00)\to (+-)$ channels, 
the top mass affects the NLO correction by up to 10\,\% 
with the largest change around $m_W r \sim 1$. At large $r$, the 
precise top-mass value employed does not matter, as
the top contribution becomes negligible in comparison to the light 
fermions. In
the $\chi^0 \chi^0 \to \chi^+ \chi^-$ channel, 
the singularity of the ratio around $m_W r \sim 5$ in the 
right panel is an artefact of showing the ratio,
as $\delta V(r,m_t)$ changes sign at slightly different 
$m_t$-dependent values of $r$. 

The top-mass uncertainty of the potential translates into a small 
effect on the Sommerfeld factor, which has already been 
investigated \cite{Beneke:2019qaa}. For example, the 
location of the first Sommerfeld resonance is shifted  due to the NLO 
potential correction from 2.283~TeV to 2.408~TeV 
instead of 2.419~TeV when the $\overline{\rm MS}$-mass 
$\overline{m}_t(\overline{m}_t) = 163.35 \,{\rm GeV}$ 
instead of $m_t = 173.1 \, {\rm GeV}$ is adopted. This effect 
is small enough to be ignored for the present, hence in the 
following we will stick with the pole mass value 
$m_t = 173.1 \, {\rm GeV}$.


\section{Wino relic abundance}
\label{sec:relic}

In this section, we compute the WIMP relic abundance under the 
thermal freeze-out assumption and discuss the importance of the 
new NLO correction to the Sommerfeld potential. 

\subsection{Technical details on the 
DM abundance calculation}

The computation divides into the calculation of the Sommerfeld 
factors for various partial-wave cross sections in all co-annihilation 
channels, followed by the thermal average and solution of the 
Boltzmann equation. Our implementation follows 
\cite{Beneke:2014gja}, to which we refer for a detailed 
description and notation employed here. 

As the freeze-out process starts for DM relative velocities
$v \sim 0.2$, the inclusion of $\mathcal{O}(v^2)$ corrections to the
annihilation cross sections is necessary to obtain percent-level 
accuracy. We therefore include $P$-wave and  
$\mathcal{O}(v^2)$-suppressed $S$-wave annihilation.  All 
required annihilation matrices $\hat f(^{2S+1}L_J)$
are conveniently tabulated in
Appendix~C of \cite{Hellmann:2013jxa}. In these short-distance 
quantities, we use the $\overline{\rm MS}$ couplings evolved with 
one-loop accuracy to the scale $\mu = 2 \mchi$. The partial-wave 
Sommerfeld factors 
for the annihilation cross section of the two-particle 
DM state $\chi_i\chi_j$ are given by 
\begin{align}
S_{ij}[\hat f(^{2S+1}L_J)] = 
\frac{ \left[ \psi^{(L,S)}_{e_4 e_3,\,ij}\right]^* \hat f^{\chi\chi \to \chi
\chi}_{\lbrace e_1 e_2 \rbrace \lbrace e_4 e_3 \rbrace}(^{2S+1}L_J) \,
\psi^{(L,S)}_{e_1 e_2, \,ij}} {\hat  f^{\chi\chi\to \chi\chi}_{\lbrace i j
\rbrace \lbrace i j \rbrace}(^{2S+1}L_J)|_{\rm LO} }\, ,
\label{eq:SFdef}
\end{align}
where $\psi^{(L,S)}_{e_1 e_2, \,ij}$ denotes the wave function at 
$\mathbf{r}=0$ for the initial state $ij$ in partial-wave 
configuration with angular momentum $L$ and spin $S$ to 
scatter into the state $e_1 e_2$ under the influence of the 
potential. The annihilation cross section in the channel $ij$
that enters the thermal average is then obtained to
$\mathcal{O}(v^2)$ accuracy by weighting each Born 
partial-wave term by its respective Sommerfeld factor, 
resulting in\footnote{Let us mention a subtlety here. 
In the computation of the Sommerfeld factors 
$S_{ij} [\hat g_\kappa(^{1}S_0)]$, $S_{ij} [\hat g_\kappa(^{1}S_0)]$ 
for the $\mathcal{O}(v^2)$ suppressed $S$-wave terms 
in \eqref{eq:SFenhancedsigma}, one uses an equation-of-motion 
identity that relates the matrix elements  
$\langle \chi_i \chi_j | \, \mathcal
P(^{2S+1}S_S) \, | \chi_i \chi_j \rangle$ of the 
$\mathcal{O}(v^2)$ suppressed derivative operator to those of the 
leading $S$-wave operator, 
$ \langle \chi_i \chi_j | \, \mathcal O(^{2S+1}S_S) \, | 
\chi_i \chi_j \rangle$ (Sec.~4.4 of \cite{Beneke:2014gja}). 
This identity contains the quantity
\begin{align}
\kappa_{\,  e_1 e_2 , e_1^\prime e_2^\prime } = \vec{p}^{\,2}_{e_1 e_2}\,
\delta_{e_1 e_2,e_1^\prime e_2^\prime} + 2 \, \mu_{e_1 e_2}\alpha_2\, \sum_a
\,  m_{\phi_a}  \,  c^{(a)}_{ e_1 e_2,  e_1^\prime e_2^\prime} \,,
\label{eq:kappa}
\end{align}
which depends through the Lippmann-Schwinger 
equation on parameters of the potential, assumed to be 
of the form $4\pi\alpha_2\sum_a 
c^{(a)}_{ e_1 e_2,  e_1^\prime e_2^\prime}/
(\mathbf{k}^2+m^2_{\phi_a})$. The second term in 
\eqref{eq:kappa} arises from a linearly divergent 
integral, and is finite but scheme-dependent in dimensional 
regularization, which 
has been used in obtaining \eqref{eq:kappa}. The scheme-dependence 
cancels with a one-loop correction to the short-distance 
annihilation matrix, but this is not available here.  
The generalization of the above identity to NLO potentials 
is not straightforward, since it would require a treatment 
of the singular short-distance behaviour in dimensional regularization. We therefore 
use the LO potentials here. This is justified, since 
the issue of the uncancelled scheme dependence for 
the  $\mathcal{O}(v^2)$ suppressed $S$-wave terms is already present 
at LO and would not be improved by adding the NLO 
correction, but more importantly since in practice, the term 
in question represents a small correction to the cross 
section, as will be discussed at the end of this section.} 
\begin{align}
    \sigma^{\chi_{i} \chi_{j} \to \,{\rm light}} \, v_\text{rel} =& \, S_{ij}
    [\hat f(^{1}S_0)] \; \hat  f^{\chi\chi \to \chi \chi}_{\lbrace i j \rbrace
    \lbrace i j \rbrace}(^{1}S_0) + \, S_{ij}[\hat f(^{3}S_1)] \; 3 \,\hat
    f^{\chi\chi \to \chi \chi}_{\lbrace i j \rbrace \lbrace i j
    \rbrace}(^{3}S_1) \nonumber\\
&+ \, \frac{\vec{p}_{ij}^{\,2}}{M_{ij}^2} \, \bigg( \, S_{ij} [\hat
g_\kappa(^{1}S_0)]  \; \hat  g^{\chi\chi \to \chi \chi}_{\lbrace i j \rbrace
\lbrace i j \rbrace}(^{1}S_0) +  S_{ij}[\hat g_\kappa(^{3}S_1)] \; 3 \, \hat
g^{\chi\chi \to \chi \chi}_{\lbrace i j \rbrace \lbrace i j
\rbrace}(^{3}S_1)
\nonumber\\
&+ \,S_{ij} \Big[\frac{\hat f(^{1}P_1)}{M^2}\Big] \; \hat  f^{\chi\chi
\to \chi \chi}_{\lbrace i j \rbrace \lbrace i j \rbrace}(^{1}P_1) + S_{ij}
\Big[\frac{\hat f({}^3P_{\cal J})}{M^2} \Big] \; \hat f^{\chi\chi \to \chi
\chi}_{\lbrace i j \rbrace \lbrace i j \rbrace}(^{3}P_{\cal J}) \bigg)\,.
\label{eq:SFenhancedsigma}
\end{align}

Technically, we determine the Sommerfeld factors using the 
variable-phase method to solve the Schr\"odinger equation developed in 
\cite{Beneke:2014gja}. This requires a fast and numerically stable 
evaluation of the NLO Sommerfeld potential in coordinate space, 
which we obtain by precalculating and interpolating the numerical 
Fourier transform where necessary. The Schr\"{o}dinger equation is
then solved from an initial value $x_0 =\mchi v r_0 = 10^{-7}$ to some
large $x_\infty = \mchi v r_\infty$, which is determined using an 
adaptive procedure, which terminates when doubling an already 
large initial $x_\infty$ changes the 
Sommerfeld factor by less than three per mille.  For 
points near the $\chi^+ \chi^-$-threshold this convergence criterion 
is sometimes hard to reach, and we abort the above procedure if 
$x_\infty > 10^4$. The behaviour around the true value is 
oscillating and since we scan the threshold accurately, 
the $1\%$ inaccuracies incurred by the abortion tend to average 
out.

We tabulate the Sommerfeld factor as a function of velocity 
in all relevant channels using 
100 velocity points distributed logarithmically between $v=10^{-4}$ 
and $1$ and additional points around the two-particle thresholds 
$v =\sqrt{2 \delta m_\chi/m_\chi}$ and $\sqrt{\delta m_\chi/m_\chi}$ 
resulting in around 150 points for every partial-wave Sommerfeld 
factor. The resulting cross-section tables are then monotonically
interpolated for use in the velocity integration to obtain the 
temperature-dependent thermally-averaged effective annihilation 
cross section including co-annihilation. The thermal average is 
calculated in the variable $x = \mchi / T$ for 160 logarithmically 
distributed points between $x =1$ and $x = 10^8$. 

These points are again monotonically interpolated. The resulting 
function forms the input to the Boltzmann equation solver. 
The differential equation is solved numerically with different 
methods, one of them simply \texttt{Mathematica}'s built-in 
\texttt{NDSolve}, with an implicit solver to determine the 
yield $Y(x)$ between $x=1$ and $x = 10^8$ with 
initial condition $Y(1)=Y_{\rm eq}(1)$.  
The relic abundance is obtained from $Y(10^8)$. As input
for the effective number of degrees of freedom in the Boltzmann 
equation we adopt the implementation from
\cite{Borsanyi:2016ksw}, extracted from the plots and tables therein,
supplemented by results of \cite{Laine:2015kra} in the regions
above $T =280 \, {\rm GeV}$ and below $T=1 \,{\rm MeV}$. 
The critical energy density value equals 
$\rho_{\rm crit.} = 1.05368 \cdot 10^{-5}
\, {\rm GeV\,cm^{-3}}\,h^2$. 


\subsection{NLO relic density for the wino model}

The first zero-energy bound-state resonance for the $\chi^0 \chi^0$
(${}^1S_0$) total annihilation cross section is located at 
$m_\chi = 2.282\,{\rm TeV}$ for the LO potential and at
$2.419\,{\rm TeV}$ for the NLO potential. The Sommerfeld factor 
for the total cross section is slightly different from 
the one for $\gamma+X$ considered in \cite{Beneke:2019qaa}, 
since in the latter case only the $\chi^+ \chi^- \to \chi^+ \chi^-$ 
component of the annihilation matrix enters. Therefore in that 
case only the wave-function components $\psi_{(00)(+-)}$ and 
$\psi_{(+-)(00)}$ are probed. On the other hand, for the relic
abundance calculation the annihilation matrix is non-zero in all 
entries
$\chi^0 \chi^0 / \chi^+ \chi^- \to \chi^0 \chi^0 / \chi^+ \chi^-$ 
and the Sommerfeld calculation is sensitive to all components of 
the wave function. Nevertheless, the resonance masses are 
the same within sub-GeV accuracy as the ones found in 
\cite{Beneke:2019qaa} for the annihilation to $\gamma+X$.

\subsubsection{Sommerfeld factors in individual channels}

\begin{figure}[t]
    \centering
    \includegraphics[width=0.61\textwidth]{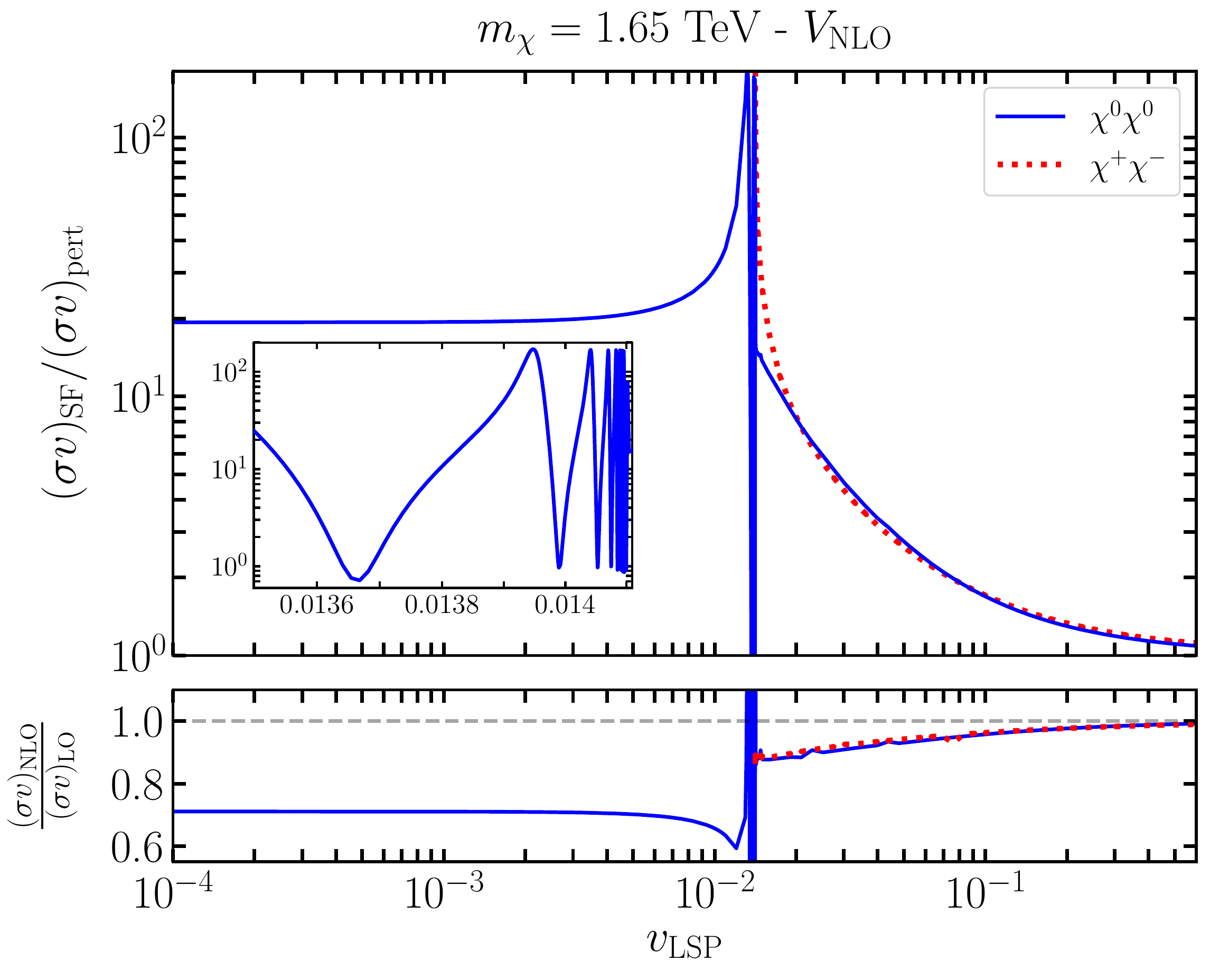}
    \includegraphics[width=0.61\textwidth]{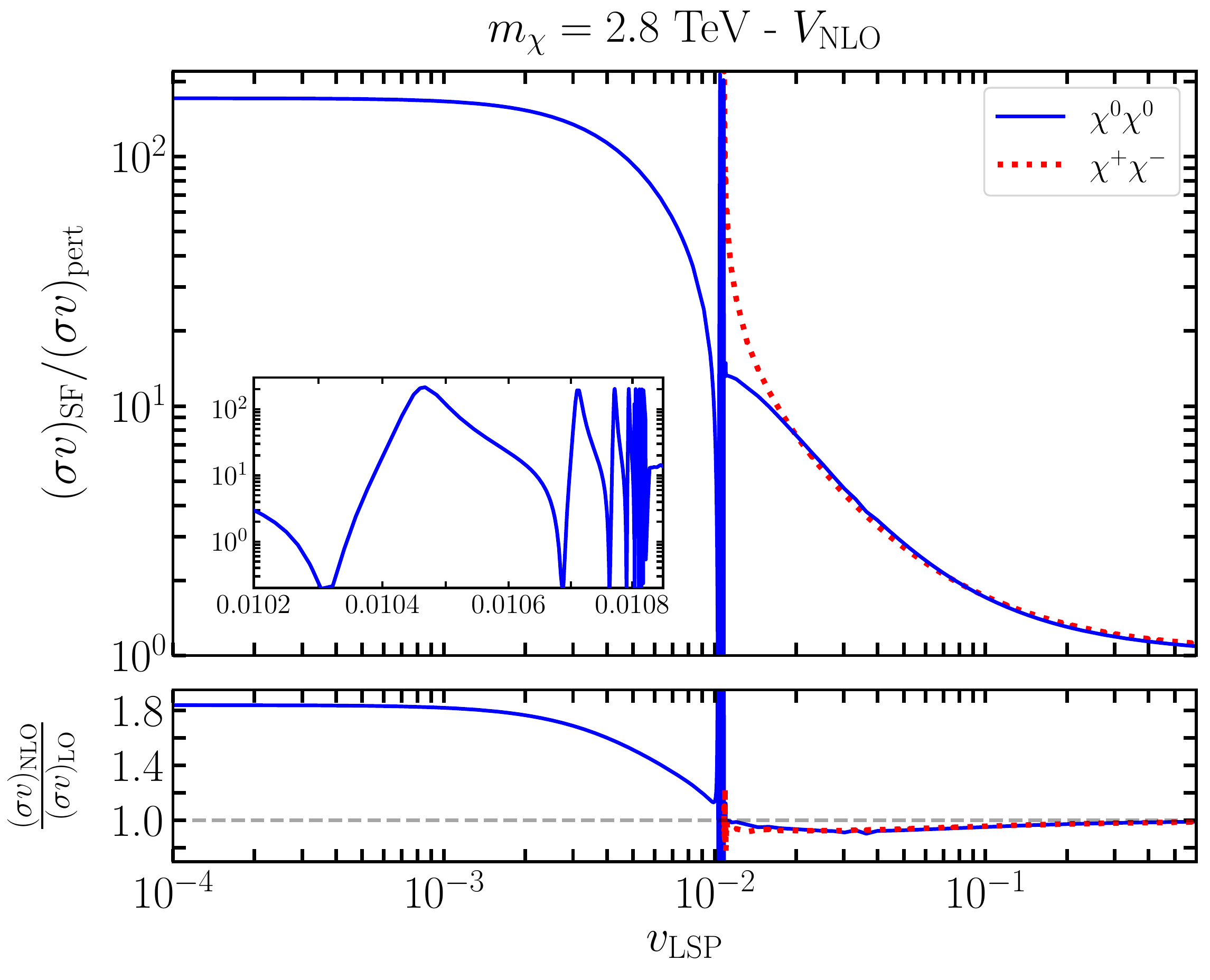}
\caption{The Sommerfeld factors for
the two channels $\chi^0 \chi^0$ and $\chi^+ \chi^-$ computed with 
the NLO potential for $m_\chi =
1.65$ TeV (upper panel) and $2.8$ TeV (lower panel). The inset 
zooms into the region
around $v_{\rm LSP}= \sqrt{2 \delta m_\chi / m_\chi}$, where Coulomb
bound states cause a rapid oscillation in the Sommerfeld factor. 
Below each panel, the effect of the NLO correction to the 
potential is highlighted by showing the ratio of the NLO to the 
LO Sommerfeld factor.}
    \label{fig:SFactor}
\end{figure}

We compare the Sommerfeld-enhanced cross section to the Born
cross section for the most important charge-neutral 
annihilation channels $\chi^0 \chi^0$ and $\chi^+ \chi^-$ in 
Fig.~\ref{fig:SFactor}. The two panels refer to two  mass values,  
one below and one above the first resonance. As the velocity 
decreases from right to left, the Sommerfeld factor 
increases, crosses the $\chi^+ \chi^-$ threshold at 
$v= \sqrt{2 \delta m_\chi/m_\chi}$ and reaches saturation 
for $v< 10^{-3}$. The spikes slightly below the
$\chi^+ \chi^-$ threshold are due to the Coulomb bound 
states \cite{Beneke:2014hja}, enlarged in the inset of
Fig.~\ref{fig:SFactor}. Formally, at the threshold the 
$\chi^+ \chi^-$ Sommerfeld factor becomes infinite as the relative 
velocity in this channel
\begin{align}
    v_{+-} = 2\, {\rm Re}\, \sqrt{\left(m_\chi v_{\rm
            LSP}^2- 2 \delta m_\chi\right)/(\mchi + \delta m_\chi) }\, \to\, 0\, ,
\end{align}
where $v_{\rm LSP}$ is the velocity of the lightest DM particle.
However, this will not spoil the abundance calculation, as in the 
thermal average, the $1/v_{+-}$ divergence is removed by the 
integration measure $d^3
\mathbf{v}_{+-} = v_{+-}^2 d v_{+-} d \Omega$. 

Below each panel, the effect of the NLO correction to the 
potential is highlighted by showing the ratio of the NLO to the 
LO Sommerfeld factor. For small velocities of the lightest 
two-particle state $\chi^0 \chi^0$, the NLO potential 
correction is very important, causing  $\mathcal{O}(1)$ 
changes in the Sommerfeld factor, as already observed in 
\cite{Beneke:2019vhz} for the $\gamma+X$ final state.  
For velocities above 0.01$-$0.1, the modification is closer to 
the few percent of a typical electroweak one-loop correction. 
Both velocity regimes are important for the relic abundance 
calculation, and their precise weight depends on 
the temperature at which freeze-out ends. In particular, 
for DM masses near the Sommerfeld resonances, the small 
velocity region is important and freeze-out is delayed.  

\subsubsection{Thermally-averaged annihilation cross section}

\begin{figure}[t]
    \centering
    \includegraphics[width=0.48\textwidth]{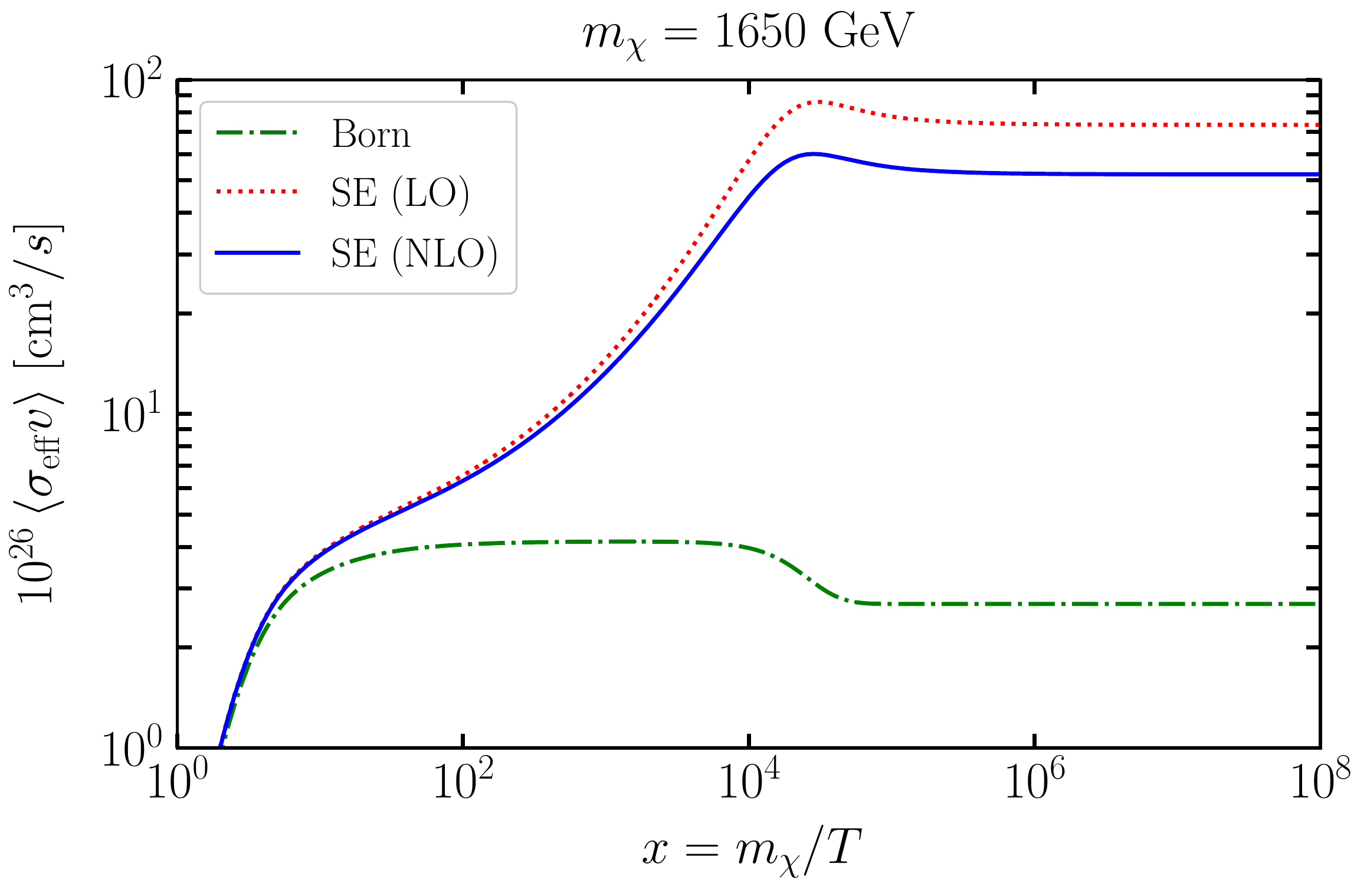}
    \includegraphics[width=0.48\textwidth]{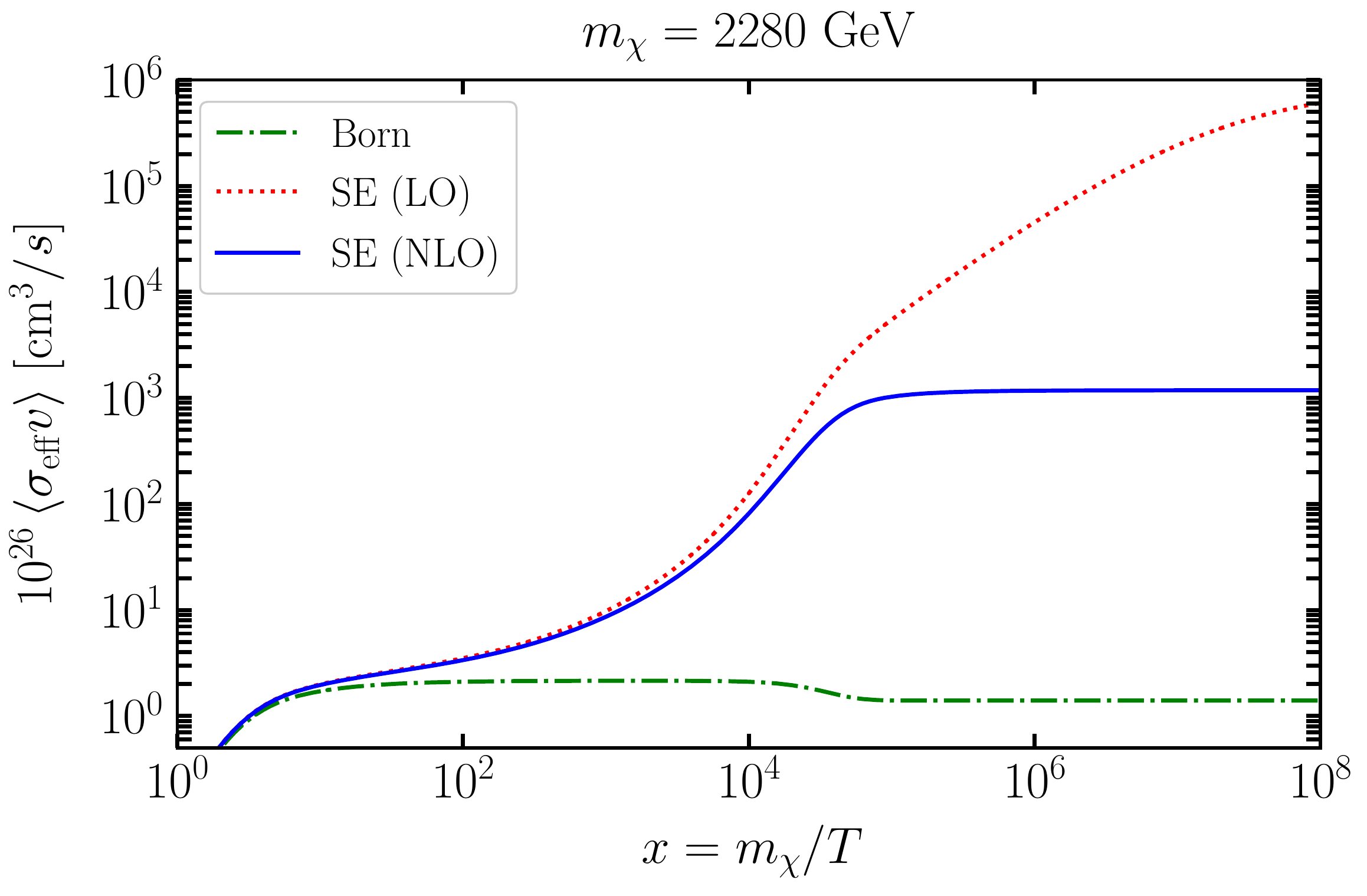}
    \includegraphics[width=0.48\textwidth]{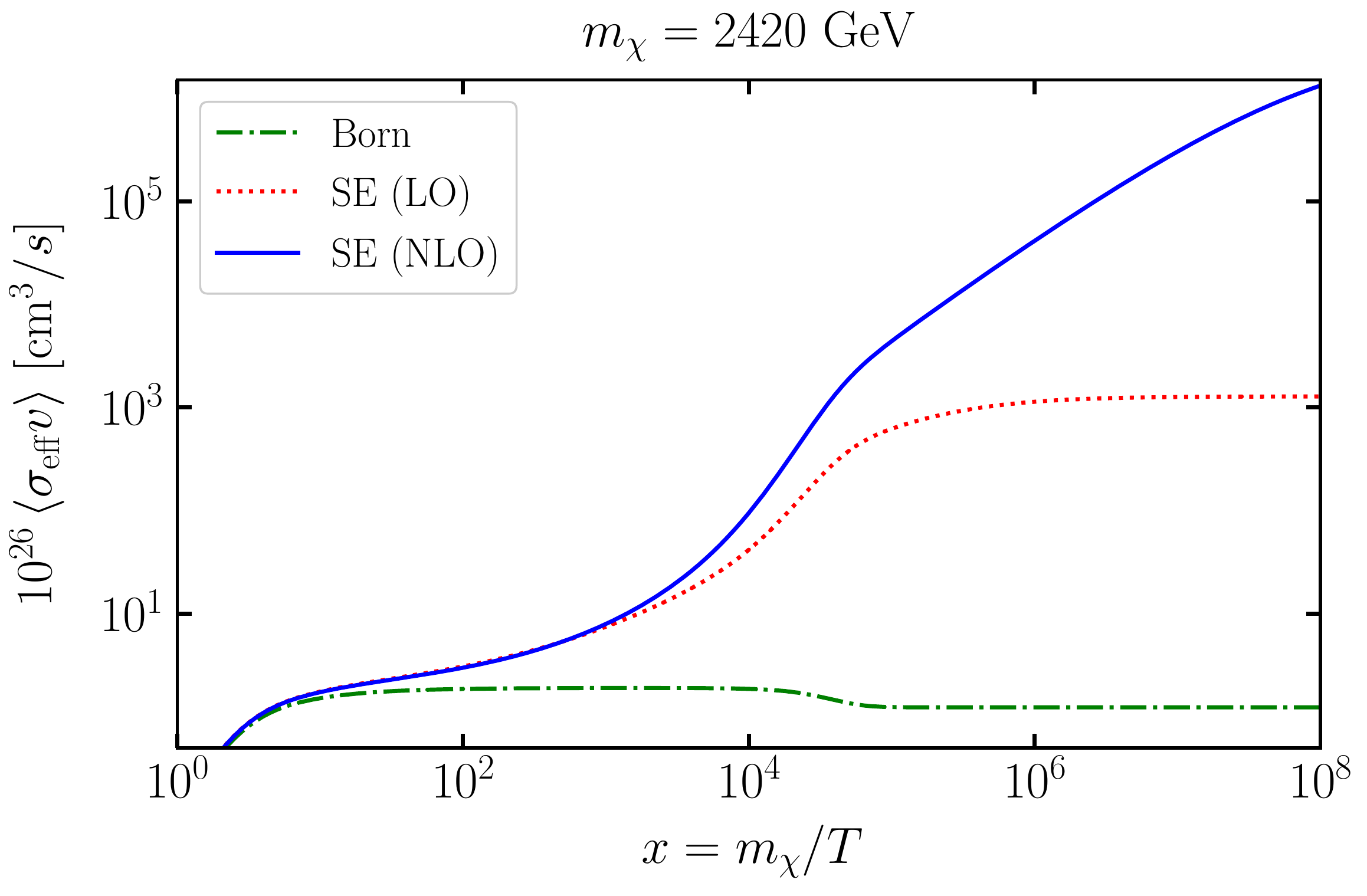}
    \includegraphics[width=0.48\textwidth]{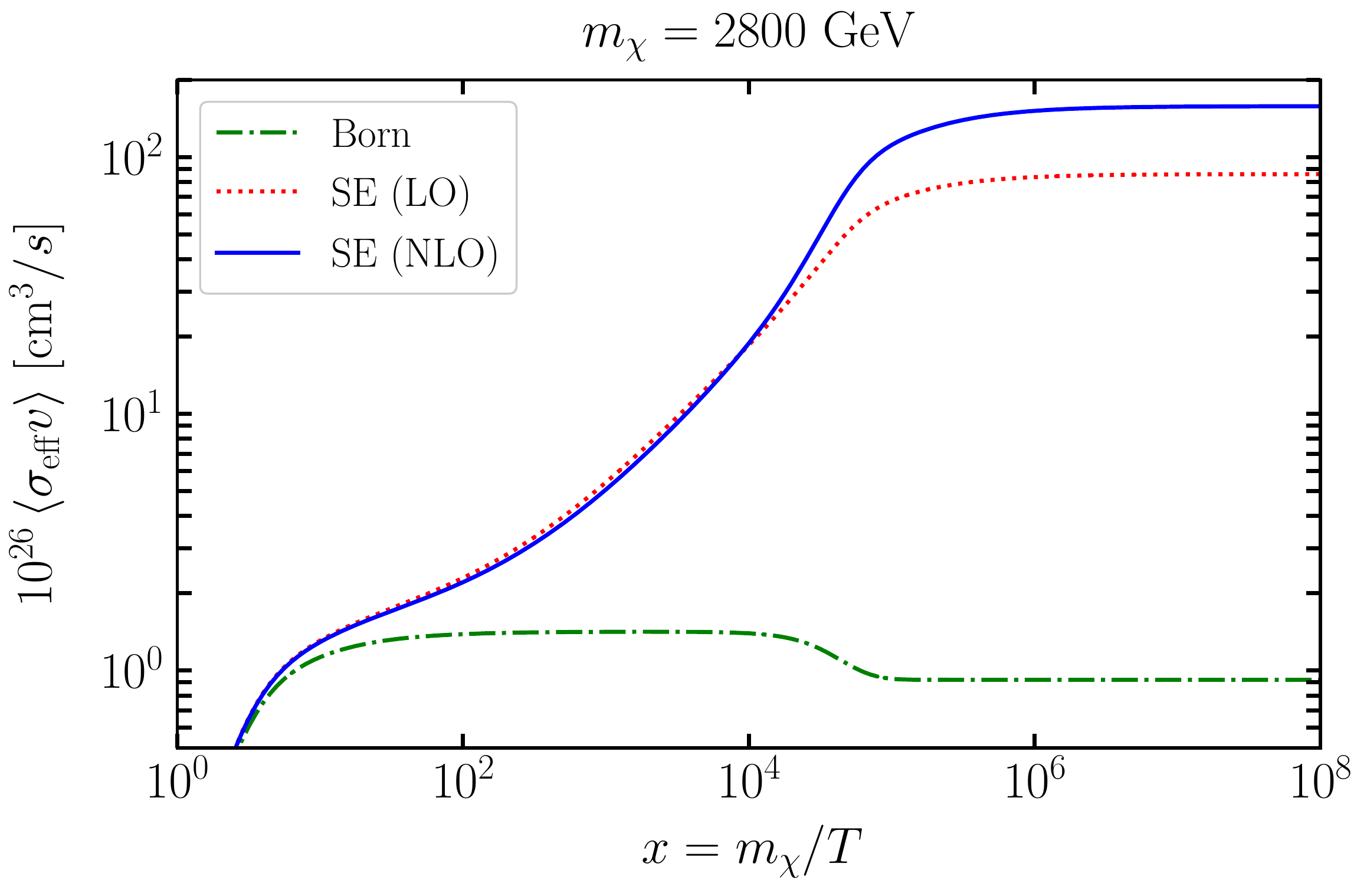}
\caption{The thermally-averaged cross section $\langle \sigma_{\rm eff} v \rangle$ in the Born approximation (green/dot-dashed), and with 
the Sommerfeld enhancement from the LO (dotted/red) and NLO 
(solid/blue) potential as a function of $x = m_\chi/T$.  We show 
$\langle \sigma_{\rm eff} v \rangle$ for mass values
below the first resonance ($m_\chi =1.65$ TeV) and
above ($m_\chi = 2.8$ TeV), and near 
the first resonance of the LO ($m_\chi = 2.28$ TeV) and
the NLO ($m_\chi = 2.42$ TeV)  potential.}
    \label{fig:thermav}
\end{figure}

The thermally-averaged cross section $\langle \sigma_{\rm eff} v 
\rangle$ is shown in Fig.~\ref{fig:thermav} for mass values   
chosen below and above, and at the resonance masses of the LO and NLO 
potential. At $x \lesssim 10$, the non-relativistic approximation is not accurate, and in
principle, a relativistic treatment is necessary. This, however, has a
negligible impact on the final abundance, as freeze-out starts for $x
\sim 20$, where the non-relativistic expansion is already very 
precise, given that $\mathcal{O}(v^2)$ terms 
have been included.  Around $x \sim m_\chi/\delta m_\chi$, a small drop 
in the Born cross section due to the decoupling of the heavier $\chi^+
\chi^-$ channel can be seen.  Except at the resonance, where the 
$1/v^2 \sim x$ enhancement persists, the cross section saturates at 
larger $x$ values and reaches a constant value.

For mass values below the first resonance, the NLO 
Sommerfeld-enhanced cross section is always smaller than the 
one computed with LO potential. After the first
resonance of the NLO potential, the late-time thermally 
averaged cross section can be larger than
for the LO potential, but even near the first resonance 
around $m_\chi=2.42~$TeV, the NLO potential cross section 
exceeds the LO one only around $x \,\grtsim \,100$. The reason is 
that the NLO correction 
weakens the potential, see Sec.~\ref{sec:NLOpot}. Therefore, 
the Sommerfeld effect is generally reduced, as seen in the small-$x$ 
regime. Due to the weaker NLO potential, the first Sommerfeld resonance 
is shifted to a larger mass value. In a significant 
mass range above the resonance, the resonant behaviour can 
outweigh the suppression due to the NLO 
potential and the NLO cross section can be larger than at LO 
(see, in particular, the lower panel of Fig.~4 
in \cite{Beneke:2019qaa}).

\subsubsection{The dark-matter yield $Y$}

\begin{figure}[t]
    \centering
    \includegraphics[width=0.48\textwidth]{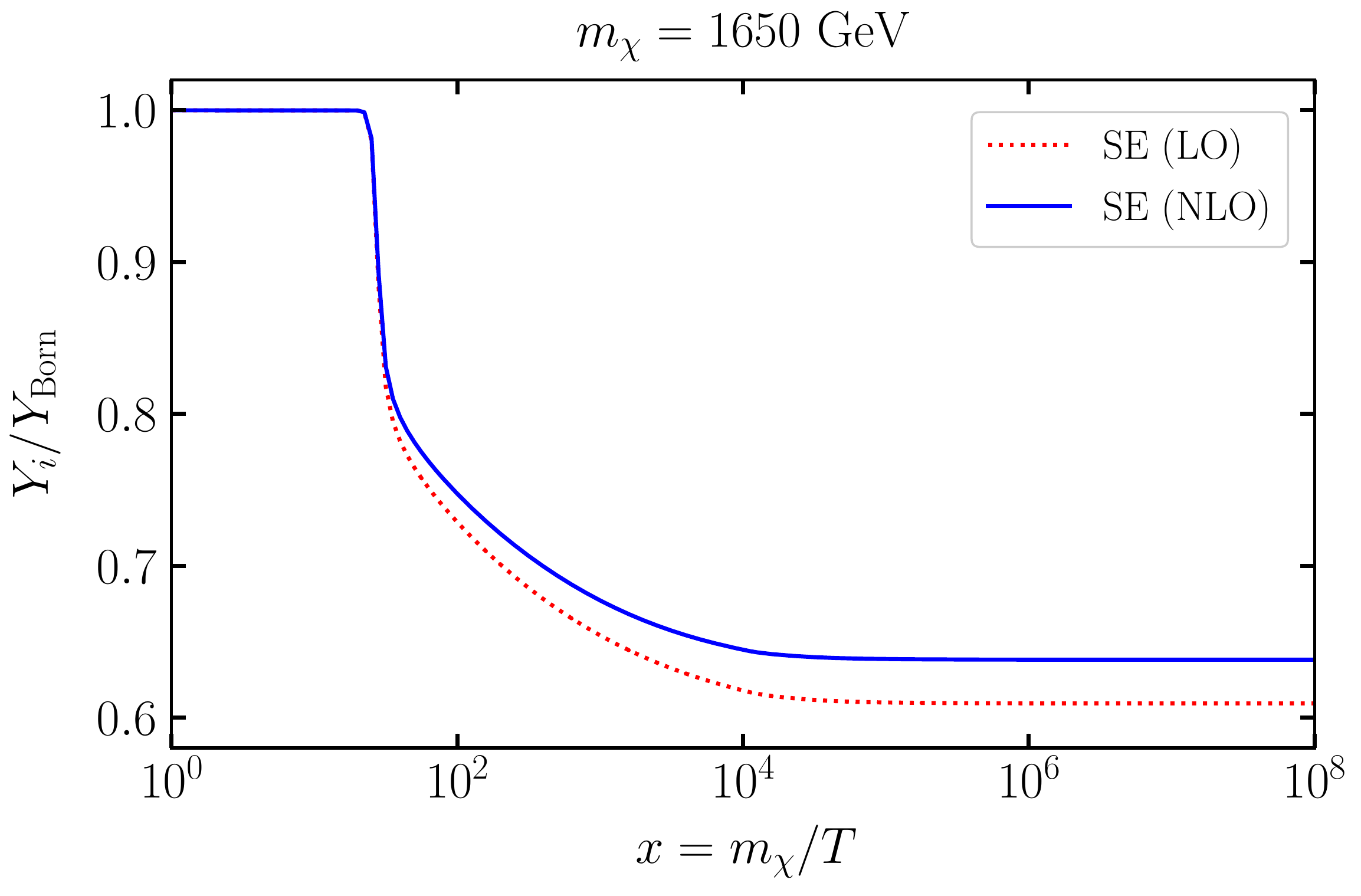}
    \includegraphics[width=0.48\textwidth]{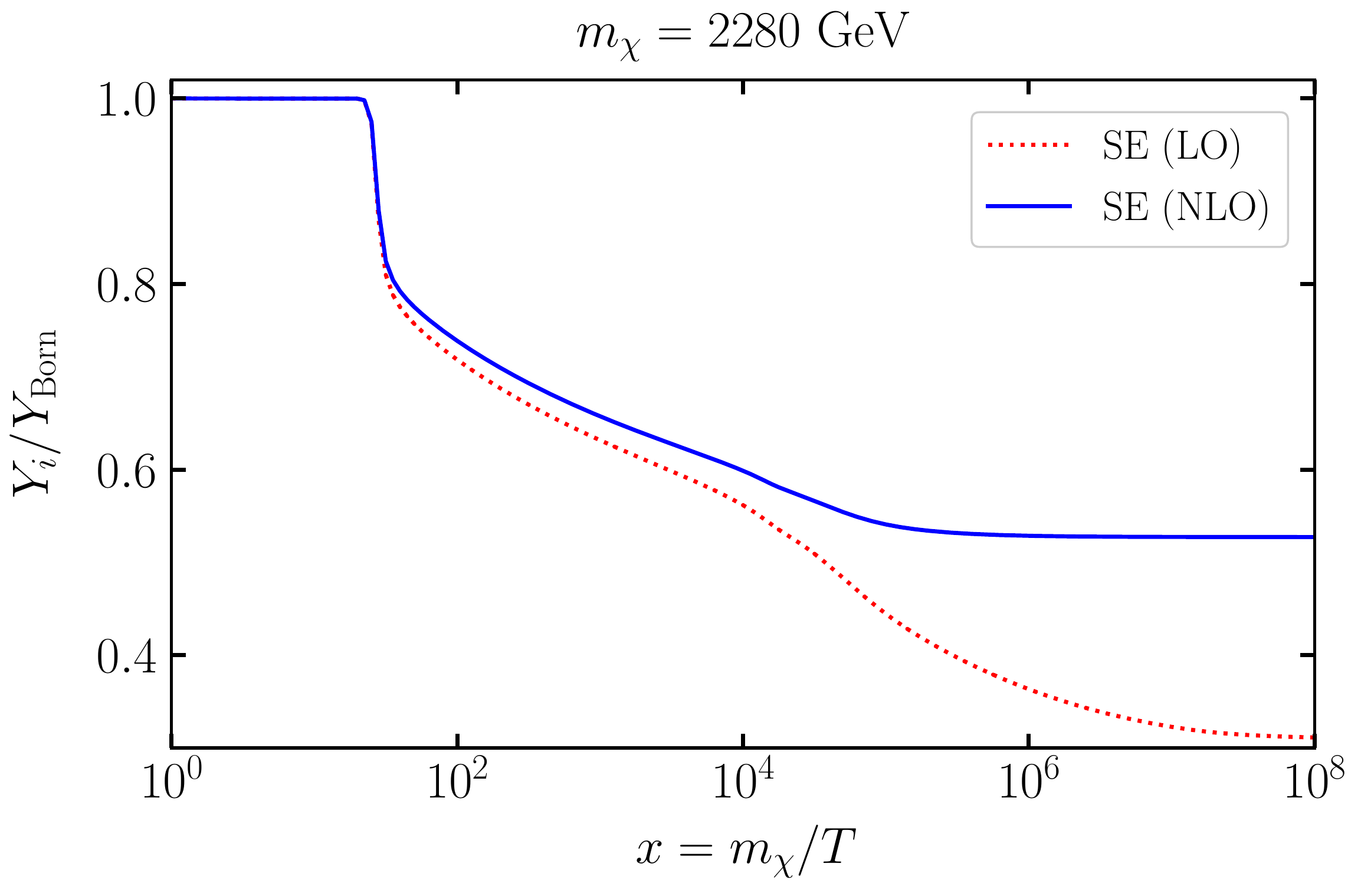}\\
    \includegraphics[width=0.48\textwidth]{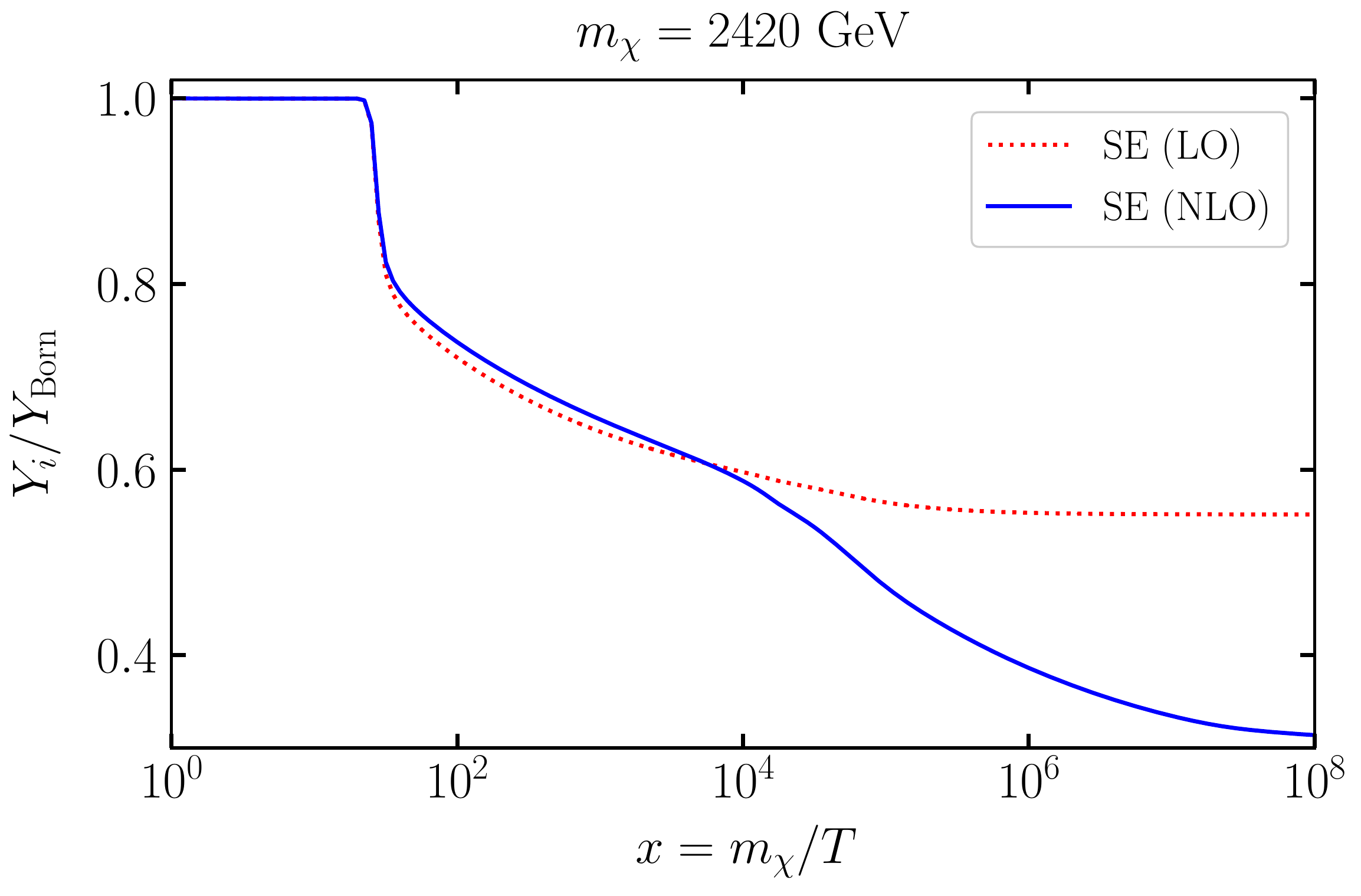}
    \includegraphics[width=0.48\textwidth]{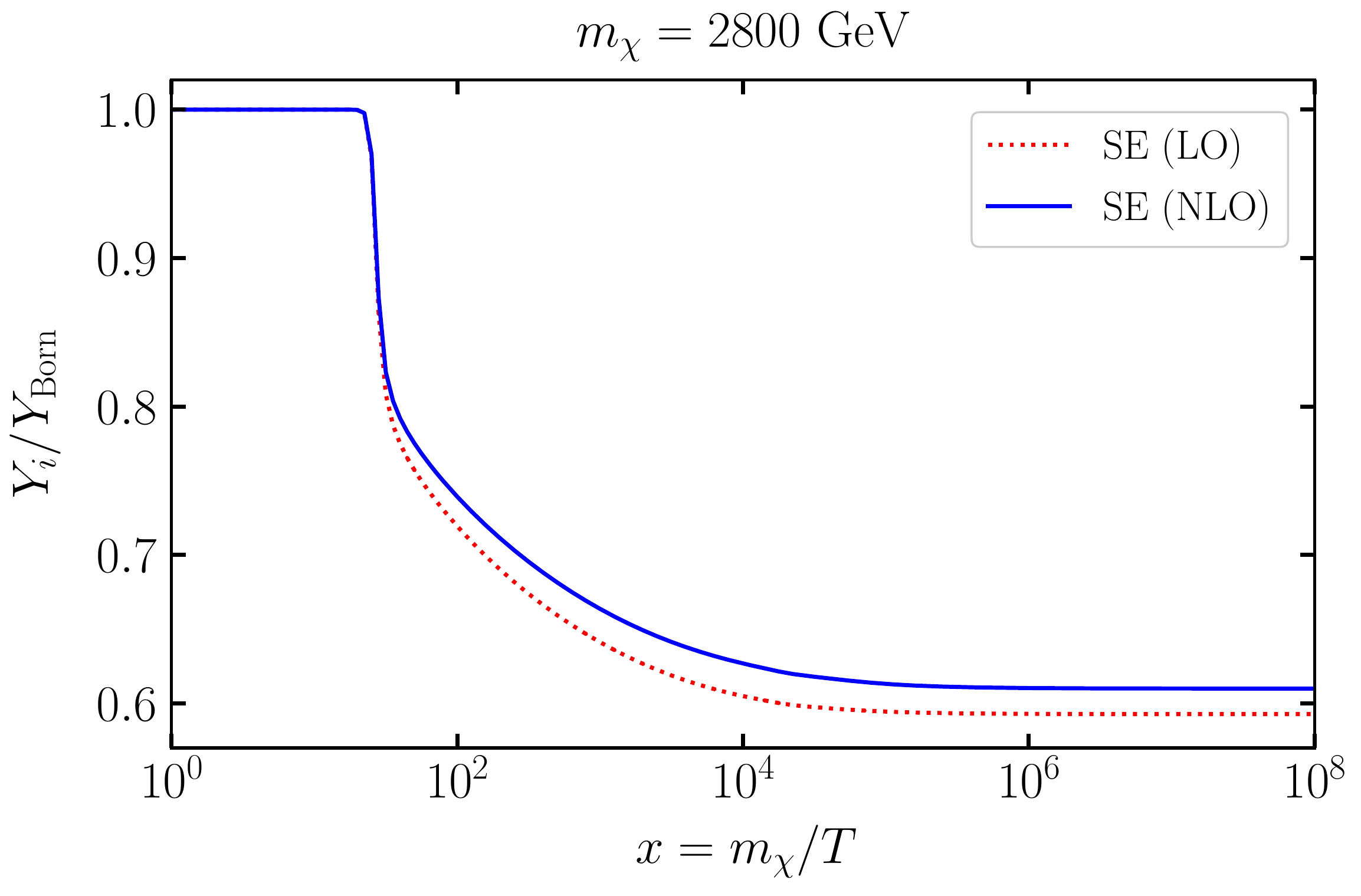}
\caption{The ratio $Y_i/Y_{\rm Born}$ as a function of $x =m_\chi/T$ 
for several values of $m_\chi$ as in Fig.~\ref{fig:thermav}. The 
perturbative yield $Y_{\rm Born}$ is computed with the Born cross 
section only, while the $Y_i$ include the LO (dotted/red) and 
the NLO Sommerfeld potential (solid/blue).}
    \label{fig:yieldvsx}
\end{figure}

Fig.~\ref{fig:yieldvsx} shows the suppression of the dark matter 
density due to the Sommerfeld effect at LO and NLO relative to 
the Born treatment. Until $x \sim 20$ all yields are the same, i.e., 
the ratio is unity, as the freeze-out process has not yet started. 
Afterwards, there is a steep drop when the Sommerfeld-enhancement 
of the annihilation cross section leads to a faster 
depletion of the DM density, and simultaneously delays the 
end of the freeze-out process. Around $x \sim 10^4$, the 
late-time annihilations cease, and the yield ratio again becomes 
constant. An exception occurs for DM masses near the resonance 
values, where the cross sections are enhanced by $1/v^2 \sim x$ in 
the low-velocity regime, leading to an additional
$\mathcal{O}(1)$ late-time change of the yield, as seen in the 
second (resonance of the LO potential at 2.28~TeV) and 
third panel (resonance of the NLO potential at 2.42~TeV).

It follows from the previous discussion of the thermally-averaged 
cross section that at early times (low $x$), the NLO yield is always 
larger than the LO one. Only at late times, the $1/v^2$ ($1/T$) 
enhancement of the (thermally-averaged) cross section for masses around the resonance value can lead to substantial late-time 
annihilations, and the NLO result drops below the LO result 
(see lower left panel of Fig.~\ref{fig:yieldvsx}). 

\subsubsection{The wino relic abundance}

\begin{figure}[t]
    \centering
    \includegraphics[width=0.75\textwidth]{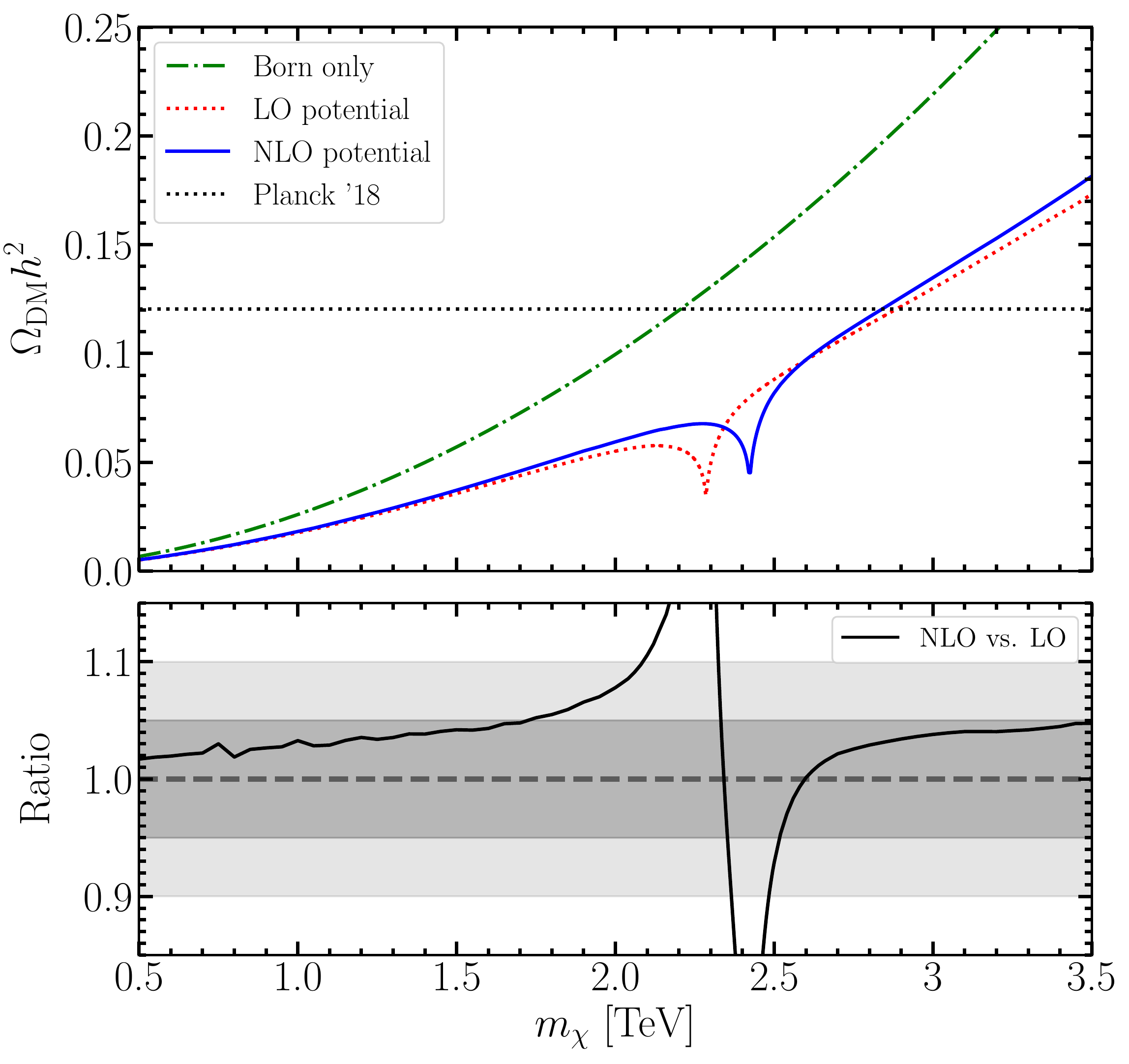}
\caption{Wino DM  relic abundance as function of 
DM mass $m_\chi$ computed with Born cross sections 
(dashed-dotted/green), including Sommerfeld enhancement with the 
LO (dotted/red), and the NLO potential (solid/blue). 
The horizontal line shows the observed relic abundance. 
In the lower panel, the ratio of the NLO to LO Sommerfeld-corrected 
relic abundance is shown. The dark (light) grey 
bands mark the 5 (10)\% variations. Except in the vicinity of the 
resonances the NLO correction is a few percent. The small wiggles 
in the ratio plot are due to numerical inaccuracies of the 
calculation.}
    \label{fig:relicvsmass}
\end{figure}

Our main result, the wino relic density as a function of wino mass 
with NLO potential is shown in Fig.~\ref{fig:relicvsmass} and 
compared to the previous LO result, and the Born approximation. 

The importance of accounting for the Sommerfeld effect is 
well-known and clearly seen in the figure.  In comparison, the 
NLO correction is moderate, except near the location of the 
resonance, whose shift is very visible. Aside from the 
resonance mass region, the NLO relic abundance is larger by about
$(2-5)\,\%$ than the LO one. 
The observed relic abundance 
$\Omega_{\rm DM} h^2 = 0.1205$ \cite{Aghanim:2018eyx}  
is attained for 
wino mass $2.842 \, {\rm TeV}$ with NLO potentials
compared to $2.886 \, {\rm TeV}$ at LO (and $2.207\,{\rm
TeV}$ for Born cross-section calculations). 
In comparison, the omission of the $\mathcal{O}(v^2)$ 
terms in the cross 
sections (\ref{eq:SFenhancedsigma}) would decrease the relic 
abundance by about $(1-3)\,\%$.

It is difficult to quantify the accuracy of the NLO relic abundance. 
Probably the largest uncertainty arises from the missing 
one-loop radiative corrections to the hard annihilation process. 
The use of running couplings in the tree-level  
process has accounted for the dominant logarithmically 
enhanced correction, which has a
considerable effect. For example, for 
$m_\chi = 3~$TeV, close to the value for the observed relic 
abundance, the relevant ratio of couplings is
$\hat{\alpha}_2^2(2 \times 3 \, {\rm TeV})/\hat{\alpha}_2^2(m_Z) = 
0.867$. In Sudakov resummation studies of wino annihilation  
into $\gamma+X$ \cite{Beneke:2018ssm,Beneke:2019vhz}, 
the non-logarithmic one-loop correction to the hard 
annihilation process was found to be at the $2\,\%$ level. 
Together with other sources of 
uncertainties, such as in the effective number of degrees 
of freedom, we estimate that the above result for the relic 
abundance is accurate to a few percent.


\section{Conclusion}
\label{sec:conclusions}

We provided a detailed description of our 
computation \cite{Beneke:2019qaa} of the NLO corrections to the 
electroweak Yukawa potential relevant to the Sommerfeld effect 
for wino dark matter. The calculation was performed in a general 
covariant gauge, and all necessary loop integrals were provided. 
The previous result was extended to include NLO effects in all 
co-annihilation channels. 

The main result of this paper is the first computation of a relic 
abundance with NLO accuracy for the Sommerfeld effect. Since the NLO 
correction weakens the potential, the value of the 
relic abundance increases by a few percent, except near the 
Sommerfeld resonances, where large effects are present and 
which are shifted towards larger masses. Our   
numerical investigations accentuate the importance of the 
NLO potential corrections for WIMP searches, and we advocate 
using the NLO Yukawa potential, which can easily be implemented 
using the fitting functions provided in Sec.~\ref{sec:fitfn}.

The computations presented here might be extended to more general 
models, such as the minimal models with general SU(2) electroweak 
multiplets. Further, it would be interesting to investigate the 
Higgsino model and DM particles with hypercharge, 
and combine the NLO potential with the state-of-the-art computations 
of the high-energy photon spectrum \cite{Beneke:2019gtg}, 
for which the missing NLO correction to the potential 
is likely to be the current largest theoretical uncertainty. 

\paragraph{Acknowledgements}
We thank M. Drees and M. Laine for correspondence. 
This work was supported in part by the DFG Collaborative Research 
Centre ``Neutrinos and Dark Matter in Astro- and Particle Physics'' 
(SFB 1258).

\appendix

\section{Expressions for the loop diagram topologies}
\label{app:loops}

In Sec.~\ref{sec:NLOpot}, we expressed the NLO correction 
to the potential in terms of functions representing diagram 
topologies. In this appendix, we provide the explicit results 
for these functions.

\subsection{Feynman gauge results}

We first collect the results for the diagram topologies
depicted in Fig.~\ref{fig:feynpmpm} in Feynman gauge. In the
following, we use the vector $v^\mu = (1,\mathbf{0})$ 
and the abbreviation
\begin{align}
  [dl] = \tilde{\mu}^{2 \epsilon} \frac{d^d l}{(2 \pi)^d} = \left(\frac{\mu^2
  e^{\gamma_E}}{4 \pi}\right)^{\!2 \epsilon} 
\,\frac{d^d l}{(2 \pi)^d} \,, 
\end{align}
where $d=4-2 \epsilon$ and $\gamma_E$ is the Euler-Mascheroni 
constant. The calculation of the diagrams
can be done with standard methods. Note that pinch poles at $v \cdot l = \pm i
\varepsilon$ must not be taken into account in the integration. The residues of
these poles are part of the Sommerfeld factor calculated with the leading-order
potential.

\subsubsection{Box topologies}
We first consider the box topologies in the first line of
Fig.~\ref{fig:feynpmpm}. The crossed-box diagrams are related trivially to the
box topologies by a minus sign in the case of the wino. We define
$\lambda(a,b,c) = (a-b-c)^2-4bc$. In the case of two unequal-mass bosons, we
find
\begin{align}
  I_{\text{box}}&(\alpha_i,m_i;\alpha_j,m_j) = i g_i^2 g_j^2 \int [dl]\ 
  \frac{-i}{l^2 - m_i^2+i \varepsilon} \frac{-i}{(l+q)^2 -m_j^2 + i
  \varepsilon} \frac{i}{v \cdot l + i \varepsilon} \frac{i}{-v \cdot l + i
\varepsilon} \nonumber \\
&=\frac{4 \alpha_i \alpha_j}{\lambda^{1/2}(-\mathbf{q}^2,m_i^2,m_j^2)}\ln
\left[\frac{m_i^2+m_j^2 + \mathbf{q}^2 +
\lambda^{1/2}(-\mathbf{q}^2,m_i^2,m_j^2)}{2 m_i m_j}\right] \, .
\label{eq:IboxuneqFeyn}
\end{align}
The box integral with masses is finite in $d=4$. 
$ \mathbf{q}^2\approx -q^2$
refers to the  exchanged potential three-momentum. If the masses of the
exchanged bosons are equal, this simplifies to
\begin{align}
    I_{\text{box}}(\alpha_i,m_i;\alpha_j, m_i)&=\frac{4 \alpha_i
      \alpha_j}{|\mathbf{q}| \sqrt{4 m_i^2 + \mathbf{q}^2}} \ln \left[\frac{2
  m_i^2+\mathbf{q}^2+\sqrt{4 m_i^2 \mathbf{q}^2 + \mathbf{q}^4}}{2
  m_i^2}\right] \, .
    \label{eq:IboxeqFeyn}
\end{align}
Finally there are the cases where one or both of the exchanged bosons are
massless. In these cases the box diagrams, expanded in $\epsilon$ including the
finite terms, result in 
\begin{align}
    I_{\text{box}}(\alpha_i,m_i;\alpha_j,0)&= -\frac{2 \alpha_i
      \alpha_j}{m_i^2+\mathbf{q}^2} \, \left(\frac{1}{\epsilon}+ \ln
      \left[\frac{\mu^2}{m_i^2+\mathbf{q}^2}\right]
    + \ln \left[\frac{m_i^2}{m_i^2 + \mathbf{q}^2}\right]\right) \, ,
    \label{eq:IboxonemassFeyn} \\
    I_{\text{box}}(\alpha_i,0; \alpha_j,0) &= -\frac{4 \alpha_i
    \alpha_j}{\mathbf{q}^2} \left(\frac{1}{\epsilon}+ \ln
  \frac{\mu^2}{\mathbf{q}^2}\right) \, . \label{eq:Ibox00Feyn}
\end{align}

\subsubsection{Vertex corrections}
The soft correction to the vertex (first diagram in the second row of
Fig.~\ref{fig:feynpmpm}) is known as it appears also in soft 
corrections to exclusive DM
annihilation \cite{Beneke:2019vhz}. We choose the convention
\begin{align}
  I_{\text{vertex}}(\alpha_i,m_i) &=g_i^2\int [dl]\ 
    \frac{-i}{l^2 - m_i^2+i \varepsilon} \frac{i}{v \cdot l + i \varepsilon}
    \frac{i}{-v \cdot l - i \varepsilon} = -\frac{\alpha_i}{2 \pi}
    \left(\frac{1}{\epsilon} + \ln \frac{\mu^2}{m_i^2}\right) \, .
    \label{eq:vertexFeyn}
\end{align}
The soft vertex correction involving the triple
gauge-vertex (cf.~Fig.~\ref{fig:feynpmpm}) vanishes in 
Feynman gauge, as the vertices project on the zero-component of the
propagator
\begin{align}
  I^{\gamma/Z/W}_{\rm{3 gauge}} &= 0 \, .
  \label{eq:tripvertexFeyn}
\end{align}

\subsubsection{DM field renormalization}
\label{app:heavyDM}
The renormalization of the heavy DM field is analogous to HQET. 
The required integral is
\begin{align}
  I_{\text{heavy DM}}(g_i,m_i) &= g_i^2\, \int [dl] \
    \frac{-i}{l^2 - m_i^2+i \varepsilon} \frac{i}{v \cdot (l+p)+ i \varepsilon}
    \nonumber \\
    &= \frac{i \, \alpha_i \, v \cdot p}{2 \pi} \left(\frac{1}{\epsilon} +
  \ln\frac{\mu^2}{m_i^2}\right)  - \frac{i \,\alpha_i\, m_i}{2} +
  \mathcal{O}\left((v \cdot p)^2\right)\, .
    \label{eq:heavyintegral}
\end{align}
From the term proportional to $v \cdot p$, we extract the on-shell 
static fermion field renormalization constants
\begin{align}
  \delta Z_{\chi^0}&=\frac{\alpha_2}{4 \pi} \left[\frac{2}{\epsilon}-2
  \ln\frac{m_W^2}{\mu^2} \right] \, , \label{eq:DeltaZ00} \\
  \delta Z_{\chi^+}&= \delta Z_{\chi^-}=\frac{\alpha_2}{4 \pi}
  \left[\frac{1+c_W^2}{\epsilon}- \ln\frac{m_W^2}{\mu^2} - c_W^2
  \ln\frac{m_Z^2}{\mu^2}\right] . \label{eq:DeltaZpm}
\end{align}
The $v \cdot p $ independent term generates the mass difference between the
neutral and charged DM fermions in the non-relativistic theory. For the wino we
find \cite{Cheng:1998hc}
\begin{align}
    \delta m_{\chi^0} &= - \alpha_2 m_W \, ,\\
    \delta m_{\chi^+} &= - \frac{\alpha_2}{2} m_W - \frac{\alpha_2 c_W^2}{2}
    m_Z \, , \\
    \delta m_\chi &= \delta m_{\chi^+} - \delta m_{\chi^-} = \frac{1-c_W}{2} \,
    \alpha_2 m_W \, .
\end{align}

\subsubsection{Gauge boson self-energies}
The self-energies in in Feynman gauge are not given explicitly for brevity, as
we give the self-energies in $R_\xi$-gauge in the following section. They
can e.g.~be found in \cite{Denner:1991kt,Beneke:2019vhz}.

\subsection{Expressions in $R_\xi$-gauge} 

All gauge-parameter dependent parts can be expressed in terms of the 
Passarino-Veltman (PV) functions \cite{Passarino:1978jh}
(expanded to $\mathcal{O}(\epsilon^0)$)
\begin{align}
  A_0 (m) &= m^2 \left[\frac{1}{\epsilon} + \ln \frac{\mu^2}{m^2} +1\right] \,,
  \nonumber \\ 
  B_0(p^2,0,0) &= \frac{1}{\epsilon} + \ln \left(-\frac{\mu^2}{p^2}\right) + 2
  \, , \nonumber \\ 
  B_0(p^2,0,m) &=\frac{1}{\epsilon} + \ln \frac{\mu^2}{m^2} + 2 - \left(1-
\frac{m^2}{p^2}\right) \ln \left(1- \frac{p^2}{m^2}\right)\, ,  \nonumber \\
B_0(p^2,m,m) &= \frac{1}{\epsilon} + \ln \frac{\mu^2}{m^2} + 2 +
\sqrt{1-\frac{4 m^2}{p^2}} \ln \left(\frac{2 m^2 - p^2 +\sqrt{p^2(p^2-4
m^2)}}{2 m^2}\right)\,, \nonumber \\ 
B_0(p^2,m_1,m_2) &= \frac{1}{\epsilon} - \frac{m_1^2 - m_2^2}{2 p^2} \ln
\frac{m_1^2}{m_2^2} + \frac{1}{2} \left(\ln \frac{\mu^2}{m_1^2} + \ln
    \frac{\mu^2}{m_2^2}\right) + 2 \nonumber \\ 
    &\quad+ \frac{\lambda^{1/2}(m_1^2,m_2^2,p^2)}{p^2} \ln\left( \frac{m_1^2 +
    m_2^2 - p^2 + \lambda^{1/2}(m_1^2,m_2^2,p^2)}{2 m_1 m_2} \right) \, ,
\end{align}
where $\lambda(a,b,c)$ is the triangle function defined above. 
Below, we use the abbreviation $\Delta_i = 1- \xi_i$, where 
$\xi_i$ denotes the gauge-fixing parameter for boson $i = A, W, Z$. 
The integrals $I_X$ provided below in general 
gauge are defined in terms of those for $\xi=1$ by 
substituting
\begin{align}
  \frac{g_{\mu \nu}}{l^2 - m_i^2} \quad  \to  \quad \frac{g_{\mu \nu}-\Delta_i
  \frac{l_\mu l_\nu}{l^2 - \xii \mi^2}}{l^2 - \mi^2}
\end{align}
in the appropriate gauge-boson propagators, e.g., in \eqref{eq:IboxuneqFeyn}.

\subsubsection{Box topologies}

The massive box evaluates to
\begin{align}
  &I^\xi_{\rm box}(\alpha_i,m_i;\alpha_j,m_j) = I^{\xi=1}_{\rm
box}(\alpha_i,m_i;\alpha_j,m_j)  \nonumber \\
  &+ \frac{\alpha_i \alpha_j}{12 \mathbf{q}^2}  \left[\delj
\frac{A_0(\mi)-A_0(\xmi)}{\mi^2} + \deli \frac{A_0(\mj) - A_0(\xmj)}{\mj^2}
\right. \nonumber \\
  &\hphantom{+\alpha_i \alpha_j \quad} -\frac{1}{\mi^2 \mj^2}
\left(\vphantom{\frac{1}{1}}\left(12 \mathbf{q}^2 (\mi^2 +\mj^2) - \lambda(
  -\mathbf{q}^2, \mi^2, \mj^2)\right) B_0(-\mathbf{q}^2,\mi,\mj) \right.
  \nonumber \\
  &\hphantom{+\alpha_i \alpha_j \quad +\frac{1}{\mW^2 \mZ^2} \quad}+ \left(
\lambda(-\mathbf{q}^2,\mi^2,\xij \mj^2) -12 \mathbf{q}^2 \mi^2\right)
B_0(-\mathbf{q}^2,\mi,\xmj) \nonumber \\
  &\hphantom{+\alpha_i \alpha_j \quad +\frac{1}{\mW^2 \mZ^2}
\quad}+\left(\lambda(-\mathbf{q}^2,\mj^2,\xii \mi^2)-12 \mathbf{q}^2 \mj^2
\right) B_0(-\mathbf{q}^2 ,\mj,\xmi) \nonumber \\
  &\hphantom{+\alpha_i \alpha_j \quad +\frac{1}{\mW^2 \mZ^2}
  \quad}- \lambda(-\mathbf{q}^2, \xii \mi^2, \xij \mj^2) \left. \left.
      B_0(-\mathbf{q}^2, \xmi, \xmj)
      \vphantom{\frac{-\mathbf{q}^2}{\mW^2}}\right) \right] \, ,
  \label{eq:boxuneqRxi}
\end{align}
where the gauge parameter with index $i$ is associated with the 
boson of mass $m_i$. For one
vanishing mass (where $\Delta_j,\xi_j$ are associated to the massless boson coupling
with $\alpha_j$)
\begin{align}
  &I^\xi_{\rm box}(\alpha_i,m_i;\alpha_j,0) = I^{\xi=1}_{\rm
box}(\alpha_i,m_i;\alpha_j,0) \hspace{8cm} \nonumber \\
  &-\alpha_i \alpha_j \left[\delj \frac{2(A_0(\mi)- \mathbf{q}^2) +
  (\mathbf{q}^2 - \mi^2) B_0(-\mathbf{q}^2,0,\mi)}{(\mathbf{q}^2+\mi^2)^2}
\right. \nonumber \\
  &\hspace{2cm} + \frac{B_0(-\mathbf{q}^2,0,\mi) -
  B_0(-\mathbf{q}^2,0,\xmi)}{\mi^2} \nonumber \\
  &\hspace{2cm} - \frac{\delj}{4\mathbf{q}^2 \mi^2} \left( A_0(\mi) -
A_0(\xmi) + (\mathbf{q}^2 - \mi^2) B_0(-\mathbf{q}^2,0,\mi) \right. \nonumber
\\
  &\hspace{4cm}-(\mathbf{q}^2-\xii \mi^2)\left.\left.
B_0(-\mathbf{q}^2,0,\xmi) \right) \vphantom{\frac{A_0
\mathbf{q}^2}{(\mathbf{q}^2)^2}} \right] \, .
  \label{eq:boxonemassRxi}
\end{align}
The massless box is given by ($\Delta_i,\xi_i$ and
$\Delta_j,\xi_j$ are associated to massless bosons coupling with $\alpha_i$ and
$\alpha_j$, respectively)
\begin{align}
  I^\xi_{\rm box}(\alpha_i,0;\alpha_j,0) &= I^{\xi=1}_{\rm
  box}(\alpha_i,0;\alpha_j,0) \nonumber \\
                                         &\quad-\frac{\alpha_i
\alpha_j}{\mathbf{q}^2} \left[\frac{\deli \delj}{2}+ (\deli +\delj)
(B_0(-\mathbf{q}^2,0,0)-2)\right] \, .
\label{eq:boxnomass}
\end{align}

\subsubsection{Vertex topologies}
For the vertex diagram without the triple gauge-boson vertex, we find
\begin{align}
  I_{\rm vertex}^{\xi}(\alpha_i,m_i) &= I_{\rm vertex}^{\xi = 1}(\alpha_i,m_i)
  - \frac{\alpha_i}{4\pi} \frac{A_0(\mi)-A_0(\xmi)}{\mi^2}  \, .
  \label{eq:vertexRxi}
\end{align}
The vertex diagram with the triple gauge interaction vanishes in
 Feynman gauge. Which particles circle in the loop depends
on the tree potential. For the Coulomb and $Z$-boson Yukawa potential,
the loop is comprised of two $W$-bosons and, therefore, only depends on
one gauge parameter $\xiW$. In the case of the $W$-boson Yukawa potential, the
loop is a mixture of $W$-boson and either photon or $Z$-boson and hence depends
on two gauge parameters and one, respectively two masses. 
The integral with $W/Z$ bosons in the loop is given by 
\begin{align}
  &I^{W Z}_{\rm 3 gauge} = -i g_2^2 c_W^2 \!\int [dl] \frac{i}{v \cdot l+i
    \varepsilon} \frac{-i \left(g_{0 \mu} - \delW \frac{l_0 l_\mu}{l^2 - \xiW
  \mW^2 + i \varepsilon}\right)}{l^2 - \mW^2 +  i \varepsilon} \, \frac{-i
  \left( g_{0\nu} - \delZ \frac{l_0 (l+q)_\nu}{(l+q)^2-\xiZ \mZ^2+i
  \varepsilon} \right)}{(l+q)^2 - \mZ^2 + i \varepsilon}  
\nonumber \\[0.2cm]
  &\hspace{4cm} \times g_{0 \rho} \, (g^{\mu \nu} (-2 l -q)^\rho + g^{\nu \rho}
  (l+2q)^\mu + g^{\rho \mu} (l-q)^\nu )\nonumber \\[0.2cm]
  &=\frac{\alpha_2 c_W^2}{4 \pi}\frac{1}{12} \left[2\left(\delW +\delZ\right) +
\left(\delZ -11 + (\delZ -1) \frac{\mW^2}{\mathbf{q}^2} +
\frac{\mZ^2}{\mathbf{q}^2}\right) \frac{A_0(\mW)}{\mW^2} \right. \nonumber \\
  &- \left(\delZ-11+(\delW -1) \frac{\mW^2}{\mathbf{q}^2} + \frac{\mZ^2}{\mathbf{q}^2}\right) \frac{A_0(\xmW)}{\mW^2} \nonumber \\
  &+ \left(\delW-11 + \frac{\mW^2}{\mathbf{q}^2} + (\delW -1) \frac{\mZ^2}{\mathbf{q}^2}\right) \frac{A_0(\mZ)}{\mZ^2} \nonumber \\
  &-\left(\delW -11 + \frac{\mW^2}{\mathbf{q}^2} + (\delZ -1) \frac{\mZ^2}{\mathbf{q}^2}\right) \frac{A_0(\xmZ)}{\mZ^2} \nonumber \\
  & +\left(c_W^2 \frac{\mathbf{q}^4}{\mW^4} - 9 (1+c_W^2) \frac{\mathbf{q}^2}{\mW^2} - \frac{9 c_W^4 -2 c_W^2 +9}{c_W^2} + \frac{s_W^4}{c_W^4} (1+c_W^2) \frac{\mW^2}{\mathbf{q}^2}\right) B_0(-\mathbf{q}^2,\mW,\mZ)\nonumber \\
  &-\left(c_W^2 \frac{\mathbf{q}^4}{\mW^4}+(2 (1-\delZ)-9 c_W^2) \frac{\mathbf{q}^2}{\mW^2} + \frac{(\delZ-1)^2 -9 c_W^4}{c_W^2} \right. \nonumber \\
  &\left. \hspace{3cm} +\frac{(c_W^2 +(\delZ-1))^2}{c_W^2} \frac{\mW^2}{\mathbf{q}^2}\right) B_0(-\mathbf{q}^2,\mW,\xmZ) \nonumber \\
  &-\left(c_W^2 \frac{\mathbf{q}^4}{\mW^4}+(2 c_W^2 (1-\delW)-9) \frac{\mathbf{q}^2}{\mW^2} + \frac{c_W^4 (\delW-1)^2 -9}{c_W^2} \right. \nonumber \\
  &\left. \hspace{3cm} +\frac{(1+c_W^2(\delW-1))^2}{c_W^4} \frac{\mW^2}{\mathbf{q}^2}\right) B_0(-\mathbf{q}^2,\xmW,\mZ) \nonumber \\
  &+ \left( c_W^2 \frac{\mathbf{q}^4}{\mW^4} +2 (c_W^2 (1-\delW) + 1-\delZ) \frac{\mathbf{q}^2}{\mW^2}\hspace{6cm} \right. \nonumber \\
  &\left. \left. \hspace{3cm}+ \frac{(c_W^2 (\delW-1)+1-\delZ)^2}{c_W^2} \right) B_0(-\mathbf{q}^2,\xmW,\xmZ) \right] \, .
\label{eq:tripvertexWZRxi}
\end{align}
If the $Z$-boson is replaced by the massless photon, 
we find
\begin{align}
  \label{eq:tripvertexWgRxi}
    &I^{W \gamma}_{\rm 3 gauge} = \frac{\alpha}{4 \pi} \left[\frac{2 \delA \mathbf{q}^2}{\mathbf{q}^2 + \mW^2} + \frac{\delW}{6}  \right. \nonumber \\
    &+\left(\frac{2 \delA \mathbf{q}^2}{\mathbf{q}^2 + \mW^2} + \frac{(3 \delA -1) \mW^2}{12 \mathbf{q}^2} - \frac{21 \delA +11}{12} \right) \frac{A_0(\mW)}{\mW^2} \nonumber \\
    &+\left(\frac{11 -3 \delA}{12} + \frac{(1-\delW) \mW^2}{12 \mathbf{q}^2}\right) \frac{A_0(\xmW)}{\mW^2}\\
    &+\left(\left[\frac{2 \delA \mathbf{q}^2}{\mathbf{q}^2 + \mW^2}- \frac{11 +21 \delA}{12} \right] \frac{\mathbf{q}^2}{\mW^2} + \frac{1+6\delA}{6} + \frac{1-3\delA}{12} \frac{\mW^2}{\mathbf{q}^2}\right) B_0(-\mathbf{q}^2,0,\mW) \nonumber \\
    &\left. +\frac{1}{12}\left((11-3\delA) \frac{\mathbf{q}^2}{\mW^2} + (1-\delW)(3 \delA -2)-(\delW-1)^2 \frac{\mW^2}{\mathbf{q}^2}\right) B_0(-\mathbf{q}^2,0,\xmW) \right] \, . \nonumber 
\end{align}
Finally, if both bosons involved are $W$-bosons, we have
\begin{align}
  &I^{W W}_{\rm 3 gauge} = \frac{\alpha}{4 \pi s_W^2} \left[\frac{\delW}{3} +\left(\frac{\delW}{6} \frac{\mathbf{q}^2 + \mW^2}{\mathbf{q}^2}- \frac{11}{6}\right) \frac{A_0(\mW)- A_0(\xmW)}{\mW^2} \right. \nonumber \\
  &+\left(\frac{1}{12} \frac{\mathbf{q}^4}{\mW^4} - \frac{3}{2} \frac{\mathbf{q}^2}{\mW^2} - \frac{4}{3}\right) B_0(-\mathbf{q}^2,\mW,\mW) \nonumber \\
  &-\frac{\mathbf{q}^2 +\mW^2}{6 \mathbf{q}^2} \left(\frac{\mathbf{q}^4}{\mW^4} - 2 (4+\delW) \frac{\mathbf{q}^2}{\mW^2} + \delW^2\right) B_0(-\mathbf{q}^2,\mW,\xmW) \nonumber \\
  &\left. + \left(\frac{1}{12} \frac{\mathbf{q}^4}{\mW^4}+ \frac{1-\delW}{3} \frac{\mathbf{q}^2}{\mW^2}\right) B_0(-\mathbf{q}^2,\xmW,\xmW) \right] \, .
  \label{eq:tripvertexWWRxi}
\end{align}

\subsubsection{DM field renormalization}

The heavy DM field renormalization constants 
in $R_\xi$-gauge are
\begin{align}
  \delta Z_{\chi^0}^\xi &= \delta Z_{\chi^0}^{\xi=1}+
  \frac{\alpha_2}{4 \pi} \frac{A_0(\mW)-A_0(\xmW)}{\mW^2} \, ,
  \label{eq:DeltaZ00Rxi} \\
  \delta Z_{\chi^+}^\xi &=  \delta Z_{\chi^+}^{\xi =1} +
  \frac{\alpha_2}{8 \pi}\frac{A_0(\mW)-A_0(\xmW)}{\mW^2} +\frac{\alpha_2
  c_W^2}{8 \pi}\frac{A_0(\mZ)-A_0(\xmZ)}{\mZ^2}  \, .
  \label{eq:DeltaZpmRxi}
\end{align}

\subsubsection{Gauge boson self-energies}

The transverse self-energies (including tadpole diagrams) are split 
into the three separately gauge-invariant pieces discussed in 
the main text according to
\begin{align}
  \Sigma_T^{ij} &= \Sigma_{T,\, {\rm light \,ferm.}}^{ij} + \Sigma_{T, \,{\rm
  3rd \, gen. \, quarks}}^{ij}+ \Sigma_{T, \, {\rm electroweak}}^{ij} \, .
\end{align}

\paragraph{The photon self-energy}\mbox{}\\
The fermionic self-energy contributions are
\begin{align}
  \Sigma^{\gamma \gamma}_{T,\,{\rm light \, ferm.}}(p^2) &= \frac{\alpha}{4
  \pi} \frac{76}{9} p^2 \left[B_0(p^2,0,0) - \frac{1}{3} \right]  \, ,
  \label{eq:SigmaggLight} \\
  \Sigma^{\gamma \gamma}_{T,\,{\rm 3rd\, gen. \, quarks}}(p^2)
  &=\frac{\alpha}{4 \pi} \left[ \frac{32}{9} \left(m_t^2 - A_0(m_t)\right)+
\frac{4}{9} p^2 \left( B_0(p^2,0,0) -\frac{5}{3}\right)\right. \nonumber \\
&\left. \hspace{2cm}+ \frac{16}{9} \left(p^2 +2 m_t^2\right) B_0(p^2,m_t,m_t)
\right] \, .
\label{eq:Sigmagg3rd}
\end{align}
The electroweak part of the self-energy reads
\begin{align}
  &\Sigma^{\gamma \gamma}_{T,\,{\rm electroweak}} (p^2) = \frac{\alpha}{4 \pi}
  \left[-4 \mW^2+\frac{2}{3} p^2 \delW  \right. \nonumber \\
  &+ \left(\frac{12 - \delW}{3} + \frac{\delW -10}{6} \frac{p^2}{\mW^2}\right)
  A_0(m_W)  \nonumber \\
  &+\left(\frac{\delW}{3} + \frac{10 - \delW}{6} \frac{p^2}{\mW^2}\right)
  A_0(\xmW) \nonumber \\
  &+ \left(-4 - \frac{17}{3} \frac{p^2}{\mW^2} + \frac{4}{3} \frac{p^4}{\mW^4}
+ \frac{1}{12} \frac{p^6}{\mW^6}\right) \mW^2 B_0(p^2,\mW,\mW) \nonumber \\
  &+\left(\frac{\delW^2}{3} + \frac{16+4 \delW - \delW^2}{6} \frac{p^2}{\mW^2}-
\frac{3+\delW}{3} \frac{p^4}{\mW^4} - \frac{1}{6} \frac{p^6}{\mW^6}\right)
\mW^2 B_0(p^2,\mW,\xmW) \nonumber \\
  &\left.+\left(\frac{\delW-1}{3} \frac{p^4}{\mW^4} + \frac{1}{12}
  \frac{p^6}{\mW^6}\right) \mW^2 B_0(p^2,\xmW,\xmW)\right] \, .
  \label{eq:SigmaggYM}
\end{align}

\paragraph{The photon-Z self-energy}
\begin{align}
  \Sigma^{\gamma Z}_{T,\,{\rm light\,ferm.}} (p^2) =& \,\frac{\alpha}{4 \pi s_W
  c_W} \frac{38 s_W^2 -38 c_W^2 +11}{9} p^2 \left(B_0(p^2,0,0) -
\frac{1}{3}\right)\, , \label{eq:SigmagZLight} \\
  \Sigma^{\gamma Z}_{T,\,{\rm 3rd\, gen.\,quarks}} (p^2) =& \, \frac{\alpha}{4
      \pi s_W c_W}\left[-\frac{2}{9}p^2 +\frac{4}{9} \left(1 + 4 s_W^2 - 4
          c_W^2 \right) \left(m_t^2 - A_0(m_t)\right)\right. \nonumber \\
  &-\frac{1}{9}\left(1+2 c_W^2 - 2s_W^2\right) p^2 \left(B_0 (p^2,0,0) -
  \frac{5}{3}\right) \nonumber \\
  &\left.+\frac{2}{9} \left(1+4s_W^2 -4 c_W^2\right) \left(p^2 + 2m_t^2\right)
B_0(p^2,m_t,m_t) \right] \, , \label{eq:SigmagZ3rd}
\end{align}

\vspace*{-0.5cm}
\begin{align}
  &\Sigma^{\gamma Z}_{T,{\rm electroweak}}(p^2) = \frac{\alpha}{4 \pi s_W c_W}
  \left[\frac{p^2}{9} \left(1-6 c_W^2 \delW\right) + \frac{m_W^2}{3}(12 c_W^2 -
  2 + \delW)\right. \nonumber \\
  &+\left(-\frac{\delW}{6} \frac{\mW^2}{p^2} -\frac{1}{6} \left(1 + c_W^2 (24
  -2 \delW)\right) + \frac{c_W^2 (10-\delW)}{6} \frac{p^2}{\mW^2} \right)
  A_0(m_W) \nonumber \\
  &+\left(\frac{\delW}{6} \frac{\mW^2}{p^2} + \frac{1}{6} \left(5 -2 c_W^2
  \delW \right) -\frac{c_W^2 (10-\delW)}{6} \frac{p^2}{\mW^2}\right) A_0(\xmW)
  \nonumber \\
  &-c_W^2 \left(p^2 - 4 \mW^2\right) \left(1 + \frac{5}{3} \frac{p^2}{\mW^2} +
\frac{1}{12} \frac{p^4}{\mW^4}\right) B_0(p^2,\mW,\mW) \nonumber \\
  &+\left(c_W^2 \left(1 - \frac{p^2}{\mW^2}\right)^2 +
  s_W^2\right)\left(\frac{\delW^2}{6} \frac{m_W^2}{p^2} + \frac{4 + \delW}{3}+
\frac{1}{6} \frac{p^2}{\mW^2}\right) \mW^2 B_0(p^2,\mW,\xmW) \nonumber \\
&\left.-\frac{c_W^2}{12} \frac{p^4}{\mW^4} \left(p^2 -4 \mW^2(1-\delW) \right)
B_0(p^2,\xmW,\xmW) \right] \, .
  \label{eq:SigmagZYM}
\end{align}

\paragraph{The Z-boson self-energy}\mbox{}\\
The fermionic contributions are given by
\begin{align}
  \Sigma^{ZZ}_{T,{\rm  light \, ferm.}}(p^2) =& \frac{\alpha}{4 \pi s_W^2
  c_W^2} \frac{49 -98 c_W^2 +76 c_W^4}{9}
  p^2\left[B_0(p^2,0,0)-\frac{1}{3}\right]\, , \label{eq:SigmaZZLight} \\ 
  \label{eq:SigmaZZ3rd}
  \Sigma^{ZZ}_{T,{\rm 3rd \, gen. \, quarks}}(p^2) =& \frac{\alpha}{4 \pi s_W^2
  c_W^2} \left[\frac{p^2 (1+8c_W^2)}{18} + \frac{17-8c_W^2 (1+4 s_W^2)}{9}
\left(m_t^2 - A_0(m_t) \right)  \right. \nonumber \\
  &+\frac{1+4s_W^2 +8 c_W^4}{18} p^2 \left(B_0(p^2,0,0) - \frac{5}{3}\right)
\\
  &\left.+ \left(\frac{9+8s_W^2 (1-4c_W^2)}{18} (2m_t^2+p^2)  -\frac{3}{2}
  m_t^2\right) B_0(p^2,m_t,m_t) \right] \, .
\nonumber
\end{align}
The $Z$- and $W$-boson self-energies receive contributions from 
tadpole diagrams. In terms of the gauge-invariant parts, the tadpoles belong to the electroweak
contributions and are related by a simple prefactor such that 
\begin{align}
  \Sigma_T^{WW, \,{ \rm tadpole}} &= c_W^2 \Sigma_T^{ZZ, \,{\rm tadpole}}
  \nonumber \\
                                  &=\frac{\alpha}{4 \pi s_W^2} \left[(2
c_W^2 \mW^2 + \mZ^2) \frac{\mZ^2}{\mh^2} - \frac{3}{4} A_0(\mh) + 6
\frac{\mt^2}{\mh^2} A_0(\mt) - 3 \frac{\mW^2}{\mh^2} A_0(\mW)\right. \nonumber
\\ 
&\hspace{2cm}\left.- \frac{3}{2} \frac{\mZ^2}{\mh^2} A_0(\mZ) - \frac{1}{2}
A_0(\xmW) - \frac{1}{4} A_0(\xmZ)\right] \, .
\end{align}
The electroweak part is then given by
\begin{align}
  &\Sigma^{ZZ}_{T,{\rm electroweak}}(p^2) =\frac{\alpha}{4 \pi s_W^2 c_W^2}
  \left[\frac{(s_W^2-c_W^2+6 c_W^4 \delW)p^2}{9} - \frac{\mh^2 +
  \mZ^2}{6}\right. \nonumber \\
  &- \frac{\mW^2}{3}\left(1+2 c_W^2 (4 c_W^2 -2 s_W^2 + \delW)\right)
  \nonumber\\
  &+\left(\frac{1}{6} - \frac{\mh^2 - \mZ^2}{12 p^2}\right) A_0(\mh) -
  \left(\frac{1}{12} - \frac{\mh^2-\mZ^2}{12 p^2}  \right) A_0(\mZ) +
  \frac{1}{4} A_0(\xmZ) \nonumber \\
  &+\frac{1}{6} \left((\delW-10) c_W^4 \frac{p^2}{\mW^2} + c_W^2-s_W^2+ 2c_W^4
(12-\delW) + (c_W^2 -s_W^2) \delW \frac{\mW^2}{p^2} \right) A_0(\mW)\nonumber
\\ &+\frac{1}{6} \left((10-\delW) c_W^4 \frac{p^2}{\mW^2} + (3-10 c_W^2 +2
c_W^4
  \delW) +(s_W^2 - c_W^2) \delW \frac{\mW^2}{p^2} \right) A_0(\xmW) \nonumber
  \\
  &+\left(\frac{1}{12} \frac{p^2}{\mZ^2} + \frac{5}{6} - \frac{\mh^2}{6 \mZ^2}
+ \frac{1}{12} \left(\frac{\mh^2}{\mZ^2}-1\right)^2 \frac{\mZ^2}{p^2} \right)
\mZ^2 B_0(p^2,\mh,\mZ) \nonumber \\
  &-\left(4 +\frac{17}{3} \frac{p^2}{\mW^2} - \frac{4}{3} \frac{p^4}{\mW^4} -
  \frac{1}{12} \frac{p^6}{\mW^6} \right) c_W^4 \mW^2 B_0(p^2,\mW,\mW) \nonumber
  \\
  &+\left(-\frac{c_W^4}{6} \frac{p^6}{\mW^6} - \frac{c_W^4(3+\delW)}{3}
\frac{p^4}{\mW^4}   +\frac{s_W^2 - c_W^2 + c_W^4 (16+4\delW-\delW^2)}{6}
\frac{p^2}{\mW^2} \right. \nonumber \\
  &\quad\left.+ \frac{(s_W^2-c_W^2)(4+\delW) + c_W^4 \delW^2}{3} +\frac{(s_W^2
  - c_W^2) \delW^2}{6} \frac{\mW^2}{p^2} \right) \mW^2 B_0(p^2,\mW,\xmW)
  \nonumber \\
  &\left. -\frac{1}{12} \left(1-\frac{p^4}{\mZ^4}\right)\left(p^2-4 \mW^2
  (1-\delW)\right) B_0(p^2,\xmW,\xmW)  \right]  + \Sigma_T^{ZZ\,{\rm tadpole}}
  \, .
  \label{eq:SigmaZZYM}
\end{align}

\paragraph{The $W$-boson self-energy}\mbox{}\\
The fermionic contributions are gauge-invariant and given by
\begin{align}
  \Sigma^{WW}_{T,{\rm light\, ferm.}}(p^2) &= \frac{\alpha}{4 \pi s_W^2} \, 3
  p^2 \left[ B_0(p^2,0,0) - \frac{1}{3}\right] \,, \label{eq:SigmaWWLight} \\
  \Sigma^{WW}_{T,{\rm 3rd \, gen.\, quarks}} (p^2) &= \frac{\alpha}{4 \pi
  s_W^2} \left[\frac{\mt^2}{2} \frac{\mt^2}{p^2} - \frac{p^2}{3} +
  \left(1-\frac{\mt^2}{2 p^2}\right) \left(\mt^2 - A_0(\mt)\right) \right.
  \nonumber \\
  &\left. \hspace{2cm}+ \left(- \frac{1}{2} \frac{\mt^2}{p^2} - \frac{1}{2 }+
  \frac{p^2 }{\mt^2}\right) \mt^2 B_0(p^2,0,\mt)\right] \,.
  \label{eq:SigmaWW3rd}
\end{align}
The electroweak contribution is 
\begin{align}
  &\Sigma^{WW}_{T,{\rm electroweak}}(p^2) = \frac{\alpha}{4 \pi s_W^2}
  \left[\frac{p^2}{9} \left(-1 + 18 s_W^2 \delA + 3 \delW + 3 c_W^2 \delZ
  \right) \right. \nonumber \\
  &-\frac{\mh^2 + \mZ^2 +18 \mW^2}{6} - \frac{\delW \mW^2 + c_W^2 \delZ
\mW^2}{3} +\left(\frac{1}{6} - \frac{\mh^2 - \mW^2}{12 p^2}\right) A_0(\mh)
\nonumber \\
&+\left(\frac{-10+3s_W^2 \delA + c_W^2 \delZ}{12} \frac{p^2}{\mW^2} +
  \frac{12 - c_W^2 \delZ +9 s_W^2 \delA}{6}\right. \nonumber \\
  &\left.\hspace{2cm}+ \frac{\mh^2 +\mZ^2 (s_W^2-c_W^2) + \mW^2 (c_W^2 \delZ +3
   s_W^2 \delA)}{12 p^2}\right) A_0(\mW) \nonumber \\
  &+\left(\frac{10 -3s_W^2 \delA -c_W^2 \delZ}{12} \frac{p^2}{\mW^2}-
\frac{2-\delW}{6} +\frac{3 s_W^2 \delA \!-2\delW + c_W^2 \delZ }{12}
\frac{\mW^2}{p^2} \right) A_0(\xmW) \nonumber \\
&+\left( \frac{c_W^2(\delW -10)}{12} \frac{p^2}{\mZ^2} + \frac{8
  c_W^4+(17-2\delW)c_W^2 - s_W^2}{12} \right. \nonumber \\
  &\left.\hspace{2cm}+\frac{c_W^4 (1+s_W^2) \delW -s_W^2(s_W^2 - c_W^2-2c_W^2
  (s_W^2-6))}{12} \frac{\mZ^2}{p^2}\right) A_0(\mZ) \nonumber \\
  &+\left( \frac{c_W^2 (10-\delW)}{12} \frac{p^2}{\mZ^2} +
\frac{3-8c_W^4+2c_W^2(\delZ-1)}{12} \right. \nonumber \\
  &\left.\hspace{2cm}+\frac{c_W^2 (2(1-\delZ)+c_W^2(\delW-2))}{12} \frac{\mW^2
  }{p^2} \right) A_0(\xmZ) \nonumber \\
  &+\left( \frac{10-3\delA}{12} \left(\frac{\mW^2}{p^2} +1 + \frac{p^2}{\mW^2}
  + \frac{p^4}{\mW^4}\right) -\frac{22+3\delA}{6}
\,\frac{\mW^2+p^2}{\mW^2}\right) s_W^2 \mW^2 B_0(p^2,0,\mW) \nonumber
\\
&+\left( \frac{10 -3 \delA}{12} \left(\frac{\mW^2}{p^2} + 1 +
  \frac{p^2}{\mW^2}-\frac{p^4}{\mW^4}\right) + \frac{(\delW -2)^2}{6}
\left(1-\frac{\mW^2}{p^2}\right)-\frac{1}{3}\frac{\mW^2 }{p^2} \right.
\nonumber \\
  &\left. \hspace{0.5cm} +\left(\frac{\delA}{2}-\frac{4}{3}\right)
  \frac{p^2}{\mW^2} +\frac{3\delA \delW-4\delW}{12} \left(\frac{\mW^2}{p^2} -
\frac{p^2}{\mW^2}\right) \right) s_W^2 \mW^2 B_0(p^2,0,\xmW) \nonumber \\
  &+\frac{1}{12}\left(\left(\frac{\mh^2}{\mW^2}-1\right)^2 \frac{\mW^2}{p^2} +
  2 \left(5-\frac{\mh^2}{\mW^2}\right)+\frac{p^2}{\mW^2}\right) \mW^2
  B_0(p^2,\mh,\mW) \nonumber \\
  &+\left( \frac{(1+ c_W^2 (10+c_W^2)) s_W^4}{12} \frac{\mZ^2}{p^2} + \frac{2
  (1+c_W^2)}{3} (c_W^4 -4 c_W^2 + s_W^2) \right. \nonumber \\
  &\left.\hspace{2cm}- \frac{9 + c_W^2 (16+9c_W^2)}{6} \frac{p^2}{\mZ^2} +
\frac{2 (1+c_W^2)}{3} \frac{p^4}{\mZ^4} + \frac{1}{12} \frac{p^6}{\mZ^6}
\right) \mZ^2 B_0(p^2,\mW,\mZ) \nonumber \\
&- \frac{c_W^2}{12} \left((c_W^2 + (\delZ-1))^2 \frac{\mZ^2}{p^2} +10 c_W^2
  +2(\delZ-1) + \frac{p^2}{\mZ^2}\right) \nonumber \\
  &\hspace{4cm} \times \left(1-\frac{p^2}{\mW^2}\right)^2  \mW^2
  B_0(p^2,\mW,\xmZ) \nonumber \\
  &- \frac{1}{12} \left((1+c_W^2 (\delW-1))^2 \frac{\mZ^2}{p^2} + 10 + 2 c_W^2
(\delW-1) + \frac{p^2}{\mZ^2}\right) \nonumber \\
  &\hspace{4cm} \times \left(\left(1-\frac{p^2}{\mZ^2}\right)^2 - s_W^4\right)
\mZ^2 B_0(p^2,\mZ,\xmW) \nonumber \\
&+\frac{1}{12} \left(- c_W^4 (c_W^2 (1-\delW) +(\delZ -1))^2
  \frac{\mZ^2}{p^2}-2 c_W^4 (c_W^2 (\delW -1) + \delZ -1 ) \right. \nonumber \\
  & \hspace{2cm}+((1-\delZ)+c_W^2 (\delW -2)) (1-\delZ+c_W^2 \delW)
  \frac{p^2}{\mZ^2} \nonumber \\
  &\left.\left. \hspace{2cm}+2((\delZ-1)+c_W^2 (\delW-1)) \frac{p^4}{\mZ^4} +
  \frac{p^6}{\mZ^6} \right) \mZ^2 B_0(p^2,\xmW,\xmZ) \right] \nonumber\\
  &+ \Sigma_T^{WW,\,{\rm tadpole}} \, .
  \label{eq:SigmaWWYM}
\end{align}


\section{Fourier transform of momentum-space potentials}
\label{app:Fourier}

In this appendix, we provide relevant expressions for the 
Fourier transformation of the potential calculated in momentum 
space to the position-space expression employed in the 
Schr\"odinger equation. Whenever possible we provide the analytic 
results.

The potentials are rotationally invariant. Therefore, we rewrite
the Fourier transform \eqref{eq:FTpotential} as 
\begin{align}
    V(r = |\mathbf{x}|) &= \int \frac{d^3 \mathbf{k}}{(2 \pi)^3}\, e^{i
    \mathbf{k} \cdot \mathbf{x}} \, \tilde{V}(\mathbf{k}) = \frac{1}{2 \pi^2 r
}\int_0^\infty d|\mathbf{k}| \, |\mathbf{k}| \,\sin(|\mathbf{k}| r)\,
\tilde{V}(|\mathbf{k}|) \, .
\label{eq:FTradial}
\end{align}
Performing the $|\mathbf{k}|$-integral analytically for as many terms as
possible is crucial, as numerical instabilities can be avoided this way. For
example, in the transforms of terms arising from Coulomb propagators, the
$1/\mathbf{k}^2$ behaviour for $\mathbf{k}\to 0$ is not always manifest, but
only in the linear combination with other terms.

\subsection{Analytic transforms}

\begin{table}[t]
\begin{center}
{\everymath{\displaystyle}
\renewcommand{\arraystretch}{2}
\begin{tabular}{lc|cl}
$\tilde{V}(\mathbf{k})$ & $\quad$ & $\quad$ & $V(r)$ \\
\hline
$\displaystyle \frac{1}{\mathbf{k}^2}$ & $\quad$ & $\quad$ &
        $\frac{1}{4 \pi r}$ \\[0.2cm]
        $\frac{1}{\left(\mathbf{k}^2\right)^n}$ & $\quad$ & $\quad$ &
        $\frac{\Gamma(2-2n) \sin (n \pi)}{2 \pi^2 r^{3-2 n}} \quad \quad
        \left(n<3/2\right)$ \\[0.2cm]
        $\frac{1}{\mathbf{k}^2 + m^2}$ & $\quad$ & $\quad$ & $\frac{1}{4 \pi r}
        e^{-m\, r}$ \\[0.2cm]
        $\frac{1}{(\mathbf{k}^2 + m^2)^2}$ & $\quad$ & $\quad$ & $\frac{1}{8
        \pi m} e^{-m \, r}$ \\[0.2cm]
        $\frac{1}{(\mathbf{k}^2+m^2)^n}$ & $\quad$ & $\quad$ &
        $\frac{2^{-n-\frac{1}{2}} m^{\frac{3}{2}-n} r^{n-\frac{3}{2}}
            K_{\frac{3}{2}-n}(m r)}{\pi ^{3/2} \Gamma (n)} \quad \quad (n \geq
            1/2)$  \\[0.4cm]
        $\frac{\ln \frac{\mathbf{k}^2 + m_W^2}{ m_W^2}}{\mathbf{k}^2+m_W^2}$ &
        $\quad$ & $\quad$ & $\frac{e^{m_W r} \, \Gamma(0, 2 m_W r)}{4 \pi r}-
        \frac{e^{-m_W r}}{8 \pi r} \ln\left( \frac{m_W^2 r^2
        e^{2\gamma_E}}{4}\right)$ 
\\[0.4cm]
$\frac{\ln \frac{\mathbf{k}^2 + m_W^2}{
        m_W^2}}{(\mathbf{k}^2+m_W^2)^2}$ & $\quad$ & $\quad$ &  $\frac{e^{-m_W
r}}{8 \pi m_W}-\frac{e^{m_W r}\Gamma(0,2 m_W r)}{8 \pi m_W}-\frac{e^{-m_W
r}}{16 \pi m_W}  \ln\left( \frac{m_W^2 r^2 e^{2\gamma_E}}{4}\right) $ \\[0.5cm]
$\frac{1}{\mathbf{k}^2 + m_W^2}
\left(\frac{\mathbf{k}^2}{m_t^2}\right)^{\!\epsilon}\hskip-0.5cm$ & 
$\quad$ & $\quad$
& $\frac{ \sin (\pi
    \epsilon ) \Gamma (2 \epsilon ) \left(m_t r\right)^{-2 \epsilon } \,
_1F_2\left(1;\frac{1}{2}-\epsilon ,1-\epsilon ;\frac{1}{4} r^2 m_W^2\right)}{2
\pi ^2 r}$\\
        $\quad$ & $\quad$ & $\quad$ & $-\frac{\sec (\pi  \epsilon )
         \left(\frac{m_W^2}{m_t^2}\right)^{\epsilon }
     \sinh \left(r m_W\right)}{4 \pi  r}$\\[0.2cm] 
        $\frac{\ln \frac{\mathbf{k}^2 + m_t^2}{m_t^2}}{\mathbf{k}^2 + m_W^2}$
        & $\quad$ & $\quad$ & $-\frac{1}{4 \pi r}
        \left(\vphantom{\frac{a^2}{b^2}} e^{m_W r} \text{Ei}(-(m_t + m_W)r)+
        e^{-m_W r} \text{Ei}((m_W-m_t)r) \right.$\\
        $\quad$ & $\quad$ & $\quad$  &\hspace{2cm} $\left.+ e^{-m_W r} \ln
        \frac{m_t^2}{m_t^2-m_W^2}\right) \quad \quad (m_t >m_W)$ 
\end{tabular}
}
\caption{Table of Fourier transforms. $K_n$ denotes the 
modified Bessel function of the second
kind, $\mathrm{Ei}$ the exponential integral function, ${}_nF_m$ the
generalized hypergeometric function.  \label{tab:FTs}}
\end{center}
\end{table}

In Table~\ref{tab:FTs} we collect Fourier transforms that are 
helpful for the NLO potential.  All NLO potential terms except 
logarithms involving the triangle function (these will be discussed in Section~\ref{app:NumTrafo}) can be transformed using these results. 
In some cases partial-fractioning identities such as
\begin{align}
    \frac{1}{A B} = \frac{1}{B-A}\left(\frac{1}{A}-\frac{1}{B}\right) 
\end{align}
are necessary to bring the expressions into manageable forms. 
Some logarithmic terms can be obtained using
\begin{align}
    \frac{1}{(\mathbf{k}^2+m^2)^n}
    \left(\frac{\mathbf{k}^2+m^2}{m^2}\right)^{\!\epsilon} =
    \frac{1}{(\mathbf{k}^2+m^2)^n} + \epsilon\ 
    \frac{\ln\left(\frac{\mathbf{k}^2+m^2}{m^2}\right)}{(\mathbf{k}^2+m^2)^n} +
    \mathcal{O}(\epsilon^2) \, . 
\end{align}
Furthermore, some Fourier transforms are related by taking derivatives, e.g.,
\begin{align}
    \frac{1}{(\mathbf{k}^2 + m_W^2)^2} = - \frac{\partial}{\partial m_W^2} \frac{1}{\mathbf{k}^2 + m_W^2} \, .
\end{align}
Lastly, the derivative of the Bessel $K_\nu$ function for
$\nu=\pm 1/2$ is (the case relevant here) is
\begin{align}
  \left. \frac{\partial K_\nu(z)}{\partial \nu}\right|_{\nu=\pm1/2} = \pm\sqrt{
      \frac{\pi}{2 z}} \, \Gamma(0,2 z) \, e^z \,, \quad \text{where} \quad
      \Gamma(s,z)= \int_z^\infty d t \, t^{s-1} e^{-t}
\end{align}
is the incomplete Gamma function. 

As an example, let us apply the above tricks to 
\eqref{eq:momasy00pmlightferm}. Using the second to last result of Table~\ref{tab:FTs} and relations among
derivatives of special functions, the $\mathcal{O}(\epsilon)$ term evaluates to
\begin{align}
    \frac{\ln \frac{\mathbf{k}^2}{m_W^2}}{\mathbf{k}^2 + \mW^2} \to -\frac{1}{4
    \pi r} \left[e^{m_W r} \mathrm{Ei}(-m_W r) + e^{-m_W r} \mathrm{Ei}(m_W r) \right]
\label{eq:FTmainexample}
\end{align}
For $r \to \infty$ the right hand side of \eqref{eq:FTmainexample} has the
behaviour
\begin{align}
    - \frac{1}{2 \pi m_W^2} \left(\frac{1}{r^3} + \frac{6}{\mW^2 r^5} +
    \mathcal{O}(r^{-7})\right) \, .
\end{align}
In \eqref{eq:momasy00pmlightferm}, we have the linear combination
\begin{align}
    \ln \frac{\mathbf{k}^2}{\mW^2}  \left[ \frac{\mW^2}{(\mathbf{k}^2 + \mW^2)^2} - \frac{1}{\mathbf{k}^2 + \mW^2}\right] = -\ln \frac{\mathbf{k}^2}{\mW^2} \frac{\partial}{\partial m_W^2} \frac{\mW^2}{\mathbf{k}^2 + \mW^2}
    \label{eq:FTmainexample2}
\end{align}
which explains why the asymptotics is proportional to $r^{-5}$. In the case
of the $\overline{\rm MS}$-scheme in momentum space (see
footnote~\ref{fn:asyMSbar}) the second term on the left-hand side of
\eqref{eq:FTmainexample2} is not present. Therefore in this case the asymptotic behaviour goes as $r^{-3}$
(cf.  \eqref{eq:MSbarasy0pm}).

\subsection{Numerical transforms}
\label{app:NumTrafo}

For certain terms involving the triangle function $\lambda(a,b,c)$, 
we were not able to perform the Fourier transformation analytically. 
In the following, we discuss an example of 
how we obtain the numerical transform in such cases. 

The example arises from the self-energies, namely from the Higgs-$Z$ loops, and
is given by
\begin{align}
    \frac{\lambda^{1/2}(-\mathbf{k}^2,m_H^2,m_Z^2)}{\mathbf{k}^2  (\mathbf{k}^2+m_Z^2)}\, \ln \left[\frac{\mathbf{k}^2 + m_H^2 + m_Z^2 +\lambda^{1/2}(-\mathbf{k}^2,m_H^2,m_Z^2) }{2 m_H m_Z}\right] \, .
  \label{eq:FTexample}
\end{align}
The numerical Fourier transform of this function is not always 
stable, depending on the value of $r$ in~\eqref{eq:FTradial} that 
determines the scales probed in the integrand. The leading 
behaviour for $\mathbf{k} \to 0$ is $1/\mathbf{k}^2$. In the final 
expression, however, it may happen that the corresponding $1/r$ 
behaviour in position space cancels against another term
in the complete expression for the loop 
integral. The subleading terms are exponentially
suppressed for large~$r$, which is hard to resolve numerically.  
We solve this issue by undoing the Feynman-parameter 
integration that led to the logarithm, writing 
\begin{align}
    &\frac{\lambda^{1/2}(-\mathbf{k}^2,m_H^2,m_Z^2)}{\mathbf{k}^2  (\mathbf{k}^2+m_Z^2)}\, \ln \left[\frac{\mathbf{k}^2 + m_H^2 + m_Z^2 +\lambda^{1/2}(-\mathbf{k}^2,m_H^2,m_Z^2) }{2 m_H m_Z}\right] \, \nonumber \\ 
    &\quad \quad= \frac{\lambda(-\mathbf{k}^2,m_H^2,\mZ^2)}{\mathbf{k}^2
    (\mathbf{k}^2 + \mZ^2)} \int_0^1 dx \, \frac{1}{2
(m_Z^2+(\mathbf{k}^2+m_H^2-m_Z^2) x - \mathbf{k}^2 x^2)}\, .
  \label{eq:FTFeyn}
\end{align}
We then subtract the Coulombic behavior in the limit $\mathbf{k} 
\to 0$ by adding the term\footnote{In some cases such a subtraction can even be necessary, e.g., for terms that
behave as $1/\mathbf{k}^4$. Only the full NLO potential is guaranteed to have a 
$\mathbf{k} \to 0$ behaviour which is not more singular than 
$(\ln \mathbf{k}^2)
/\mathbf{k}^2$.  Sometimes, the integrand of the Feynman-parameter
representation is more singular than $1/\mathbf{k}^2$ even though 
the integral is not, and to perform the
analytic transform the subtraction of an expression that vanishes after
integration over Feynman parameters is necessary.}
\begin{align}
    -\frac{m_H^2 - m_Z^2}{2 m_Z^2 \mathbf{k}^2} \, \ln \frac{m_H^2}{m_Z^2} \, .
  \label{eq:FTsubt}
\end{align}
The Fourier integral over $|\mathbf{k}|$ can now be performed 
analytically, resulting in 
\begin{align}
\int_0^1 dx\,  \frac{1}{8 \pi r (m_Z^2 (1-x)^2 + m_H^2 x)} &\left[e^{-m_Z r} \frac{m_H^2 (4
  m_Z^2 - m_H^2)}{m_Z^2} \right. \nonumber \\
&\left.\hspace*{-2cm} + \,e^{-r \sqrt{\frac{m_Z^2}{x}+\frac{m_H^2}{1-x}}} \frac{\left(m_Z^2
(1-x)^2 -m_H^2 x^2\right)^2}{(1-x) x (m_Z^2 (1-x) + m_H^2 x)}\right] \, .
\end{align}
The above expression looks complicated, however, the numerical integration
that has to be performed is now only an integral from $x=0$ to $1$ instead of an integral from $|\mathbf{k}|=0$ to $\infty$. This stabilizes the large-$r$ tail as the exponential suppression of the final result is already captured
in the integrand before the Feynman-parameter integration, and the numerical
difficulty of accurately determining the exponential tail is thus circumvented.


\section{Expressions for the asymptotic behaviour}
\label{app:asy}

In Sec.~\ref{sec:asymptotics} we discussed the asymptotic behaviour of the NLO correction to the potential in the limits $r \to 0$ and $r \to \infty$. In this appendix, we provide the mass-dependent functions appearing in the main text. The  $\arctan$ terms in the 
expressions below stem from simplifying the real parts of 
self-energies. 

\subsection{The $r \to 0$ asymptotics}

\begin{align}
  A(m_W, m_Z,m_t) =&-\frac{80 s_W^2}{27} + \frac{(64 c_W^2 -16)}{9}
  \frac{m_t^2}{m_Z^2}-\frac{1}{2 c_W^2 s_W^2} \frac{m_t^4}{m_Z^4} - \frac{3+ 2
  s_W^2}{6 s_W^2} \ln \frac{m_t^2}{m_Z^2}  \nonumber \\
                    &+\left(\frac{c_W^2}{s_W^2}-\frac{m_t^2}{2 s_W^2 m_Z^2}
\left(3-\frac{m_t^4}{m_W^4}\right)\right) \ln \frac{m_t^2}{m_t^2-m_W^2}
\nonumber \\
                    &-\sqrt{\frac{4 m_t^2}{m_Z^2} -1} \  \arctan\left[
                    \frac{\sqrt{4 m_t^2 - m_Z^2} m_Z}{2 m_t^2 - m_Z^2}\right]
                    \nonumber \\
                    &\quad\times\left(\frac{4}{3} - \frac{1}{2 s_W^2} -
                  \frac{16 s_W^2}{9} + \left(\frac{8}{3} + \frac{1}{2 s_W^2}-
                \frac{32 s_W^2}{9}\right)\frac{m_t^2}{m_Z^2}\right)
\end{align}

\begin{align}
  B(m_W,m_Z,m_H)=&\, \frac{125}{6}-\frac{1}{12 c_W^2} -\frac{80 s_W^2}{3}- 8
  c_W^2 s_W^2- \frac{m_H^4}{12 m_W^2 m_Z^2}\nonumber \\
  &+\left(-\frac{s_W^2}{c_W^2}+\frac{1-c_W^4}{6c_W^4}
\frac{m_H^2}{m_Z^2}\right) \frac{m_H^4}{4 s_W^2 m_Z^4} \ln \frac{m_H^2}{m_Z^2}
\nonumber \\
                 &+\left(-\frac{1+11 c_W^2}{2 c_W^2}+\frac{13+10c_W^2}{6 s_W^2}
                 - \frac{1+2 c_W^2}{24 c_W^4 s_W^2} \right. \nonumber \\
                 &\quad\quad \left.-\frac{3 m_H^2}{4 m_Z^2 s_W^2}
               +\frac{m_H^4}{4 m_W^2 m_Z^2 s_W^2} - \frac{m_H^6}{24 m_W^4 m_Z^2
             s_W^2}\right)\ln \frac{m_W^2}{m_Z^2} \nonumber \\
             &+\sqrt{\frac{4 m_W^2}{m_Z^2}-1} \arctan \left[\frac{\sqrt{\frac{4
       m_W^2}{m_Z^2} -1}}{\frac{2 m_W^2}{m_Z^2} -1}\,\right]\!
       \left(-\frac{1}{12}-\frac{29 c_W^2}{3} -4 c_W^4 +
       \frac{33 c_W^2}{4 s_W^2}\right) \nonumber \\
       &+\sqrt{\frac{4 m_W^2}{m_Z^2} -1} \arctan \left[\sqrt{\frac{4
 m_W^2}{m_Z^2} -1}\, \right]\! \left(4+\frac{4}{3 c_W^2} - \frac{25}{3 s_W^2} +
 \frac{1}{12 c_W^4 s_W^2}\right) \nonumber \\
                 &-\sqrt{\frac{4 m_Z^2}{m_H^2}-1} \arctan\left[ \sqrt{\frac{4
 m_Z^2}{m_H^2}-1}\, \right]\! \left(\frac{m_H^2}{m_Z^2 s_W^2} - \frac{m_H^4}{3
 m_Z^4 s_W^2} + \frac{m_H^6}{12 m_Z^6 s_W^2}\right) \nonumber \\
                 &+\sqrt{\frac{4 m_W^2}{m_H^2}-1} \arctan\left[ \sqrt{\frac{4
                 m_W^2}{m_H^2}-1}\,\right] \nonumber \\ 
                 &\quad\times\left(\frac{m_H^2}{m_Z^2 s_W^2} - \frac{m_H^4}{3
             m_W^2 m_Z^2 s_W^2} + \frac{m_H^6}{12 m_W^4 m_Z^2 s_W^2}\right)
\end{align}

\subsection{The $r\to \infty$ asymptotics}

\begin{align}
  C&(m_W,m_Z,m_H)= \frac{109}{8} - \frac{299 s_W^2}{72} - \frac{3029 s_W^4}{90}
  + \frac{104 s_W^6}{3} - 8 s_W^8 - \frac{3 c_W^2 m_H^2}{8 m_Z^2}- \frac{s_W^2
  m_H^4}{12 m_Z^4} \nonumber \\
                 &+ \ln \frac{m_H^2}{m_Z^2} \frac{c_W^2 m_H^2}{m_H^2-m_Z^2}
  \left(\frac{3}{2} + \frac{(s_W^2 - 3 c_W^2)m_H^2 }{4 m_W^2}- \frac{(1+7 c_W^2
  s_W^2) m_H^4}{24 m_W^4} +\frac{c_W^2 (1+ c_W^2 s_W^2) m_H^6}{24 m_W^6}\right)
  \nonumber \\
                 &-\frac{\ln \frac{m_W^2}{m_Z^2}}{24 c_W^2 s_W^2}
  \left(1+14c_W^2 -106 c_W^4+2 c_W^6+18 c_W^4 \frac{m_H^2}{m_Z^2} -6 c_W^2
  \frac{m_H^4}{m_Z^4}+\frac{m_H^6}{m_Z^6}  \right) \nonumber \\
  &+\sqrt{\frac{4 m_W^2}{m_Z^2} -1} \arctan \sqrt{\frac{4 m_W^2}{m_Z^2} -1}
  \left(\frac{4}{3}+ 4 c_W^2 +\frac{1}{12 c_W^2 s_W^2}-\frac{25
  c_W^2}{3s_W^2}\right) \nonumber \\
  &+\sqrt{\frac{4 m_W^2}{m_Z^2} -1} \arctan\left[\frac{\sqrt{\frac{4
  m_W^2}{m_Z^2} -1}}{\frac{2 m_W^2}{m_Z^2} -1}\,\right] c_W^2
  \left(-\frac{55}{4}+\frac{33}{4 s_W^2} - \frac{13 s_W^2}{3} + \frac{41
  s_W^4}{3} - 4 s_W^6\right) \nonumber \\
                 &+\sqrt{\frac{4 m_W^2}{m_H^2}-1} \frac{c_W^2 m_H^2}{s_W^2
             m_Z^2} \arctan\left[ \sqrt{\frac{4 m_W^2}{m_H^2}-1}\,\right]
             \left(1-\frac{m_H^2}{3 m_W^2} + \frac{m_H^4}{12 m_W^4}\right)
             \nonumber \\
                 &-\sqrt{\frac{4 m_Z^2}{m_H^2}-1} \frac{m_H^2}{m_Z^2}
             \frac{c_W^2+s_W^2 c_W^2}{s_W^2} \arctan\left[ \sqrt{\frac{4
             m_Z^2}{m_H^2}-1}\,\right] \left(1-\frac{m_H^2}{3 m_Z^2} +
           \frac{m_H^4}{12 m_Z^4}\right)
\nonumber
\end{align}

\begin{align}
  D&(m_W,m_Z,m_t) = \frac{80 s_W^2}{27} + \left(\frac{16}{9}-\frac{64
  c_W^2}{9}-\frac{1}{4 c_W^2} \right) \frac{m_t^2}{m_Z^2} + \frac{c_W^2 -
s_W^2}{2 s_W^2} \frac{m_t^4}{m_W^4} \nonumber \\
   &+\frac{c_W^2-s_W^2}{s_W^2} \ln \frac{m_t^2}{m_t^2 - m_W^2} \left(-1 +
   \frac{3 m_t^2}{2 m_W^2} - \frac{m_t^6}{2 m_W^6}\right) + \frac{3 c_W^2 -
 s_W^2}{6 s_W^2}\ln \frac{m_t^2 }{m_Z^2}  \nonumber \\
   &+\sqrt{\frac{4 m_t^2}{m_Z^2}-1} \arctan \left[\frac{\sqrt{\frac{4
   m_t^2}{m_Z^2}-1}}{\frac{2 m_t^2}{m_Z^2}-1}\,\right] \nonumber \\
   &\quad\quad\times\left(\frac{16 c_W^2}{9} - \frac{17}{18 s_W^2}+ \frac{4
   c_W^2}{9 s_W^2} + \left(\frac{32 c_W^2}{9} - \frac{7}{18 s_W^2} + \frac{8
   c_W^2}{9 s_W^2}\right) \frac{m_t^2}{m_Z^2}\right)
\nonumber
\end{align}

\begin{align}
  E&(m_W,m_Z,m_H) = -\frac{145}{12}+ \frac{2+29c_W^2}{24 c_W^4}+\frac{56
  c_W^2}{3} + 8 c_W^4-\frac{3}{8} \frac{m_H^2}{m_W^2} + \frac{s_W^2}{12}
  \frac{m_H^4}{m_W^4} \nonumber \\
   &+\ln \frac{m_H^2}{m_Z^2} \, \frac{m_H^2}{m_H^2 - m_W^2} \left(\frac{3}{2} +
\frac{c_W^2-4}{4} \frac{m_H^2}{m_W^2}+ \frac{7 s_W^2 -c_W^4}{24}
\frac{m_H^4}{m_W^4} + \frac{c_W^4 - s_W^2}{24} \frac{m_H^6}{m_W^6}\right)
\nonumber \\
   &+\ln \frac{m_W^2}{m_Z^2}\left(\frac{65}{12} \frac{c_W^2 -s_W^2}{c_W^2
s_W^2}+ \frac{1}{24 s_W^2 c_W^6} + \frac{1}{2 s_W^2 c_W^4}-\frac{3}{4}+
\frac{3}{4} \frac{s_W^2 - c_W^2}{s_W^2} \frac{m_H^2}{m_W^2}\right.
\nonumber \\
   &\hspace{2cm}\left.+ \frac{c_W^2 -s_W^2}{4 s_W^2} \frac{m_H^4
   }{m_W^4}+ \frac{s_W^2-c_W^2}{24 s_W^2} \frac{m_H^6}{m_W^6} - \frac{3}{4}
   \frac{m_W^2}{m_H^2 - m_W^2} \right) \nonumber \\
   &+ \sqrt{\frac{4 m_W^2}{m_H^2}-1} \arctan \left[\sqrt{\frac{4
   m_W^2}{m_H^2}-1}\,\right] \frac{c_W^2 - s_W^2}{s_W^2}\left(
 \frac{m_H^2}{m_W^2}-\frac{1}{3}\frac{m_H^4}{m_W^4} + \frac{1}{12}
 \frac{m_H^6}{m_W^6}\right) \nonumber \\
   &-\sqrt{\frac{4 m_Z^2}{m_H^2}-1} \arctan \left[\sqrt{\frac{4
   m_Z^2}{m_H^2}-1}\,\right] \frac{1}{s_W^2}
   \left(\frac{m_H^2}{m_Z^2}-\frac{1}{3} \frac{m_H^4}{m_Z^4}+ \frac{1}{12}
   \frac{m_H^6}{m_Z^6}\right) \nonumber \\
   &+\sqrt{\frac{4 m_W^2}{m_Z^2} -1} \arctan \left[\sqrt{\frac{4 m_W^2}{m_Z^2}
   -1}\, \right] \frac{c_W^2 - s_W^2}{c_W^2} \left(\frac{4-17 c_W^2}{3 c_W^2
   s_W^2} - \frac{4 c_W^2}{s_W^2} + \frac{1}{12 c_W^4 s_W^2} \right) \nonumber
   \\
 &+\sqrt{\frac{4 m_W^2}{m_Z^2} -1} \arctan \left[\frac{\sqrt{
 \frac{4 m_W^2}{m_Z^2} -1}}{\frac{2 m_W^2}{m_Z^2} -1}\,\right]
 \left(\frac{33}{4 s_W^2}-22 + \frac{53 s_W^2}{3}-4 s_W^4 \right)
\end{align}


\input{paper.bbl}


\end{document}

%% file: paper.bbl
\providecommand{\href}[2]{#2}\begingroup\raggedright\endgroup